\documentclass[submission, Phys]{SciPost}

\pdfoutput=1
\usepackage{euscript,amsmath,amsthm,amsfonts,amssymb}
\usepackage{mathtools}
\usepackage{float}
\usepackage{outlines}
\usepackage{adjustbox}
\usepackage{longtable}

\usepackage{makecell}
\usepackage[table]{colortbl}
\usepackage{color}

\usepackage{tabularx}

\newcommand{\be}{\begin{equation}}
\newcommand{\ee}{\end{equation}}
\newcommand{\bea}{\begin{eqnarray}}
\newcommand{\eea}{\end{eqnarray}}

\newcommand{\de}{\partial}
\newcommand{\ba}{\begin{eqnarray}}
\newcommand{\ea}{\end{eqnarray}}

\newcommand{\f}{\frac}
\newcommand{\s}{\sqrt}

\newcommand{\ti}{\tilde}

\newcommand{\no}{\nonumber \\}
\newcommand{\la}{\langle}
\newcommand{\lb}{\rangle}
\newcommand{\ep}{\epsilon}

\newcommand*\pFq[2]{{}_{2}F_{1}(#1;#2)}

\DeclareMathOperator{\area}{area}

\DeclareMathOperator{\Tr}{Tr}

\hypersetup{
    colorlinks,
    linkcolor={red!50!black},
    citecolor={blue!50!black},
    urlcolor={blue!80!black}
}

\usepackage[bitstream-charter]{mathdesign}
\usepackage{orcidlink}
\usepackage{comment}

\DeclareSymbolFont{usualmathcal}{OMS}{cmsy}{m}{n}
\DeclareSymbolFontAlphabet{\mathcal}{usualmathcal}

\begin{document}

\begin{center}{\Large \textbf{ Multi-entropy at low Renyi index in 2d CFTs\\
}}\end{center}

\begin{center}
Jonathan Harper\,\orcidlink{0000-0003-4913-6470}\textsuperscript{1},
Tadashi Takayanagi\, \orcidlink{0000-0001-7187-3130}\textsuperscript{1,2} and
Takashi Tsuda\textsuperscript{1}
\end{center}

\begin{center}
{\bf 1} Center for Gravitational Physics and Quantum Information,\\ Yukawa Institute for Theoretical Physics, Kyoto University,\\ Kitashirakawa Oiwakecho, Sakyo-ku, Kyoto 606-8502, Japan
\\
{\bf 2} Inamori Research Institute for Science,\\
620 Suiginya-cho, Shimogyo-ku,
Kyoto 600-8411, Japan
\\
Email:  {\small \sf \href{mailto:jonathan.harper@yukawa.kyoto-u.ac.jp}{jonathan.harper@yukawa.kyoto-u.ac.jp}, \href{mailto:takayana@yukawa.kyoto-u.ac.jp}{takayana@yukawa.kyoto-u.ac.jp}, \href{mailto:takashi.tsuda@yukawa.kyoto-u.ac.jp}{takashi.tsuda@yukawa.kyoto-u.ac.jp} }
\end{center}

\section*{Abstract}
{ For a static time slice of AdS$_3$ we describe a particular class of minimal surfaces which form trivalent networks of geodesics. Through geometric arguments we provide evidence that these surfaces describe a measure of multipartite entanglement. By relating these surfaces to Ryu-Takayanagi surfaces it can be shown that this multipartite contribution is related to the angles of intersection of the bulk geodesics. 

A proposed boundary dual \cite{2022PhRvD.106l6001G,2023JHEP...08..202G,2022arXiv221116045P}, the multi-entropy, generalizes replica trick calculations involving twist operators by considering monodromies with finite group symmetry beyond the cyclic group used for the computation of entanglement entropy. We make progress by providing explicit calculations of Renyi multi-entropy in two dimensional CFTs and geometric descriptions of the replica surfaces for several cases with low genus. 

We also explore aspects of the free fermion and free scalar CFTs. For the free fermion CFT we examine subtleties in the definition of the twist operators used for the calculation of Renyi multi-entropy. In particular the standard bosonization procedure used for the calculation of the usual entanglement entropy fails and a different treatment is required.}

\vspace{10pt}
\noindent\rule{\textwidth}{1pt}
\tableofcontents\thispagestyle{fancy}
\noindent\rule{\textwidth}{1pt}
\vspace{10pt}

\section{Introduction}
\label{sec:intro}

The AdS/CFT correspondence has been one of our most successful tools for improving our understanding of quantum gravity \cite{Maldacena:1997re,Gubser:1998bc,Witten:1998qj}. A particular hallmark of this duality is the connection between spacetime geometry and measures of entanglement. More specifically the entanglement entropy, a measure of bipartite entanglement is dual to Ryu-Takayanagi (RT) surfaces: bulk minimal area surfaces which are homologous to the boundary region \cite{Ryu:2006bv,Ryu:2006ef,Hubeny:2007xt}. This provides a notion of entanglement between different regions of the boundary theory ``building up" the geometry of the bulk spacetime.

However, by definition the entanglement entropy is only able to diagnosis bipartite correlations and one may wonder if it is possible to generalize and identify measure of multipartite entanglement between multiple boundary regions which are dual to specific classes of optimal surfaces in the bulk spacetime. 

In this paper we will be focused on one such class of surfaces which are defined by partitioning the entire boundary into $q$ regions we then look for the collection of bulk surfaces with smallest total area which partition the \emph{entire} bulk in $q$ regions each of which is homologous to one of the boundary regions. Notably compared to RT surfaces this allows for the possibility of bulk intersections of surfaces. For AdS$_3$ where constant time slices are the hyperbolic disk the solution comes in the form of ``Steiner trees" which are networks of geodesics which meet at equiangular trivalent vertices.

The authors of \cite{2022PhRvD.106l6001G,2023JHEP...08..202G,2022arXiv221116045P} have proposed a dual boundary quantity which has been named the ``multi-entropy". This is calculated by considering particular $q$-point functions of twist operators. For the usual entanglement entropy, which corresponds to $q=2$, one considers a two-point function where the monodromies of the twist operators are chosen from the cyclic group $\mathbb{Z}_n$ and act to cyclically permute one through the $n$ copies. Via the Euclidean path integral this is equivalent to the trace of $n$ copies of the reduced density matrix \cite{Holzhey:1994we,Calabrese:2004eu,Calabrese:2009qy}. This can be viewed as a particular set of ``contractions" between the reduced density matrices which are determined by the monodromies of the twist operators.

The multi-entropy generalizes this procedure by considering twist operators with \emph{other} finite group monodromies. Specifically for the $q$-party multi-entropy one chooses the Abelian group $\mathbb{Z}_n^{q-1}$. The reason for this is two fold: Within this group it is possible to find non-zero $q$-point functions of twist operator for which the cycle structure of the mondromies of the operators is the same. This is in turn guarantees that the conformal dimension of the operators is the same. While not necessary it is reasonable to expect that if one is interested in a quantity that is sensitive to $q$-party entanglement that this would be enhanced by considering a quantity symmetric with respect to all parties. Secondly, in the $n\rightarrow1$ replica limit the operators have the right scaling to be identified with tension-less surfaces in the bulk geometry.

Just like the entanglement entropy the path integral on Riemann surfaces can be used to relate the monodromies of the chosen twist operators to contractions of copies of reduced density matrices. In turn this defines a information quantity which can in practice be calculated for any quantum state. 

The purpose of this paper is further this proposal by providing additional evidence for the duality. Specifically we focus on the case of low Renyi which is amenable to direct calculation. This is accomplished by utilizing the uniformization method in which one constructs a replica manifold of higher genus by gluing copies of the CFT together according to the monodromies of the twist operators. Since the replica manifold has no operator insertions one can calculate the partition function (which is theory dependent) and then use properties of the uniformization map from the replica surface back to the single copy of the theory to determine the desired $q$-point function of twist operators. While the construction of the replica surface can be done for any choice of operators. In practice this method is difficult to use when the genus of the replica surface is $g\geq 2$. This is because both the partition function and uniformization map are generally unknown. For the multi-entropy unfortunately the genus of the replica surface increases with Renyi index so we focus specifically on several case with low Renyi index where we have the means to fully complete the calculation. Of course one would desire more robust methods that would allow for the full calculation regardless of the choice twist operators and the resulting genus of the replica manifold.

Beyond holographic theories it is also interesting to study the multi-entropy for other CFTs. As a first example we consider the free fermion CFT. In this theory the entanglement entropy can be directly calculated via bosonization \cite{Casini:2005rm,Azeyanagi:2007bj,Takayanagi:2010wp}. Regrettably, we find that this method can not be extended directly for the twist operators used for the calculation of multi-entropy. Our calculations of Renyi multi-entropy do however provide an alternative means of generating the correct twist operators however as we discuss they are unusual in that they break the symmetry originally present in the original twist operators.

We also explore the multi-entropy for thermal states in the free fermion CFT.
The result have interesting implications for the multi-entropy in holographic CFT thermal setups, particularly regarding the shape of the corresponding RT surface.

Additionally, we looked into Renyi multi-entropy in the context of a free scalar CFT with local excitations. The calculation accurately matches the expected results from computations in discrete systems. This agreement confirms the effectiveness of our uniformization method in constructing a replica manifold.

The structure of the rest of the paper is as follows: In section \ref{sec:review} we review the definition of the multi-entropy and proposed holographic dual. Then in section \ref{sec:reny} we proceed with the explicit calculation of the Renyi multi-entropy for several cases. 
Next in section \ref{sec:FF} we discuss the free fermion CFT. We start by considering the vacuum state, proceed to thermal setups, and finally compare the results to the holographic calculation. 
In section \ref{sec:LQ} we discuss Renyi multi-entropy in a free scalar CFT with local excitations.
Finally we conclude in section \ref{sec:con}. The appendices \ref{sec:failure} and \ref{detailLQ} contain additional details of the free theory calculations. Specifically the failure of the bosonization of twist operators and local operator excitation respectively.

\section{Review of multi-entropy}\label{sec:review}

\subsection{Multi-trace measures}
In this paper we will be interested in a particular class of \emph{multi-trace measures} which can be defined in terms of partial traces of copies of density matrices with respect to permutations from a choice of finite group symmetry.

To illustrate this we first consider a particularly well studied example: the entanglement entropy. Given a pure state $\rho=|\psi\rangle\langle\psi|$ we partition the system into two regions $A$ and its purifier $O=A^c$. Tracing out $O$ the entanglement entropy is given by
\be\label{eq:EE}
S=-\Tr\rho_{A}\log(\rho_A)
\ee
and its renyi counterpart
\be
S_{n}=\frac{1}{1-n}\log(\Tr\rho_A^n).
\ee
This can be re-written in terms of a multi-trace as follows: The density matrix on $q$ parties can be viewed a rank $2q$ tensor where each bra and ket corresponds to a different index. In this case we have
\be
\rho_{a'o'}^{ao}=|A\rangle| O \rangle\otimes\langle A|\langle O|
\ee
where we have organized the indices so that each column corresponds to a different party and the kets (bras) are the lower (upper) indices.
In this language the reduced density matrix is given by the contraction of the bra and ket of the same party
\be
\rho_A=Tr_{O}\rho=\rho_{ai}^{a'i}.
\ee
In the case of the Reyni-4 entropy we need to take a trace of four copies of the reduced density matrix. Considering the cyclic group of order 4
\be
\mathbb{Z}_4:\langle a|
a^4=e\rangle
\ee
we make a choice for each region a group element and consider its permutation representation. Here we make the choice
\be
\tau_O=e, \quad (1)(2)(3)(4); \quad \tau_A=a^3, \quad (1,4,3,2)
\ee
and then perform the contractions of the tensor indices on each party according to the corresponding $\tau$:
\be
-3S_4=MTr(\rho^4)_{\tau_A,\tau_O}=Tr(\rho^4_A)=\rho^{\delta i}_{\alpha i}\rho^{\gamma j}_{\delta j}\rho^{\beta k}_{\gamma k}\rho^{\alpha l}_{\beta l}
\ee

As another example we can consider a three party state
\be
\rho_{a'b'o'}^{abo}=|A\rangle|B\rangle| O \rangle\langle A|\langle B|\langle O|
\ee
and the direct product of two cyclic groups of order 2
\be
\mathbb{Z}_2^2:\langle a,b|a^2=b^2=e\rangle.
\ee
Making the choice
\be
\tau_O=e, \quad (1)(2)(3)(4);\quad \tau_A=a, \quad (1,2)(3,4); \quad \tau_B=b, \quad (1,4)(2,3)
\ee
corresponds to the measure
\be
-2S_2^{(3)}=MTr(\rho^4)_{\tau_A,\tau_B,\tau_O}=\rho^{\beta\nu i}_{\alpha \mu i}\rho^{\alpha \theta j}_{\beta\phi j}\rho^{\delta \mu k}_{\gamma \nu k}\rho^{\gamma \phi l}_{\delta \theta l}.
\ee
This is the renyi-2 \emph{multi-entropy} which will discussed in more detail later.

 A general multi-trace measure up to normalization for a group $G$ on $q$ parties $\{A_1,\cdots A_q\}$ is specified by the choice of $q$ $\tau$s:
\be
\log(MTr(\rho^n)_{\{\tau_i\}}), \quad \tau_{A_1},\cdots,\tau_{A_q}\in G, \quad |G|=n
\ee
In our definition we have chosen to work with identical copies of the density matrix $\rho$. More generally one could consider measures defined as multi-traces of different states potentially with different numbers of parties. Furthermore we will be interested in those quantities for which each copy $\rho$ is treated symmetrically. These restrictions place a few natural constraints on the choice of $\tau s$\footnote{These constraints have a more clear interpretation in the context of the 2d CFT. Each density matrix is a copy of the theory which in order to construct the replica manifold is conformally mapped to a polygonal region. Together these constraints imply that each copy is mapped using the same conformal map; that is to identical polygonal regions each with the same number and length of sides and measure of angles.}:
\begin{itemize}
\item For a measure on $n$ copies of the density matrix the order of the symmetry group must be $n$ and the $\tau$s should be associated with permutation representations of the regular action on $n$ points.

\item Furthermore, for each copy to be treated the same it is necessary that for a given $\tau$ the length of all cycles in the permutation representation must be the same. This immediately implies the relation $n=lm$ where $l$ is the length of the cycles and $m$ is the number of cycles.
\end{itemize}

A few comments (for more details and information see \cite{2022arXiv221116045P,2022PhRvD.106l6001G}):
\begin{itemize}

\item For a choice of $\tau$s a new equivalent set of $\tau$s can be constructed by acting by either left multiplication or conjugation by a group element. Using this freedom it is always possible to choose one of the $\tau$s to be the identity. This amounts to instead working with the reduced density matrix on $q-1$ parties. 

\item Besides this freedom different choices of $\tau$s will generally correspond to different multi-trace measures.
\end{itemize}

\subsection{Multi-entropy in the boundary theory}

In this paper we will be interested in the determination of multi-trace measures in 2d CFTs which can be accomplished through the calculation of correlation functions of twist operators. We consider $n$  copies of our CFT with fields $X_{I}, \; I\in n$. Associated to a twist operator is a monodromy which dictate non-trivial boundary conditions between the different copies. Given a twist field $\sigma_g(x_i)$ at a point $x_i$ with monodromy $g\in G$ (we assume $G$ is an Abelian group) we have the relation
\be
X_I(e^{2\pi i}(x-x_i))\sigma_g(x_i)=X_{g(I)}(x-x_i)\sigma_g(x_i).
\ee

The correlation function
\be
\langle\sigma_{g_1}(x_1)\cdots \sigma_{g_n}(x_n)\rangle, \quad g_i\in G
\ee
can be non-zero only if the product $g_1\cdots g_n=e$ where $e$ is the identity element.

There is a direct relationship between the $\tau$s of the previous section and the mondromies of the twist operators. In the CFT language $\tau_{A_i}$ provide instructions for how copies of the theory are identified across the region $A_i$ by reflection. The action of cycling around the twist operators is then given by a reflection across one region and the inverse reflection across the next region. Concretely given two regions $A_1$ and $A_2$ with $\tau_{A_1},\tau_{A_2}$ the monodromy of the twist operator at their shared boundary at a point $p$ is given by
\be
\sigma(p)_{\tau_{A_1}\tau^{-1}_{A_2}}.
\ee

As an example we consider again the entanglement entropy \eqref{eq:EE}. Given a pure state $\rho=|\psi\rangle\langle\psi|$ we partition the boundary into two regions $A=[x_1,x_2]$ and its purifier $O=A^c$. Tracing out $O$ the bra and ket of the reduced density matrix can be described using the euclidean path integral as the lower and upper half plane while the partial trace identifies the two everywhere along the real axis except the interval $A$. The product of $n$ density matrices is then given by a cyclic gluing of $n$ copies along the upper and lower boundaries. This determines monodromies at each of the end points of $A$. As such this can equally be described by the insertion of two twist operator at the boundaries. In the literature the resulting two point function is typically written
\be
\langle\sigma_n(x_1)\sigma_{-n}(x_2)\rangle
\ee
where the subscripts $n$, $-n$ indicate that the two twist operators have inverse mondromies and implement this cyclic identification. In this work we will find it useful to be slightly more precise. The monodromies of these operators are associated with group elements $a,a^{n-1}\in\mathbb{Z}^n$ of the cyclic group with presentation $\langle a|
a^n=e\rangle$ where $a$ is the generator of the group and $e$ the identity element. Representations of these group elements can be written in terms of permutations which allows us to define the twist operators and their monodromies as
\be
\begin{split}
\sigma_{a}(x_1):& \quad (1,2,3,\cdots n)\\
\sigma_{a^{n-1}}(x_2):& \quad (1,n,n-1,\cdots 2)
\end{split}
\ee
which corresponds to the choice
\be
\tau_O=e, \quad \tau_A=a^{n-1}
\ee
we thus have
\be
S_n=\frac{1}{1-n}\log(\langle\sigma_a(x_1)\sigma_{a^{n-1}}(x_2)\rangle).
\ee

The calculation of q-point functions of twist operators is accomplished by utilizing the uniformization method \cite{Lunin:2000yv}. Each bra and ket (one copy of the upper half plane) can be mapped to a polygonal region with sides determined by the cycle structure of the twist operators. In particular a cycle of length $l$ will lead to an angle $\frac{\pi}{l}$ in the polygonal region.

For example in the case of three twist operators suppose a copy is present in cycles of length $\frac{1}{\mu},\frac{1}{\nu},\frac{1}{\lambda}$. The correct map from the upper half plane to a triangle with straight lines or circular arcs for edges and internal angles $\lambda\pi<\nu\pi<\mu\pi$ is given by the Schwarz triangle function
\be\label{eq:schw_tri}
z(x)=x^\lambda\frac{\pFq{a',b',c'}{x}}{\pFq{a,b,c}{x}}
\ee
where
\be
\begin{split}
&a=\frac{1}{2}(1-\lambda-\nu-\mu), \quad a'=\frac{1}{2}(1+\lambda-\nu-\mu)\\
&b=\frac{1}{2}(1-\lambda+\nu-\mu), \quad b'=\frac{1}{2}(1+\lambda+\nu-\mu)\\
&c=1-\lambda, \qquad\qquad\qquad c'=1+\lambda
\end{split}
\ee
The cycles then determine the correct procedure for gluing of the polygonal regions (the bras and kets) together. The result is a Riemann surface $\Sigma$ with no operator insertion. The genus of $\Sigma$ is given by the Riemann-Roch theorem just from the data of the cycle structure
\be
g=\frac{1}{2}\sum_{j=1}^q\sum_{i=1}^{c_j}(l_i-1)-n+1
\ee
here there are $q$ operators each with $c_{j}$ cycles of lengths $l_i$.

The inverse map or "uniformization map" $\Gamma$ is multivalued and maps from the Riemann surface back the single copy of the theory on the sphere. This map is meromorphic and the locations and structure of the singularities contain all of the necessary information about the original twist operator insertions. In particular the conformal dimensions of each twist operator are fixed to be
\be
\Delta_j=\frac{c}{12}\sum_{i}^{c_j}\left(l_i-\frac{1}{l_i}\right).
\ee

Using $\Gamma$ it is possible to relate the partition functions of the original theory and the Riemann surface. as a result of the conformal transformation there is a conformal anomaly given by the Liouville action $S_L$:

\be
\langle\sigma_{g_1}(x_1)\cdots \sigma_{g_n}(x_n)\rangle=Z=e^{S_L}Z_{\Sigma},\quad S_L=\frac{c}{96\pi}\int_{\Sigma}d^2z\sqrt{g}\left[\partial_\mu\phi\partial_\nu\phi g^{\mu\nu}+2R\phi\right] \label{LVAB}
\ee
The field $\phi$ is determined explicitly by $\Gamma$ as
\be
\phi=2\log(|\partial_z \Gamma|).
\ee
Typically these calculations are challenging to perform as the Liouville action requires careful regularization and the exact uniformization map can be difficult to determine especially for $g\geq2$.

\begin{figure}[H]
\begin{tabular}{cc}
\centering
\quad \includegraphics[width=.38\textwidth,page=33]{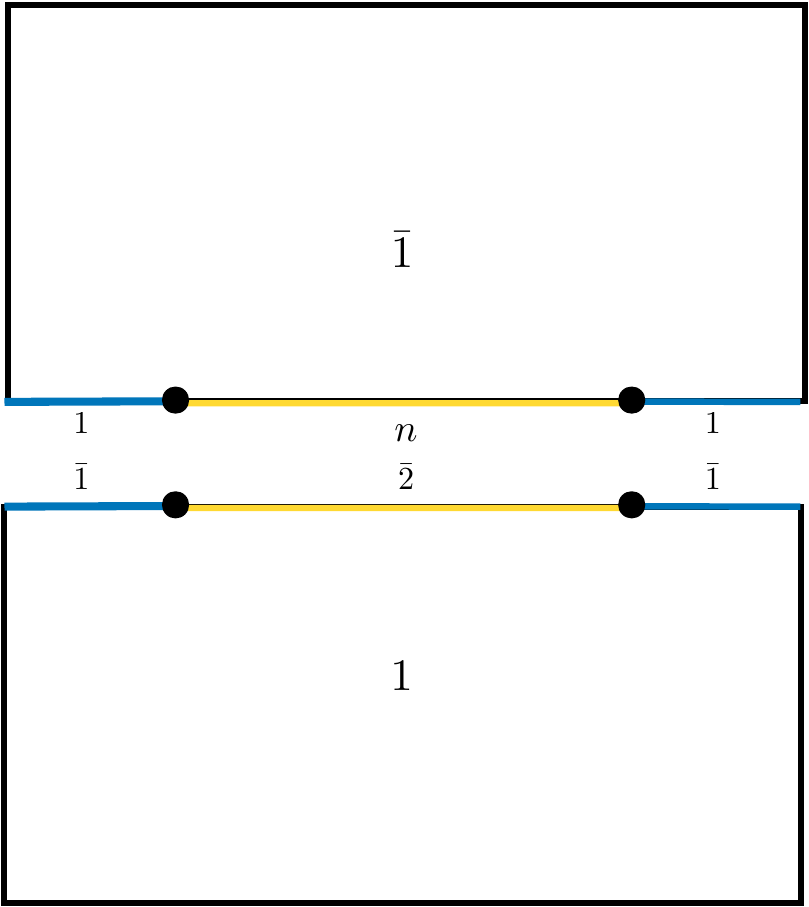}&\qquad\qquad\includegraphics[width=.34\textwidth,page=34]{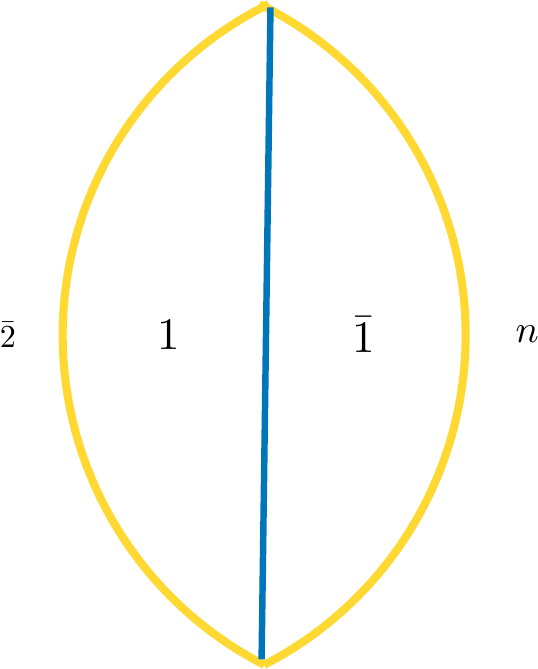}
\end{tabular}
\caption{\label{fig:digon} L: A single copy of the CFT split into its bra and ket in the gravitational path integral. The twist operator insertions are shown at the real axis. The twist operators divide the real axis into two regions. We choose to glue the region $O$ (shown in blue) such that the gravitation path integral corresponds to the reduced density matrix $\rho_{A}$. Along the two remaining intervals are shown the corresponding copy it will be glued to to form the replica manifold. These are done in accordance with the monodromies of the twist operators. R: After a mobius transformation and application of the map $x^{\frac{1}{n}}$ each half plane is mapped to an digon with angle $\frac{\pi}{n}$. }
\end{figure}

Returning to the entanglement entropy each ket and bra is mapped to two-sided ``digon" with angle $\frac{\pi}{n}$ (see figure \ref{fig:digon}). These are arranged with the endpoints at $z=0,\infty$ forming a sphere. The explicit uniformization map up to mobius transformations is in this case
\be
\Gamma(z)=z^n
\ee
from which the entanglement entropy can be explicitly calculated
\be
S=\frac{c}{3}\log\left(\frac{|x_2-x_1|}{\epsilon}\right).
\ee

The renyi multi-entropy defined in \cite{2022arXiv221116045P,2022PhRvD.106l6001G} is the generalization of this procedure to a pure state with $q$ intervals where the monodromies of the twist operators are chosen from the group $\mathbb{Z}_n^{(q-1)}$ in such a way that the resulting cycle structure of all operators is the same consisting of $n^{q-2}$ cycles each of length $n$. In the language of the previous sections this corresponds to a particular multi-trace measure with appropriately chosen $\tau$s. That is we consider the $q$ point function
\be
\langle\sigma_{g_1}(x_1)\sigma_{g_2}(x_2)\cdots\sigma_{g_q}(x_q)\rangle)_n, \quad  g_i\in \mathbb{Z}_n^{(q-1)} \text{ s.t. } \prod_i^qg_i=e
\ee
where the conformal dimension of each operator is
\be
\Delta=\frac{cn^{q-2}}{12}\left(n-\frac{1}{n}\right)
\ee
The renyi multi-entropy and multi-entropy are then given by
\be
S_n^{(q)}=\frac{1}{1-n}\frac{1}{n^{q-2}}\log\left(\langle\sigma_{g_1}(x_1)\sigma_{g_2}(x_2)\cdots\sigma_{g_q}(x_q)\rangle)_n\right)
\label{sqn}
\ee
and
\be\label{eq:me_def}
S^{(q)} =\lim_{n\to 1}S_n^{(q)}= -\partial_n\log \left(\langle\sigma_{g_1}(x_1)\sigma_{g_2}(x_2)\cdots\sigma_{g_q}(x_q)\rangle)_n\right)|_{n=1}.
\ee

\subsubsection*{Example
: Constructing the replica manifold}
As a concrete example let's consider the case $q=3$ $n=4$. Here the monodromies of the twist operators are chosen from $\mathbb{Z}_4\times\mathbb{Z}_4$ with group presentation $\langle a,b|a^4=b^4=e\rangle$. For the two generators $a,b$ we pick an explicit representation in terms of permutations to be 
\be
\begin{split}
   a:\quad (1,2,3,4)(5,6,7,8)(9,10,11,12)(13,14,15,16)\\
   b:\quad (1,10,15,7)(2,11,16,8)(3,12,13,5)(4,9,14,6)
\end{split}
\ee
from which using the group multiplication we can determine the permutation representation of all other group elements.

\begin{figure}[H]
\begin{tabular}{cc}
\centering
\quad \includegraphics[width=.45\textwidth,page=31]{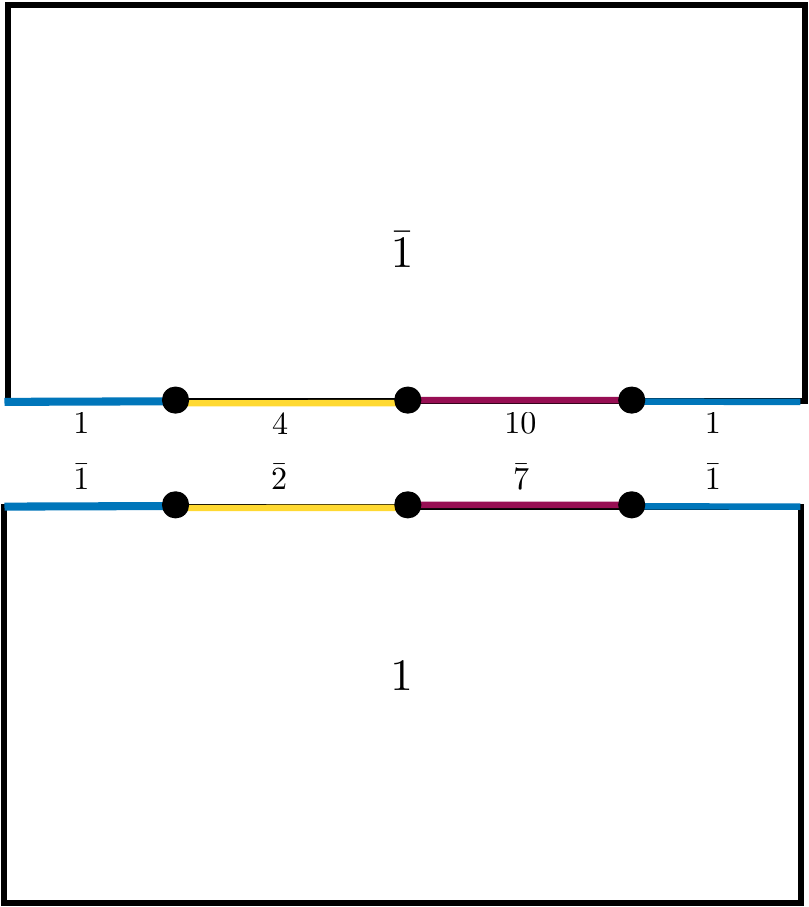}&\qquad\qquad\includegraphics[width=.25\textwidth,page=32]{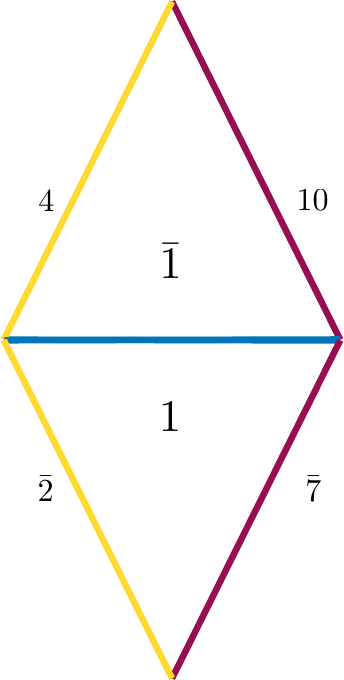}
\end{tabular}
\caption{\label{fig:n4triangle} L: A single copy of the CFT split into its bra and ket in the gravitational path integral. The twist operator insertions are shown at the real axis. The twist operators divide the real axis into three regions. We choose to glue the region $O$ (shown in blue) such that the gravitation path integral corresponds to the reduced density matrix $\rho_{AB}$. Along the center between the bra and ket we label for each interval the copy that the it will be glued to. These identifications are done in accordance with the monodromies of the twist operators. R: After applying the map \eqref{eq:schw_tri} each half plane is mapped to an equiangular hyperbolic triangle with angle $\frac{\pi}{4}$. }
\end{figure}

 Now we choose the three twist operators to be charged under the group elements $a,a^3b^3,b$ which determines the monodromies:
\be
\begin{split}
\sigma_{a}(x_1):& \quad (1,2,3,4)(5,6,7,8)(9,10,11,12)(13,14,15,16)\\
\sigma_{a^3b^3}(x_2):& \quad (1,6,13,11)(2,7,14,12)(3,8,15,9)(4,5,16,10)\\
\sigma_{b}(x_3):& \quad (1,10,15,7)(2,11,16,8)(3,12,13,5)(4,9,14,6).\\
\end{split}
\ee
Each twist operator has a cycle structure consisting of four cycles of length four.
Equivalently this corresponds to the choice of $\tau$s
\be
\tau_O=e, \quad \tau_A=a^3, \quad \tau_B=b.
\ee



\begin{figure}[H]
  \centering
   \includegraphics[width=.65\textwidth,page=1]{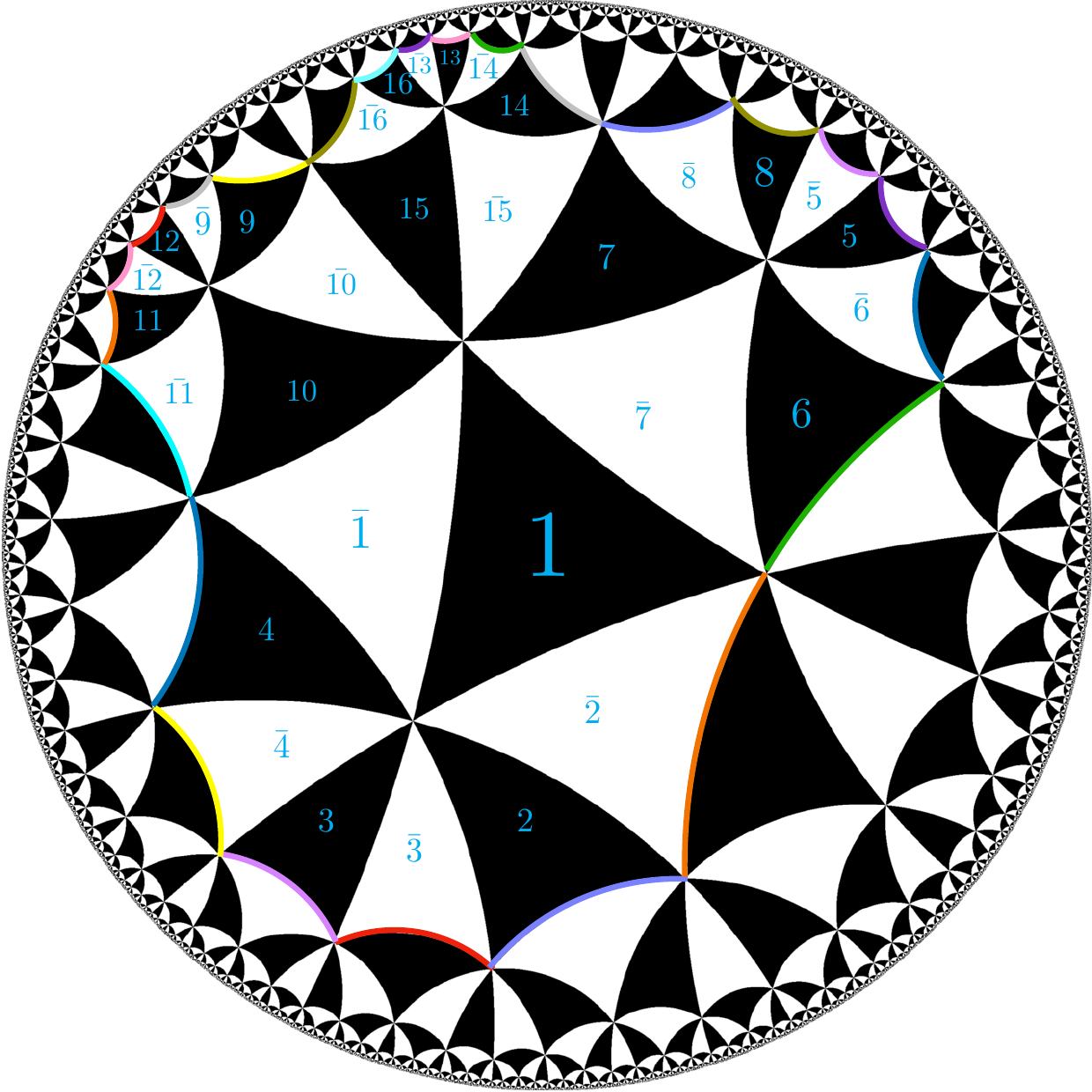}
  \caption{The replica manifold associated with the renyi multi-entropy $S_4^{(3)}$. Each fundamental region consists of two equiangular hyperbolic triangles each with with internal angles $\frac{\pi}{4}$. The sixteen fundamental regions are glued according to the explicit permutation representation for the mondromies of the twist operators. They tile a subregion of the hyperbolic plane where the final pairs of gluings at the boundary are indicated by the different colored subregion boundary segments. After performing these the resulting Riemann surface is genus $g=3$. This Riemann surface is equivalently constructed by the quotient of the hyperbolic disk by the Fuschian triangle group (4,4,4).} 
\label{fig:n4replica}
\end{figure}

Using the map \eqref{eq:schw_tri} each ket can be mapped from the upper half plan to an equilateral hyperbolic triangle with angles $\frac{\pi}{4}$ and then reflected to form the fundamental region (see figure \ref{fig:n4triangle}). This is equivalent to tracing out the region $O$ such that each fundamental region corresponds to a copy of $\rho_{AB}$.  The replica manifold is constructed as the gluing of these 16 regions following the prescribed cycle structure (see figure \ref{fig:n4replica}). 

\subsection{Steiner trees and proposed holographic dual}
We start with a static time slice of pure AdS$_3$ and partition the boundary into $q$ connected regions. We label the collection of regions $\{A_i\}$: $A,B,\cdots$ but label the final region $O$ signifying that it is the purifier and that we have divided the entire boundary such that we have a pure state. We consider the following minimization problem: find the collection of geodesics of minimal length that partitions the entire bulk into $q$ regions each of which is homologous to a different boundary region. We will call the set of such surfaces which satisfy this homology constraint to be $\{\Sigma\}$ and the surface of minimal length $\Sigma^*$. We define the holographic multi-entropy to be the length of this surface
\be
S^{(q)}(\{A_i\})= \frac{1}{4G_N}\area(\Sigma^*).
\ee

\begin{figure}[H]
  \centering
   \includegraphics[width=.45\textwidth,page=9]{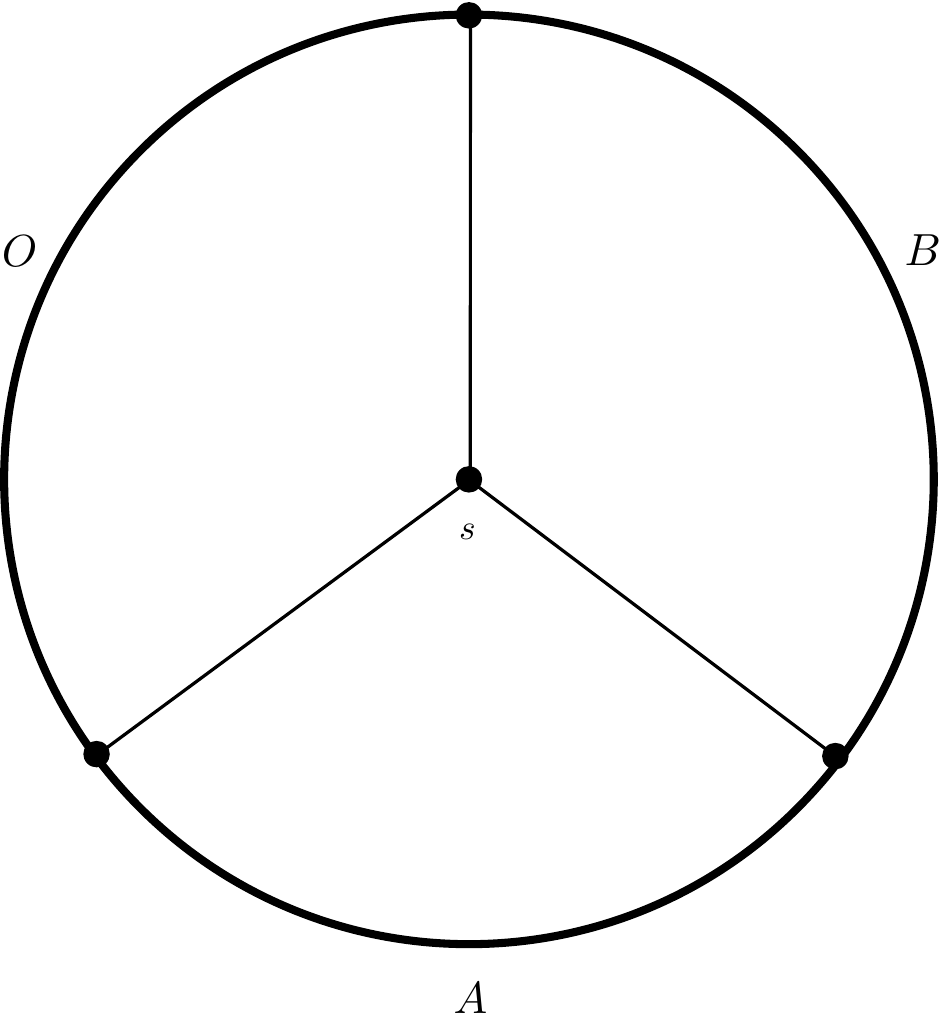}
  \caption{An example of a minimal Steiner tree for three boundary regions.} 
\label{fig:steinerexample}
\end{figure}

The solution to this problem is given by a Steiner tree \cite{2019JHEP...02..023A,Gueron2002TheFP} these are networks of geodesics which meet at equiangular trivalent vertices . 

Of particular interest to us will be the case $n=3$ in which case we expect a single bulk vertex along with geodesics from this point to the boundary. We start by considering the most symmetric case where the boundary division are equally spaced around the circle at infinity with angular separation $\frac{2\pi}{3}$. By symmetry the minimal configuration consists of radial geodesics from the boundary to the bulk intersection at the center. These geodesics also meet at angles of $\frac{2\pi}{3}$ (see figure \ref{fig:steinerexample}).

\begin{figure}[H]
  \centering
   \includegraphics[width=.45\textwidth,page=30]{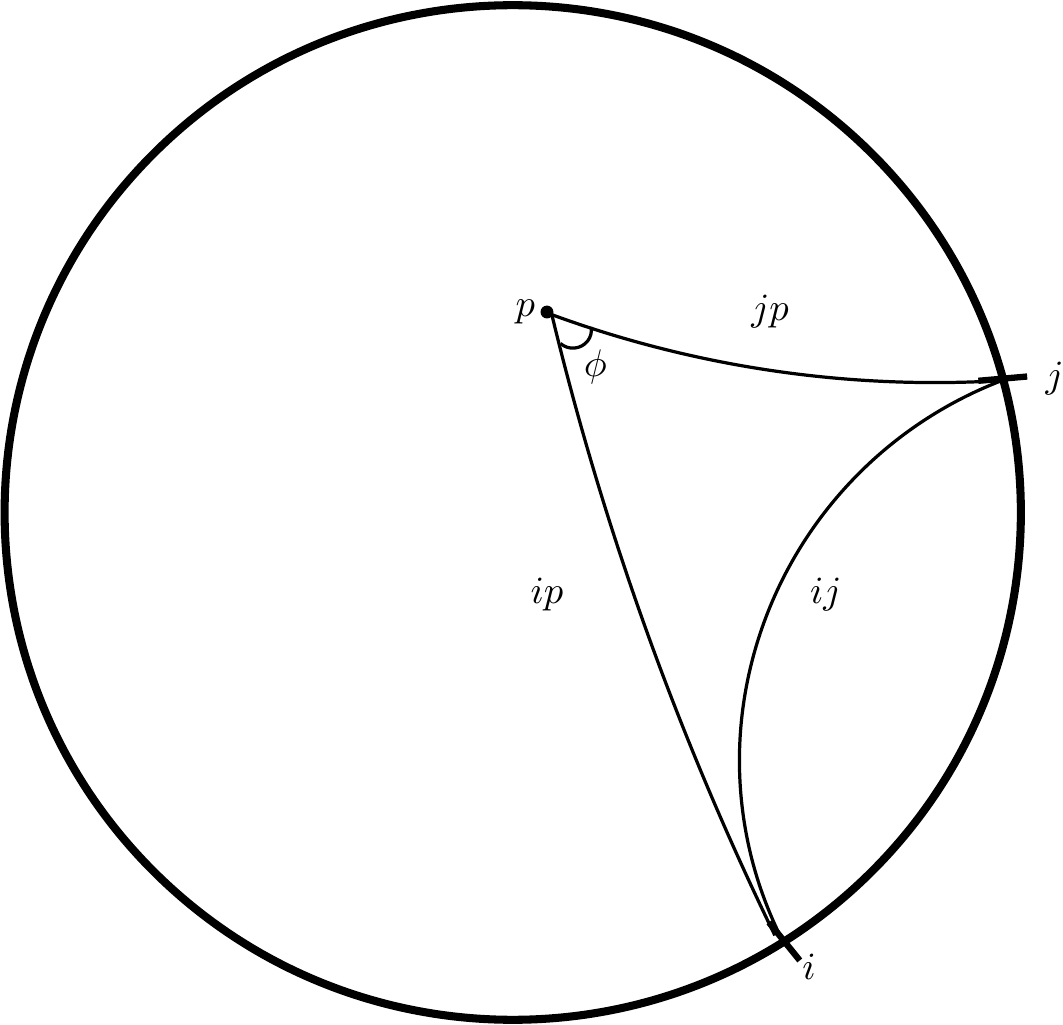}
  \caption{Given a geodesic with end points on the boundary (such as an RT surface) we can pick any point in the bulk and draw geodesics from the two boundary points to this bulk point. Together the three geodesics form a hyperbolic triangle with an interior angle of $\phi$ (the others are zero being ideal points). The length of the two new geodesics together have the same length as the original plus and additional increase which depends only on the angle $\phi$.} 
\label{fig:hcos}
\end{figure}

For the hyperbolic disk it is possible to relate the lengths of sides of a triangle using the hyperbolic law of cosines. We consider a triangle with points $i,j,p$. If $g_{i_1,i_2}$ is the geodesic connecting the points $i_1$ and $i_2$ and define 
\be
i_1i_2=\area(g_{i_1,i_2})
\ee
With this in place we have
\be
\cosh(ij)=\cosh(ip)\cosh(jp)-\sinh(ip)\sinh(jp)\cos(\phi)
\ee
where $\phi$ is the angle the angle between the geodesics $g_{i,p}$ and $g_{j,p}$.
Now taking the points $i$ and $j$ to be ideal (located on the boundary) this simplifies to the following relation (see figure \ref{fig:hcos})
\be\label{eq:hcos2}
ij+2\log\left(\csc\left(\frac{\phi}{2}\right)\right)=ip+jp.
\ee

Next, we consider the surface $\Sigma^*$ which is comprised of three geodesics $g_{a,s}$, $g_{b,s}$, $g_{o,s}$ where the intersection point $s$ is at the center. We also include the geodesics $g_{a,b}$ $g_{b,o}$ and $g_{a,o}$ which are the RT surfaces of the boundary regions $A,B,O$ and whos areas calculate the holographic entanglement entropies $S_A,S_B,S_{AB}$. Each RT surface can be related to two of the geodesics of $\Sigma^*$ using \eqref{eq:hcos2}. Doing this three times, one for each RT surface we are led to the relation
\be
2\frac{1}{4G_N}\area(\Sigma^*)-(S_{A}+S_{B}+S_{AB})=3\frac{c}{3}\log\left(\frac{2}{\sqrt{3}}\right).
\ee
Now for $q=3$ we have the equivalence 
\be
S_{A}+S_{B}+S_{AB}=I_{AB}+I_{AO}+I_{BO}
\ee
where
\be
I_{A_1A_2}=S_{A_1}+S_{A_2}-S_{A_1A_2}
\ee
is the mutual information between regions $A_1$ and $A_2$. Thus
\be\label{eq:hme_3region}
\kappa\coloneqq\frac{1}{4G_N}\area(\Sigma^*)-\frac{1}{2}(I_{AB}+I_{AO}+I_{BO})=\frac{c}{2}\log\left(\frac{2}{\sqrt{3}}\right).
\ee
In particular the quantity $\frac{1}{2}I_{A_1A_2}$ is precisely the maximum number of distillable bell pairs between the regions $A_1$ and $A_2$. The result \eqref{eq:hme_3region} states that if we remove all possible distillible bipartite entanglement the holographic multi-entropy is still positive, so the remaining contribution must be multipartite. In addition the multipartite contribution is completely characterized by the angles the geodesics make at the bulk intersection point\footnote{This is commensurate with claims related to another measure the reflected entropy that such corner terms should be associated with the presence multipartite entanglement. In this context a similar relation between the difference of the reflected entropy and one half the mutual information, the "Markov gap", was related to a particular information theoretic recovery protocol. It would be interesting to pursue this similarity further \cite{2021JHEP...10..047H,2023arXiv230714434V}.}.

\begin{figure}[H]
\begin{tabular}{cc}
\centering
\includegraphics[width=.45\textwidth,page=29]{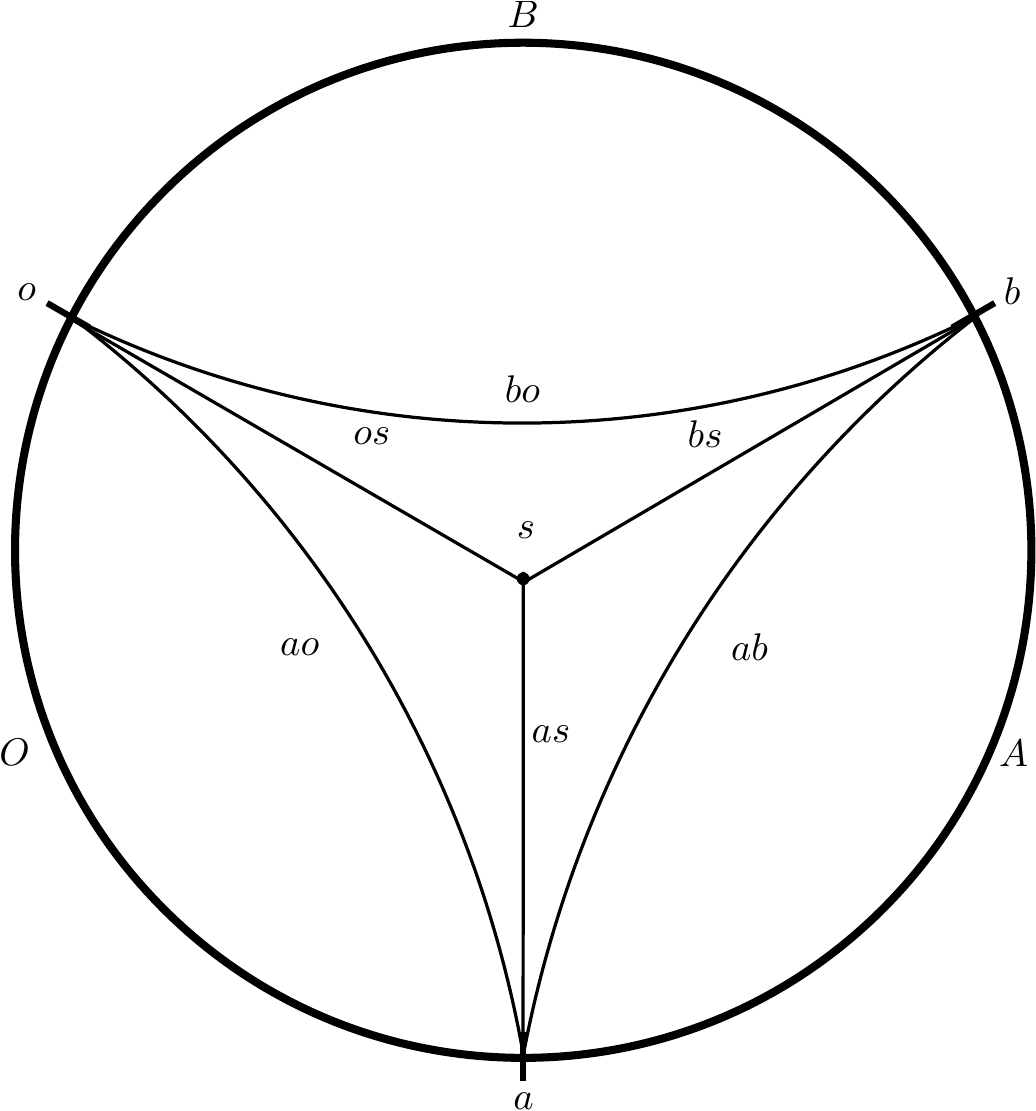}&
\includegraphics[width=.45\textwidth,page=28]{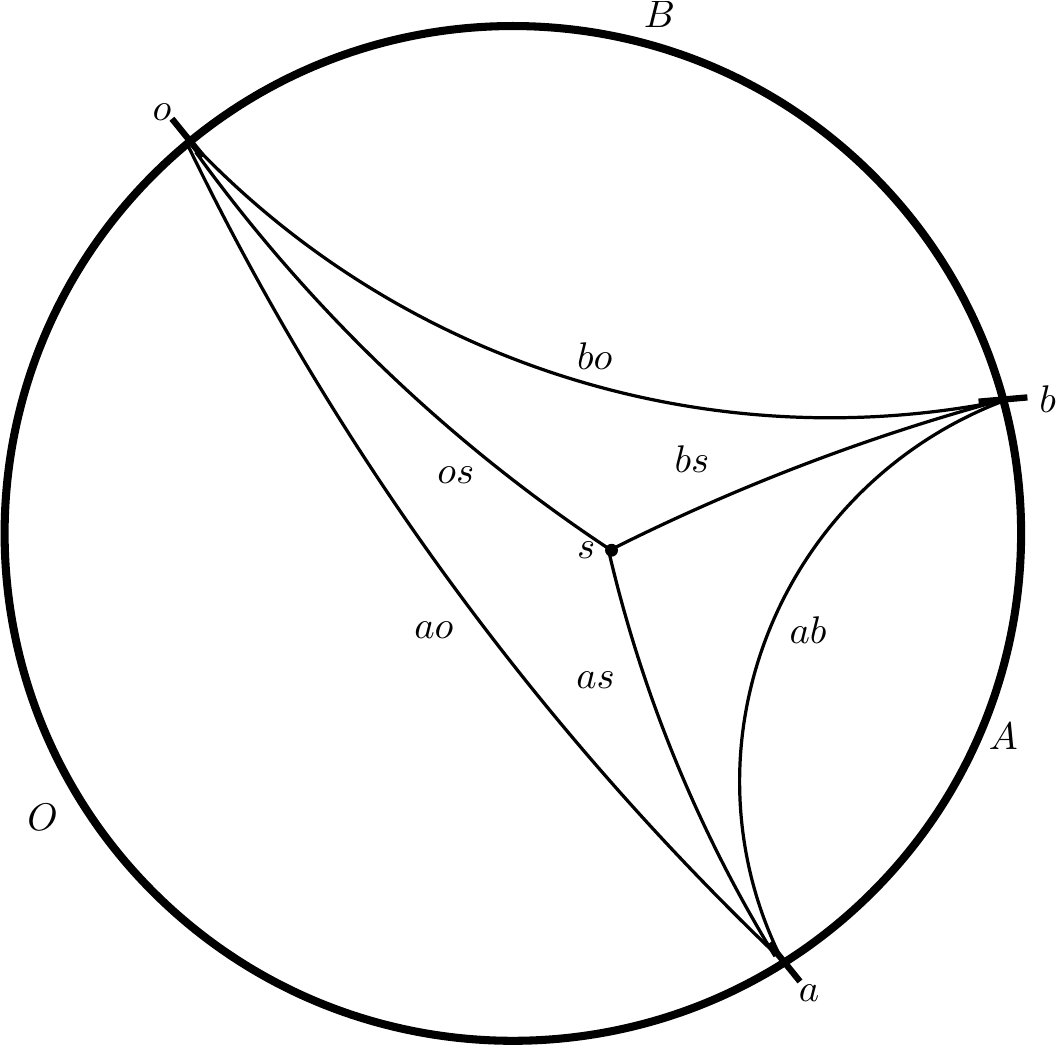}
\end{tabular}
\caption{\label{fig:hme3} The three RT surfaces along with $\Sigma^*$ form three hyperbolic triangles. Using this the surfaces can be related. The increase in area due to the three angles of $\phi=\frac{2\pi}{3}$ which we call $\kappa$ gives the contribution which is multipartite. For different choices of boundary regions the holographic multi-entropy will change, but this change is entirely encapsulated by the corresponding changes in the three RT surfaces; $\kappa$ remains the same. }
\end{figure}

\begin{figure}[H]
\begin{tabular}{cc}
\centering
\includegraphics[width=.45\textwidth,page=14]{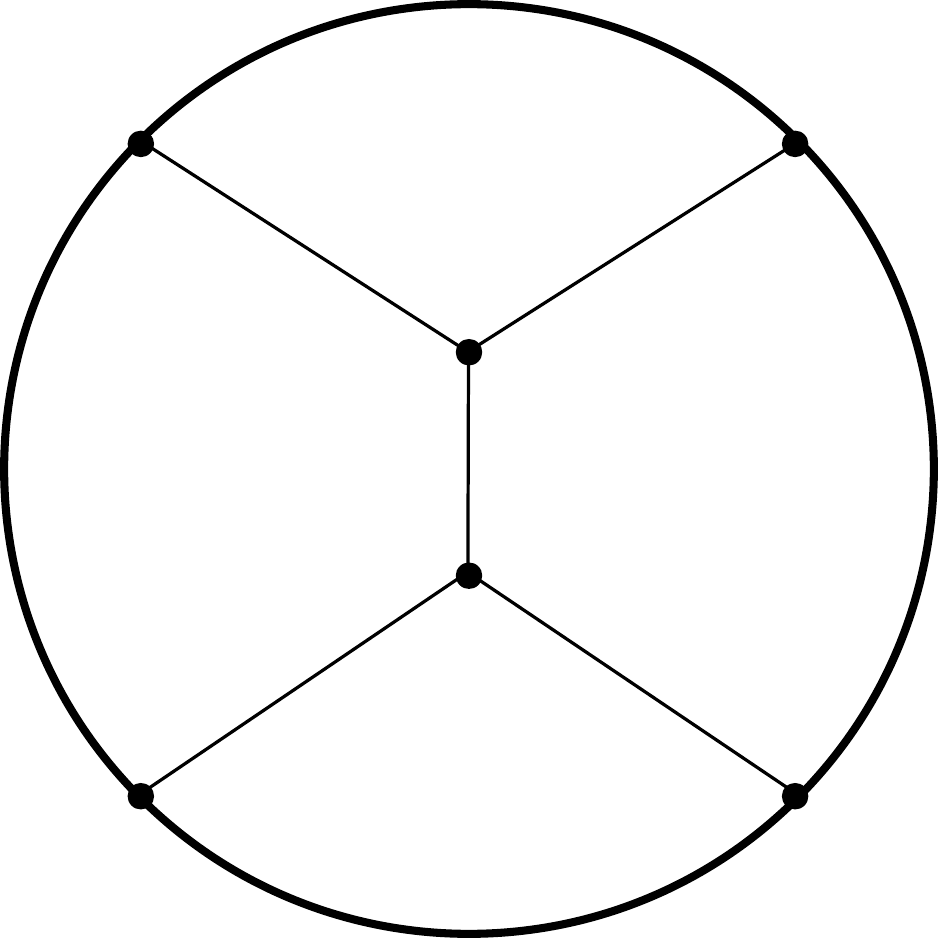}&
\includegraphics[width=.45\textwidth,page=15]{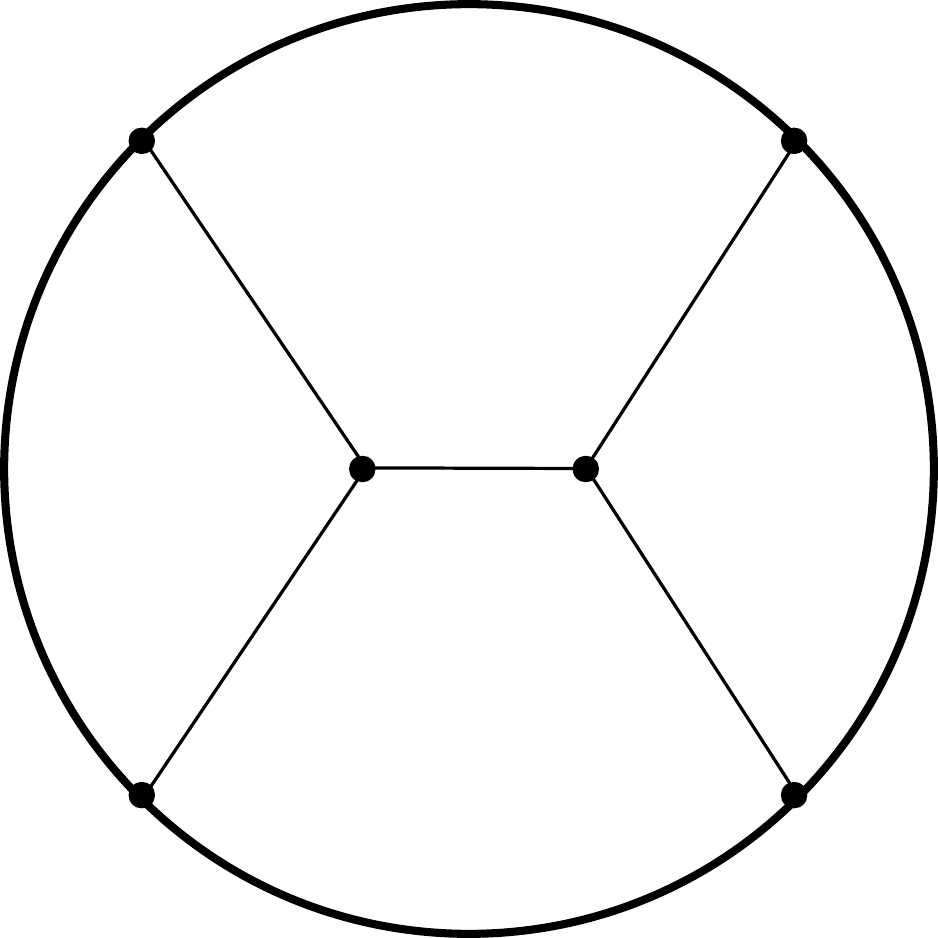}
\end{tabular}
\caption{\label{fig:hme4} Examples of Steiner trees for four boundary regions. For four or more regions there are phase transitions between different configurations depending on the relative positions of the boundary vertices. Here the surfaces exhibit a s,t channel-like transition depending on the orientation of the internal geodesic segment. The constant $\kappa$ remains important as each trivalent vertex will induce a contribution proportional to $\kappa$. Configurations with higher valence intersections (i.e. not Steiner trees) will never be minimal.}
\end{figure}

For other divisions of the boundary we note that the points $a,b,o$ and always be brought to the symmetric configuration via a mobius transformation. Such transformations do not preserve the lengths of bulk geodesics, but do preserve angles between them. Thus for any division of the bulk the geodesics of the minimizing surface will always meet at angles $\phi=\frac{2\pi}{3}$. Furthermore, the argument leading up to \eqref{eq:hme_3region} remain unchanged so that $\kappa$ is the same for all choices of boundary region (see figure \ref{fig:hme3}). Generalizations of these relationships to $q>3$ have already been considered in \cite{2023arXiv230816247G} (see figure \ref{fig:hme4}).

Now comparing with the boundary proposal and specializing to $q=3$ we have
\be
S_n^{(3)}=\frac{1}{1-n}\frac{1}{n}\log\left(\langle\sigma_{g_1}(x_1)\sigma_{g_2}(x_2)\sigma_{g_3}(x_3)\rangle)_n\right)
\label{sqn}
\ee
where the three point function can be fixed by conformal symmetry to the form
\be
\langle\sigma_{g_1}(x_1)\sigma_{g_2}(x_2)\sigma_{g_3}(x_3)\rangle_n=\frac{C_n}{(x_{21}x_{13}x_{23})^{\frac{c}{12}(n^2-1)}}
\ee
and we have used that the conformal dimension of all of the twist operators is given by $\frac{c}{12}(n^2-1)$ since they each have monodromies consiting of $n$ cycles of length $n$.
We thus find
\be
\begin{split}
S_n^{(3)}&=-\frac{c}{12}\frac{1}{1-n}\frac{1}{n}(n^2-1)\log(x_{21}x_{13}x_{23})+\frac{1}{1-n}\frac{1}{n}\log(C_n)\\
&=\frac{c}{12}\left(1+\frac{1}{n}\right)\log(x_{21}x_{13}x_{23})+\frac{1}{1-n}\frac{1}{n}\log(C_n).
\end{split}
\ee
The scaling of the first term is the same as that of the Renyi entanglement entropy $S^{(2)}_n$. This suggests for all $n$ we should consider the quantity
\ba
\kappa^{(3)}_{n}=S^{(3)}_{n}-\frac{1}{2}\left(S^{(2)}_n(A)+S^{(2)}_n(B)+S^{(2)}_n(C)\right).
\ea
Now when we take the $n\rightarrow 1$ limit
\be
S^{(3)}=\frac{c}{6}\log(x_{21}x_{13}x_{23})-\partial_n\log(C_n)|_{n=1}
\ee
where the first term is the same as half the area of the three RT surfaces. In particular the extra term coming from the intersection of the geodesics is directly related to the three point coefficient $C_n$. The holographic calculation gives the prediction
\be
\kappa=-\partial_n\log(C_n)|_{n=1}\overset{?}{=}\frac{c}{2}\log\left(\frac{2}{\sqrt{3}}\right).
\ee
The confirmation of this equality by direct calculation of $C_n$ in the boundary CFT would provide strong evidence the duality between the multi-entropy and the area of minimal Steiner trees in the bulk geometry.

As mentioned previously to our knowledge the only method for the exact computation of this quantity relies on the uniformization method \cite{Lunin:2000yv}. In order to complete the calculation for a specific $n$ two pieces of information are needed: the uniformization map between the replica surface and the original theory and the partition function of the replica surface. In the current case of interest the genus of the replica surface increases with $n$ such that beyond $n=3$ these quantities are unknown. In what follows we content ourselves with exact calculations of $\kappa_n^{(3)}$ for those cases which are tractable. We also discuss some cases with $q=4$ and the fusion of operators in the coincident limit.

\section{Renyi multi-entropy}\label{sec:reny}

In this section we present explicit calculations of multi-entropy for $(n,q)=(2,3),(2,4)$ and $(3,3)$, starting from the standard entanglement entropy i.e. $q=2$.

\subsection{Entanglement Entropy from Liouville field method}
  We take the conformal transformation $w=f(z)$,
\begin{align}
    ds^2 = dw\overline{dw} = \left|\frac{df}{dz}\right|^2 dz\overline{dz}.
\end{align}
Here we set the classical Liouville field $\varphi$ as $e^{2\varphi}=\left|\frac{df}{dz}\right|^2$. Notice that this Liouville field $\varphi$ is related to the previous Liouville field $\phi$ in (\ref{LVAB}) via 
$\phi=2\varphi$. The partition function in the coordinate $w$ is given by 
\begin{align}
    Z_w = Z_z e^{I^L},
\end{align}
where $I_L$ is the Liouville action
\begin{align}
    I^L = \frac{c}{24\pi} \int dz^2 \left(
    4\partial\varphi \Bar{\partial}\varphi + \mu e^{2\varphi}
    \right).
\end{align}

As a warm up example, we briefly explain the calculation of entanglement entropy (i.e. $q=2$) via the the replica trick using the Liouville field theory.
We take the map as $z^n = \frac{w-a}{w-b}$ and calculate the $n$-th Renyi entropy on $w$-plane. Then the Liouville field is
\begin{align}
    e^{2\varphi} = n^2 |b-a|^2 \frac{|z|^{2(n-1)}}{|z^n-1|^4},
\end{align}
and the Liouville action $I_L^{(n)}$ is 
\begin{align}
    I^L_{n} = \frac{c}{24\pi} \int dz^2 \left(
    \frac{|(1-n)-(1+n)z^n|^2}{|z^n-1|^2|z|^2}
    +\mu \frac{n^2 |b-a|^2|z|^{2(n-1)}}{|z^n-1|^4}
    \right).
\end{align}
$I^L_{n}$ diverges at $z=0, \infty$ because of the UV divergence. We set the UV cut-off in the $w$-plane to be $\epsilon$.
\begin{align}
    \begin{cases}
        z=0, w=a: & \displaystyle |z|>\left|\frac{\epsilon}{b-a}\right|^{1/n}, \\
        z=\infty, w=b: & \displaystyle |z|<\left|\frac{b-a}{\epsilon}\right|^{1/n}.
    \end{cases}
\end{align}
The integral near $z\to 0$ is evaluated as 
\begin{align}
    I^L_{n} \simeq \frac{c}{24\pi} \int_{z\simeq 0} \frac{(1-n)^2}{|z|^2} \simeq \frac{c}{12} \frac{(1-n)^2}{n} \log \left|\frac{b-a}{\epsilon}\right|,
\end{align}
and $z\to \infty$ is 
\begin{align}
    I^L_n \simeq \frac{c}{24\pi} \int_{z\to \infty} \frac{(1+n)^2}{|z|^2} \simeq \frac{c}{12} \frac{(1+n)^2}{n} \log \left|\frac{b-a}{\epsilon}\right|.
\end{align}

$\operatorname{Tr}_A \left[(\rho_A)^n\right]$ is obtained by introducing the normalization factor $nI^L_{1}$,
where
\begin{align}
\label{eq:NormalizationLiouville2interval}
    I^L_{1}=\frac{c}{12}\left(\frac{(1-1)^2}{1}+ \frac{(1+1)^2}{1}\right) \log \left|\frac{b-a}{\epsilon}\right|=\frac{c}{12}\cdot4\cdot\log \left|\frac{b-a}{\epsilon}\right|.
\end{align}
$\operatorname{Tr}_A \left[(\rho_A)^n\right]$ is,
\begin{align}
    \operatorname{Tr}_A \left[(\rho_A)^n\right] =&~ e^{I^L_{n}-nI^L_{1}} \nonumber\\
    =&~ \left|\frac{b-a}{\epsilon}\right|^{ \frac{c}{12} \frac{(1-n)^2}{n}+\frac{c}{12} \frac{(1+n)^2}{n}-n\cdot\frac{c}{12}\cdot4} \nonumber\\
    =&~ \left|\frac{b-a}{\epsilon}\right|^{-\frac{c}{6}\left(n-\frac{1}{n}\right)}.
\end{align}
This result reproduces the well-known result of $n-$th Renyi entropy and the (von-Neumann) entanglement entropy \cite{Holzhey:1994we,Calabrese:2004eu}:
\begin{align}
    S^{(2)}_n =&~ \frac{1}{1-n} \log\operatorname{Tr}_A \left[(\rho_A)^n\right] \nonumber\\
    =&~ \frac{c(n+1)}{6n} \log\left|\frac{x_1-x_2}{\epsilon}\right|, \\
    S^{(2)} =&~ \lim_{n\to 1} S^{(2)}_n \nonumber\\
    =&~ \frac{c}{3} \log\left|\frac{x_1-x_2}{\epsilon}\right|.
\end{align}

\subsection{Analysis of $n=2$ and $q=3$ in the coincident limit}

Let us move onto the first non-tirivial example, namely three-partite entropy $q=3$. Here we choose the subsystem $A$, $B$ and $C$ are all intervals which are situated next to each other in this order. For simplicity, we choose $n=2$ i.e. $S^{(3)}_2$, so that the whole geometry of path-integral is mapped into a genus zero surface. 

The essential part of this entropy calculation is the evaluation of three point function of $Z_2\times Z_2$ twist operators  $\sigma_1\equiv \sigma_A$, $\sigma_2\equiv\sigma_{A^{-1}B^{-1}}$ and $\Sigma_3\equiv \sigma_3$ at 
$w=w_1,w_2$ and $w_3$. The $n=2$ replicated partition function is given by 
\ba
\la\sigma_1(w_1)\sigma_2(w_2)\sigma_3(w_3)\lb.
\ea
Since we assume that the two dimensional CFT is on an infinite line, the subsystem $A$ becomes infinitely in both directions Re$[w]\to\pm\infty$. Thus the end points of the intervals $B$ and $C$ are $(w_1,w_2)$ and $(w_2,w_3)$, respectively. 

\subsubsection{Conformal Anomaly Analysis}

One way to evaluate this three point function is to insert an energy stress tensor as in \cite{Calabrese:2004eu}
and consider its expectation value 
\ba
\la T(w)\lb=\frac{\la T(w)\sigma_1(w_1)\sigma_2(w_2)\sigma_3(w_3)\lb}
{\la\sigma_1(w_1)\sigma_2(w_2)\sigma_3(w_3)\lb}.
\ea

\begin{figure}[hhh]
  \centering
   \includegraphics[width=7cm]{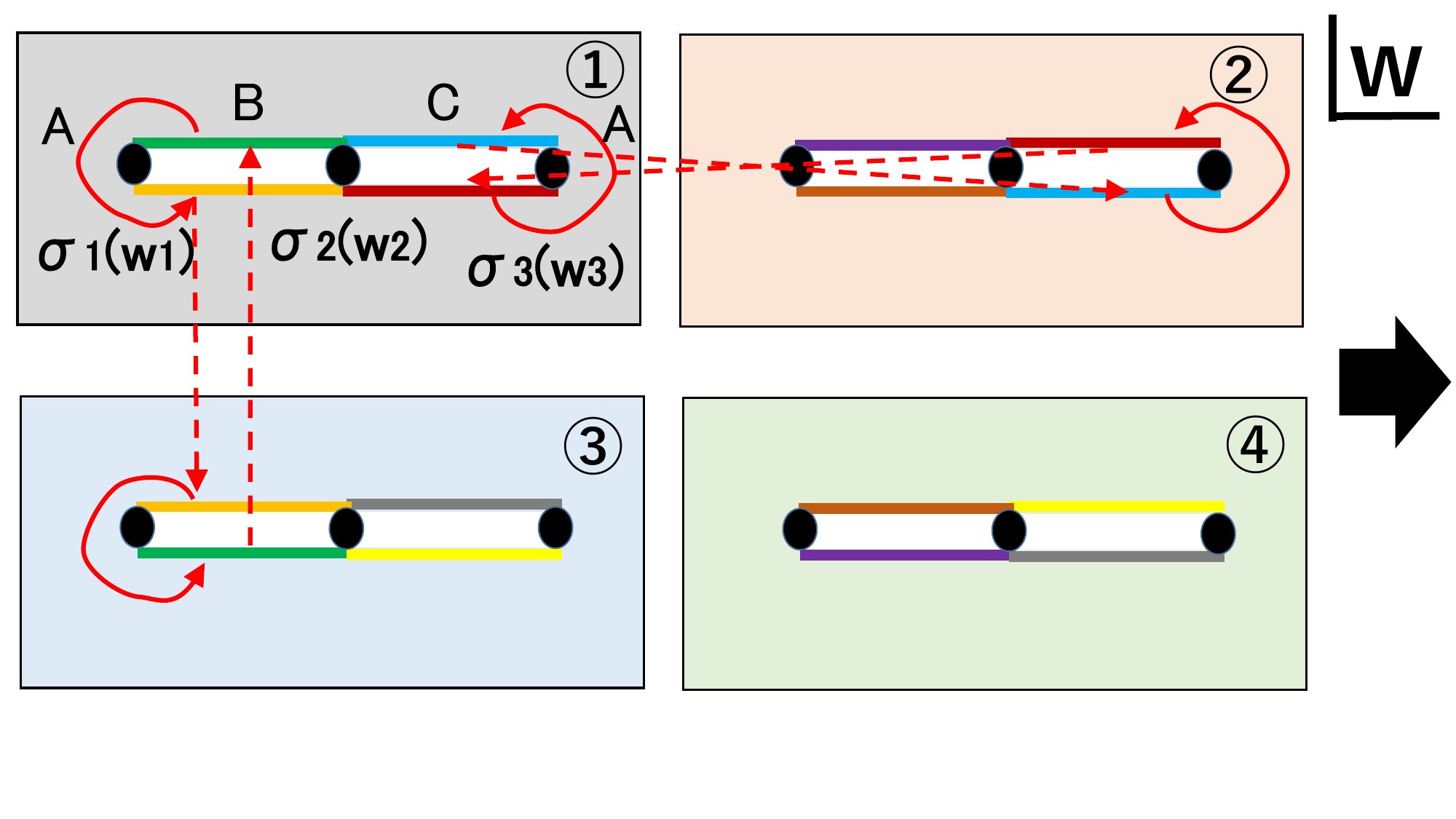}
     \includegraphics[width=7cm]{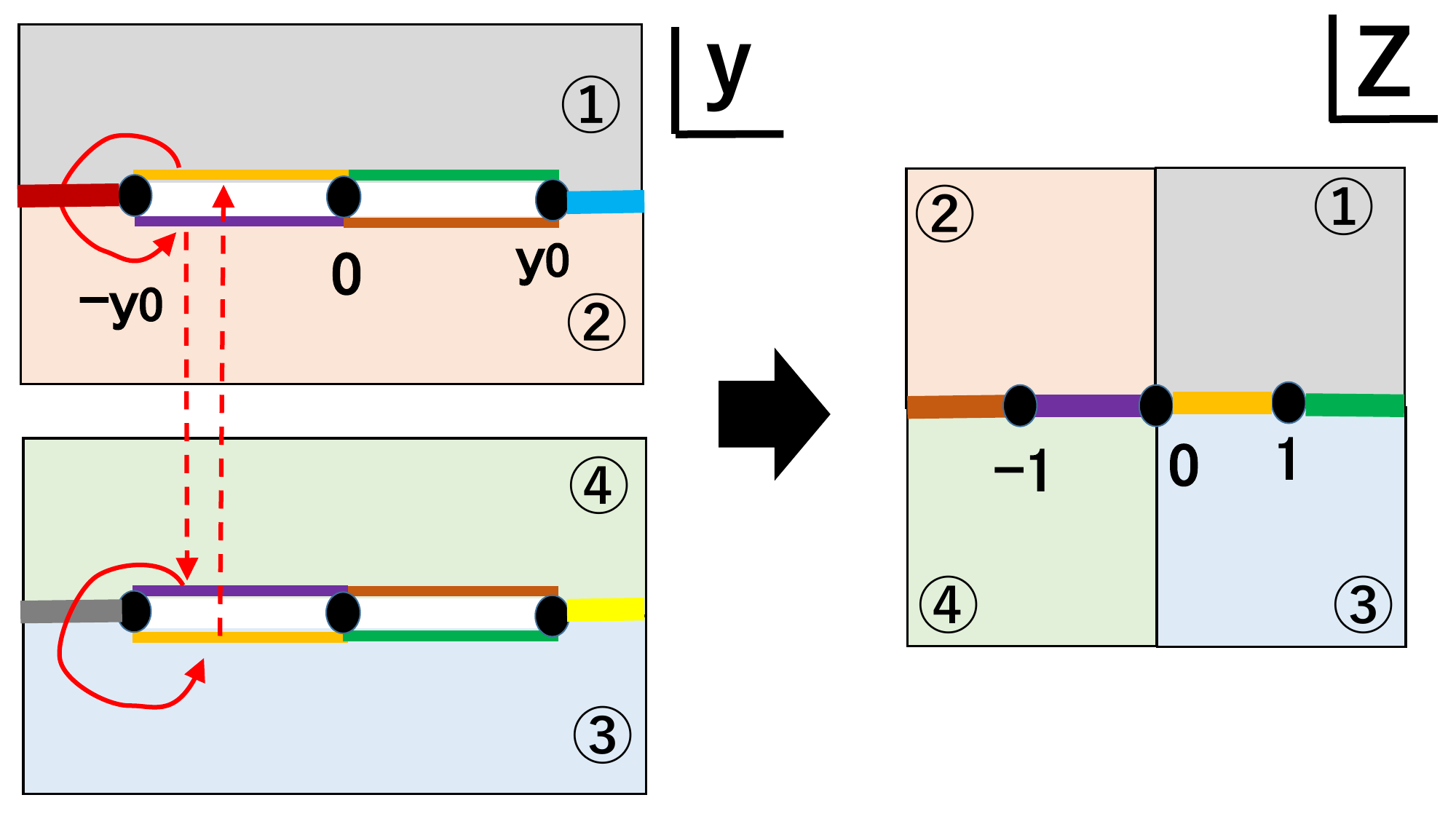}
  \caption{The conformal map $w\to z$ given by (\ref{cmapr}) .This is decomposed into two transformations 
  $y^2=\frac{w-w_1}{w_3-w}$ and $z^2=\frac{y+y_0}{y_0-y},$
  where we defined $y_0=\sqrt{\frac{b-a}{c-b}}$.} 
\label{fig:replicam}
\end{figure}

To evaluate this, we first note that the conformal transformation defined by
\ba
\frac{(w_3-w_2)(w-w_1)}{(w_2-w_1)(w_3-w)}=\left(\frac{z^2-1}{z^2+1}\right)^2,\label{cmapr}
\ea
maps the 4-sheeted $w$ plane, which describes the $n=2$ replicated manifold for multi-entropy, into a complex plane ${\bf C}$, whose coordinate is called $z$. Note that $w=w_1$,$w=w_2$ and $w=w_3$ corresponds to $z=\pm 1$, $z=0,\infty$ and $z=\pm i$. This map is explained in Fig.\ref{fig:replicam}.

We can calculate $\la T(w)\lb$ by the conformal transformation of energy stress tensor:
\ba
\la T(w)\lb=\left(\frac{dz}{dw}\right)^{-2}\left[\la T(z)\lb-\frac{c}{12}\{w:z\}\right],
\ea
where $\{w:z\}\equiv \frac{\de^3_z w}{\de w}-\frac{3}{2}\left(\frac{\de^2_z w}{\de_z w}\right)^2$ is the Schwarz derivative.
By noting $\la T(z)\lb=0$ on the flat plane, we obtain
\ba
\la T(w)\lb=\frac{c}{32}\cdot \frac{P(w)}{(w-w_1)^2(w-w_2)^2(w-w_3)^2},
\ea
where $P(w)$ is defined by 
\ba
P(w)&=&(w_1^2+w_2^2+w_3^2-w_1w_2-w_1w_3-w_2w_3)w^2\no
&&-(w_1^2w_2+w_1w_2^2+w_1^2w_3+w_1w_3^2+w_2^2w_3+w_2w_3^2-6w_1w_2w_3)w\no
&& +(w_1^2w_2^2+w_2^2w_3^2+w_1^2w_3^2-w_1^2w_2w_3-w_1w_2^2w_3-w_1w_2w_3^2).
\ea
However note that this $T(w)$ is the energy stress tensor on one of the four sheets. To obtain the total contribution for the replicated CFT with the central charge $4c$ on a plane instead of the individual CFT with the central charge $c$, we need to multiply the factor $4$:
\ba
\la T_{tot}(w)\lb=\frac{c}{8}\cdot \frac{P(w)}{(w-w_1)^2(w-w_2)^2(w-w_3)^2}.\label{ttot}
\ea

On the other hand, if we defined the conformal dimension of the twist operators $\sigma_i$ to be $h_i$  ($i=1,2,3$) the Ward identifty tells us 
\ba
&& \la T_{tot}(w)\sigma_1(w_1)\sigma_2(w_2)\sigma_3(w_3)\lb \no
&&=\left[\sum_{i=1}^3\frac{h_i}{(w-w_i)^2}+\frac{1}{w-w_i}\frac{\de}{\de w_i}\right]\la\sigma_1(w_1)\sigma_2(w_2)\sigma_3(w_3)\lb.\label{ward}
\ea
If we set 
\ba
h_1=h_2=h_3=\frac{c}{8}\left(=\frac{c}{24}(n^2-1)\right),
\ea
then the $\la T_{tot}\lb$ computed from
\ba
\la T_{tot}(w)\lb=\frac{\la T_{tot}(w)\sigma_1(w_1)\sigma_2(w_2)\sigma_3(w_3)\lb}
{\la\sigma_1(w_1)\sigma_2(w_2)\sigma_3(w_3)\lb}.,
\ea
by plugging the right hand side of (\ref{ward}) precisely agree with (\ref{ttot}). This confirms that $\sigma_i$ $(i=1,2,3)$ are primary fields with the conformal dimension $h_i=\frac{c}{8}$. Thus, the replicated partition function is given by 
\ba
\la\sigma_1(w_1)\sigma_2(w_2)\sigma_3(w_3)\lb=C^{(3)}_{2}\cdot  \left|(w_1-w_2)(w_2-w_3)(w_3-w_1)\right|^{-\frac{c}{4}}.
\ea
This gives the expected multi-entropy (see (\ref{sqn} for its definition)
\ba
S^{(3)}_{2}=\frac{c}{8}\log \frac{\left|(w_1-w_2)(w_2-w_3)(w_3-w_1)\right|}{\ep^3}-\frac{1}{2}\log C^{(3)}_{2},\label{q3c}
\ea
though we cannot fix the constant $C^{(3)}_{2}$ using this method.

\subsubsection{Liouville field analysis}
We can obtain the same result from the Liouville field theory analysis as follows.
The conformal map can be written as
\begin{align}
    w=\frac{w_3(w_1-w_2)(z^2-1)^2+w_1(w_3-w_2)(z^2+1)^2}{(w_2-w_1)(z^2-1)^2+(w_3-w_2)(z^2+1)^2}.
\end{align}
The Liouville action becomes
\begin{align}
    I_L = \frac{c}{24} \int dz^2 \left(
    \left|\frac{4w_2z^2(3+z^4) -w_1(z^2-1)^2(1+8z^2+3z^4) +w_3(z^2+1)^2(1-8z^2+3z^4) 
    }{z(z^4-1)\bigl((w_2-w_1)(z^2-1)^2+(w_3-w_2)(z^2+1)^2\bigr)}\right|^2 
    + \mu e^{2\varphi}
    \right).
\end{align}
$I_L$ diverges at $z=0, \pm 1, \pm i , \infty$ because of the UV divergence. We set the UV cut-off in the $w$-plane to be $\epsilon$.
\begin{align}
    \begin{cases}
        z=0, w=w_2: & \displaystyle |z|>
        \left|\frac{(w_3-w_1)\epsilon}{(w_2-w_1)(w_3-w_2)}\right|^{1/2}, \\
        z=\pm 1, w=w_1: & \displaystyle |z|>\left|\frac{(w_3-w_2)\epsilon}{(w_2-w_1)(w_3-w_1)}\right|^{1/2}, \\
        z=\pm i, w=w_3: & \displaystyle |z|>\left|\frac{(w_2-w_1)\epsilon}{(w_3-w_2)(w_3-w_1)}\right|^{1/2}, \\
        z=\infty, w=w_2: & \displaystyle |z|<
        \left|\frac{(w_2-w_1)(w_3-w_2)}{(w_3-w_1)\epsilon}\right|^{1/2}.
    \end{cases}
\end{align}
$I_L$ is evaluated as
\begin{align}
    I_L \simeq \frac{c}{12}\log\frac{|b-a|^5|c-b|^5}{|c-a|^3\epsilon^7}.
\end{align}
$\langle\sigma_1(a)\sigma_2(b)\sigma_3(c)\rangle$ is obtained by introducing the normalization factor $2^2I_L^{(1)}$, where we set $I_L^{(1)}$ as similar to (\ref{eq:NormalizationLiouville2interval}), 
\begin{align}
    I_L^{(1)} =&~ \frac{c}{12}\cdot4\cdot\frac{\log \left|\frac{w_3-w_2}{\epsilon}\right|+\log \left|\frac{w_2-w_1}{\epsilon}\right|}{2} \nonumber\\
    =&~\frac{c}{12}\log \left|\frac{\epsilon^4}{(w_3-w_2)^2(w_2-w_1)^2}\right|.
\end{align}
The result agrees with the previous calculation
\begin{align}
    \langle\sigma_1(w_1)\sigma_2(w_2)\sigma_3(w_3)\rangle \propto & e^{I_L-4I_L^{(1)}} \nonumber\\
    =&~ \left|\frac{(w_2-w_1)(w_3-w_2)(w_3-w_1)}{\epsilon^3}\right|^{-\frac{c}{4}}.
\end{align}

\subsection{Analysis of $q=3$ and $n=2$ in the disconnected case}
\label{sec:q=3n=2general}

\begin{figure}[hhh]
  \centering
   \includegraphics[width=10cm]{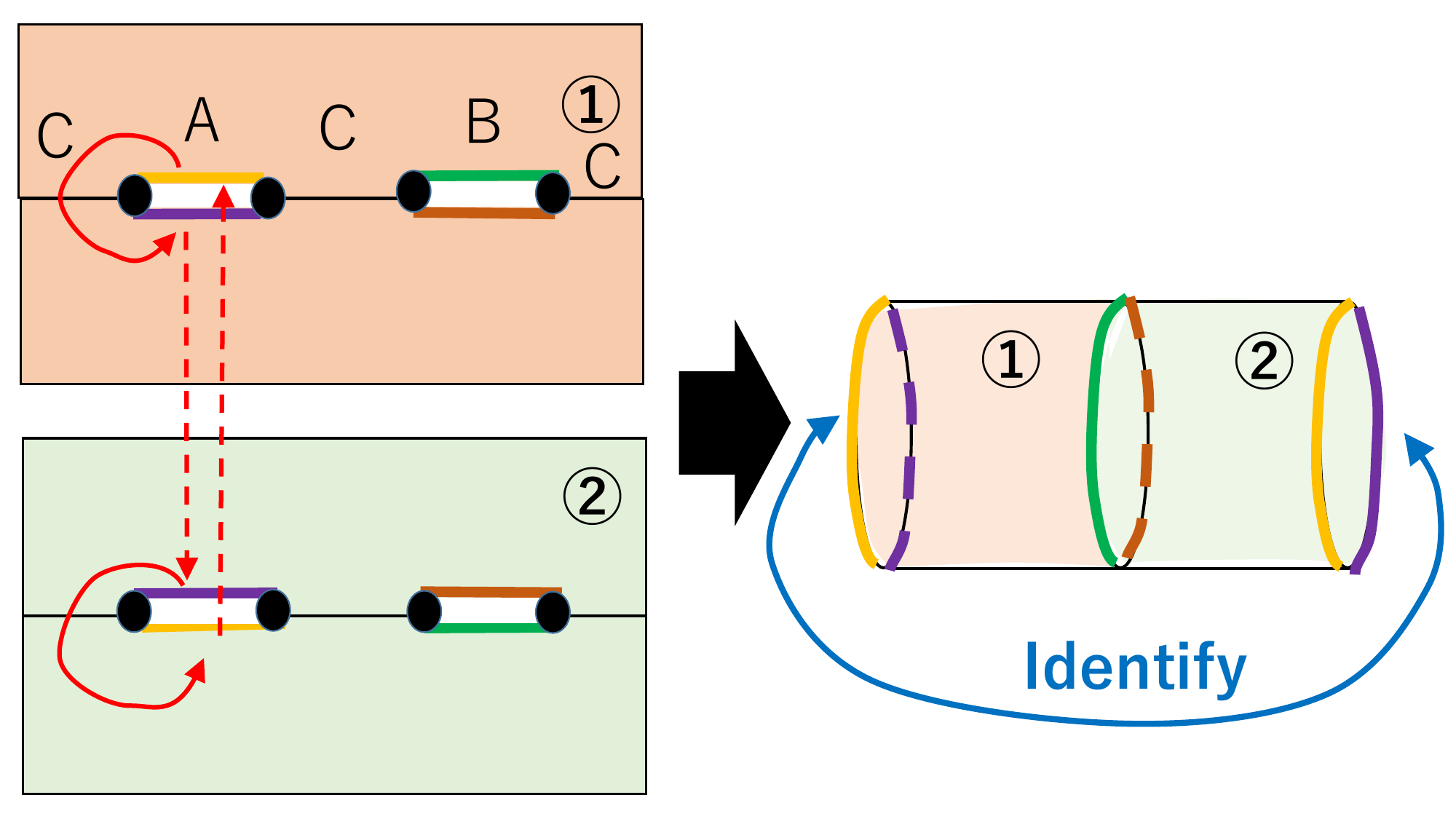}
  \caption{The conformal map to a torus for the replica method calculation for the second Renyi entropy with the disconnected subsystem.} 
\label{fig:replicamani}
\end{figure}

\begin{figure}[hhh]
  \centering
     \includegraphics[width=10cm]{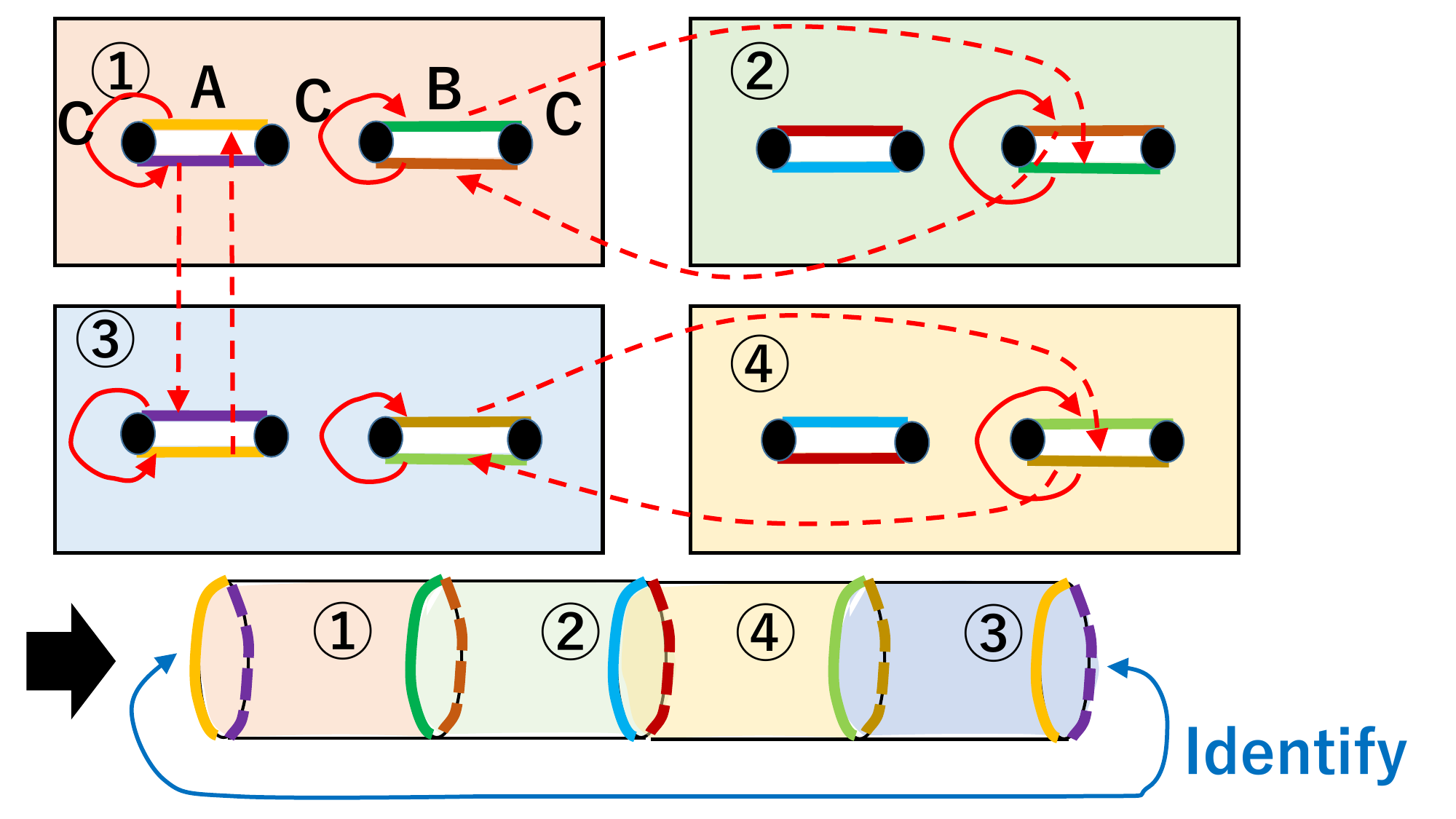}
  \caption{The conformal map to a torus for the replica method calculation for the $n=2$ multi entropy with the separated subsystems.} 
\label{fig:replicamania}
\end{figure}

Now we consider a generalization of the previous result to the case where the subsystem $A$ and $B$ are disconnected and their complement $C$ consists of two disconnected intervals. 
We note that the Riemann surface for this $n=2$ three-partite entropy is obtained by making double the length of the Riemann surface (see Fig.\ref{fig:replicamani}) for the $n=2$ replica method of Renyi entropy, as depicted in 
Fig.\ref{fig:replicamania}.

In the 2nd Renyi entropy we know that the twist operator four point function is expressed in terms of the torus partition function \cite{Lunin:2000yv,Headrick:2010zt}
\ba
\la\sigma(0)\sigma(\eta)\sigma(1)\sigma(\infty)\lb
=\left[2^8\eta(1-\eta)\right]^{-\frac{c}{12}}\cdot Z_{torus}(\tau), \label{renee}
\ea
where $\tau$ is the moduli of (rectangilar) torus and is related to the cross ratio of the twist operator insertions via
\ba
\eta=
\left[\frac{\theta_2(\tau)}{\theta_3(\tau)}\right]^{4},\ \ \ \ 
1-\eta=\left[\frac{\theta_4(\tau)}{\theta_3(\tau)}\right]^{4},
\label{modlitet}
\ea
where $\theta_{2},\theta_3$ and $\theta_4$ are standard elliptic theta functions (see e.g. the textbook \cite{Yellow}).

Now, in our replica method for the $n=2$ multi entropy, we have four sheets instead of two ones. Thus we can obtain the four point function by doubling the kinematical factor from the Weyl anomaly. Also the torus moduli is doubled as $\ti{\tau}=2\tau$. Thus we obtain
\ba
\la\sigma_1(0)\sigma_2(\eta)\sigma_3(1)\sigma_4(\infty)\lb
=\left[2^8\eta(1-\eta)\right]^{-\frac{c}{6}}\cdot Z_{torus}(\ti{\tau}),
\label{twisft}
\ea
such that 
\ba
\eta=
\left[\frac{\theta_2(\ti{\tau}/2)}{\theta_3(\ti{\tau}/2)}\right]^{4},\ \ \ \ 
1-\eta=\left[\frac{\theta_4(\ti{\tau}/2)}{\theta_3(\ti{\tau}/2)}\right]^{4}.
\ea
Note that for our $q=3$ multi entropy calculation, we can choose $\sigma_2=\bar{\sigma}_1$ and $\sigma_4=\bar{\sigma}_3$.

It is also straight forward to generalize this to the general four points $x_1,x_2,x_3,x_4$  with the cross ratio 
\ba
\eta=\frac{x_{12}x_{34}}{x_{13}x_{24}}.
\ea
We can use the standard formula (see e.g. \cite{Yellow} or appendix A of \cite{Kusuki:2017upd}):
\ba
&& \la O_1(x_1)O_2(x_2) O_3(x_3)O_4(x_4)\lb \no
&& =\left[x_{12}x_{13}x_{14}x_{23}x_{24}x_{34}\right]^{-\frac{4h}{3}}\cdot\eta^{\frac{4}{3}h}(1-\eta)^{\frac{4}{3}h}\cdot
\la O_1(0) O_2(\eta) O_3(1) O_4(\infty)\lb,
\ea
where we assumed the chiral conformal dimensions of $O_i$ $(i=1,2,3,4)$ are all given by $h$.

By applying this formula to (\ref{twisft}), we eventually find the final formula
\ba
\la \sigma_1(x_1)\sigma_2(x_2) \sigma_3(x_3) \sigma_4(x_4)\lb =2^{-\frac{4}{3}c}\cdot\left[x_{12}x_{13}x_{14}x_{23}x_{24}x_{34}\right]^{-\frac{c}{6}}
\cdot Z_{torus}(\ti{\tau}).
\ea

We can evaluate the multi-entropy $S^{(3)}_{2}$ by setting $n=2$ in the original definition
\ba
S^{(3)}_{n}=\frac{1}{n(1-n)}\log\la\sigma^{(n)}_1(x_1)\sigma^{(n)}_2(x_2)\sigma^{(n)}_3(x_3)\sigma^{(n)}_4(x_4)\lb,
\ea
as follows
\ba
S^{(3)}_{2}
&=&-\frac{1}{2}\log\la\sigma_1(x_1)\sigma_2(x_2)\sigma_3(x_3)\sigma_4(x_4)\lb,\no
&=&\frac{2c}{3}\log 2 +\frac{c}{12}\log\left[x_{12}x_{13}x_{14}x_{23}x_{24}x_{34}\right]-\frac{1}{2}\log Z_{torus}(\ti{\tau}).
\ea
Below we will study its behavior in various limits.

\subsubsection{The limit $x_2\to x_3$}
If we take the limit $x_2\to x_3$, which is equivalent to 
\ba
\tau\to 0, \ \ \ \eta\to 1,
\ea
we obtain
\ba
\la \sigma_1(x_1)\sigma_2(x_2) \sigma_3(x_3) \sigma_4(x_4)\lb 
\simeq 2^{-\frac{4}{3}c}\cdot x_{23}^{-\frac{c}{4}}\cdot\left(x_{12}x_{24}x_{14}\right)^{-\frac{c}{4}},
\label{xxw}
\ea
where we employed 
\ba
\eta\simeq 1-2^4 e^{-\frac{2\pi i}{\ti{\tau}}},\ \ \ Z_{torus}(\ti{\tau})\simeq 
e^{\frac{\pi ci}{6\ti{\tau}}}\simeq 2^{\frac{c}{3}}(1-\eta)^{-\frac{c}{12}}.
\ea

Thus, in the limit, the multi-entropy is computed as follows:
\ba\label{eq:q3_n2_23limit}
S^{(3)}_{2}\simeq \frac{c}{2}\log 2 +\frac{c}{8}\log \frac{x_{23}}{\ep}+ \frac{c}{8}\log \left[\frac{x_{12}x_{24}x_{14}}{\ep^3}\right],\label{xxxa}
\ea
where we insert the cut off $\ep$ dependence as usual. 

Even though we are not expecting to have a simple gravity dual because we are working with the Renyi quantity ($n=2$), it is still instructive to give a holographic interpretation of $S^{(3)}_{2}$ qualitatively, just ignoring the back-reaction issue. As depicted in the left of Fig.\ref{fig:phase}, we can interpret (\ref{xxxa}) as a sum of the contribution from the geodesic which connects $x_2$ and $x_3$ and the one from geodesics with three-way intersection as computed in (\ref{q3c}), plus a certain constant.

\subsubsection{The limit $x_1\to x_2$}
\begin{figure}[hhh]
  \centering
     \includegraphics[width=10cm]{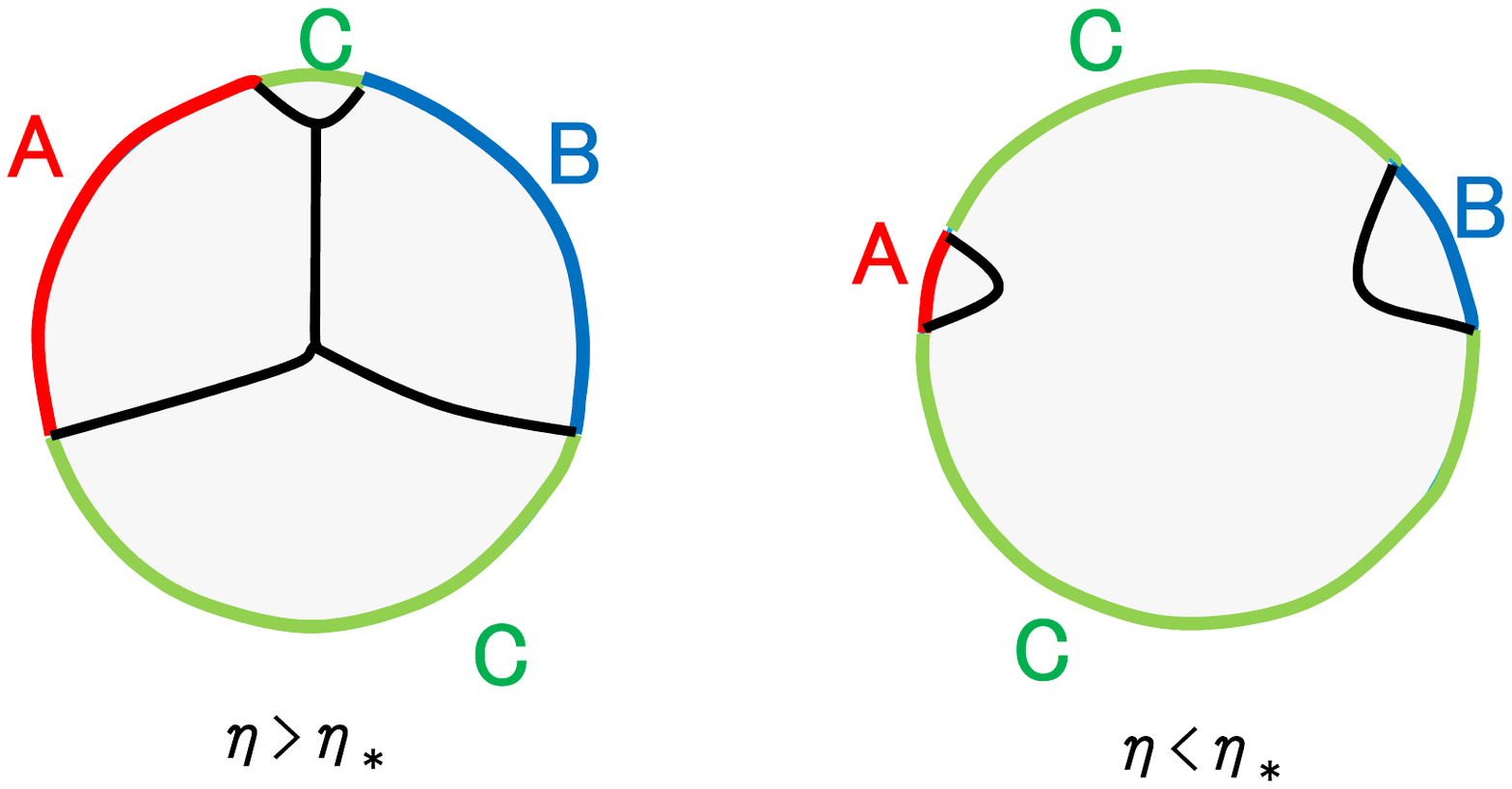}
  \caption{The holographic interpretation of the two phases of the multi-entropy $S^{(3)}_{2}$. The left is in the phase $\eta>\eta_*$. The right is in the phase $\eta>\eta_*$.} 
\label{fig:phase}
\end{figure}

On the other hand, in the opposite limit $x_1\to x_2$, we find 
\ba
\eta\simeq 2^4 e^{\frac{\pi i}{2}\ti{\tau}}\to 0,\ \ \ Z_{torus}(\ti{\tau})\simeq e^{-\frac{\pi c i}{6}\ti{\tau}}
=2^{\frac{4}{3}c}\eta^{-\frac{c}{3}}.
\ea
This leads to the multi-entropy 
\ba
S^{(3)}_{2}\simeq \frac{c}{4}\log \frac{x_{12}}{\ep}+\frac{c}{4}\log \frac{x_{34}}{\ep}. \label{xxxb}
\ea
This may be easily understood by taking OPE between $\sigma_1$ and $\sigma_2=\bar{\sigma}_1$ and focusing on the identity sector as the intermediate states.  The qualitative holographic interpretation for this is depicted in the right of Fig.\ref{fig:phase}.

\subsubsection{Phase Transition in holographic CFT}

In holographic CFTs with large $c$ limit, the torus partition function experiences the phase transition \cite{Witten:1998qj,Witten:1998zw} as follows
\ba\label{eq:torusPT}
Z_{torus}(\tau)=\mbox{Min}\left[e^{-\frac{\pi i}{6}c\ti{\tau}},\ e^{\frac{\pi i}{6\ti{\tau}}c}\right],
\ea
where the phase transition occurs at $\tau=i$, which is equivalent to 
\ba
\eta_*=\left(\frac{\theta_2(i/2)}{\theta_3(i/2)}\right)^4\simeq 0.9705.
\ea
Thus for $\eta>\eta_*$, we find 
\ba
S^{(3)}_{2}=\frac{2c}{3}\log 2 +\frac{c}{12}\log\left[x_{12}x_{13}x_{14}x_{23}x_{24}x_{34}\right]-\frac{\pi ic}{12\tau}.
\ea
On the other hand, for $\eta<\eta_*$, we obtain 
\ba
S^{(3)}_{2}=\frac{2c}{3}\log 2 +\frac{c}{12}\log\left[x_{12}x_{13}x_{14}x_{23}x_{24}x_{34}\right]+\frac{\pi ic\tau}{12}.
\ea

\subsubsection{Reduction to Coincident Limit}

\begin{figure}[H]
\centering
 \includegraphics[width=10cm]{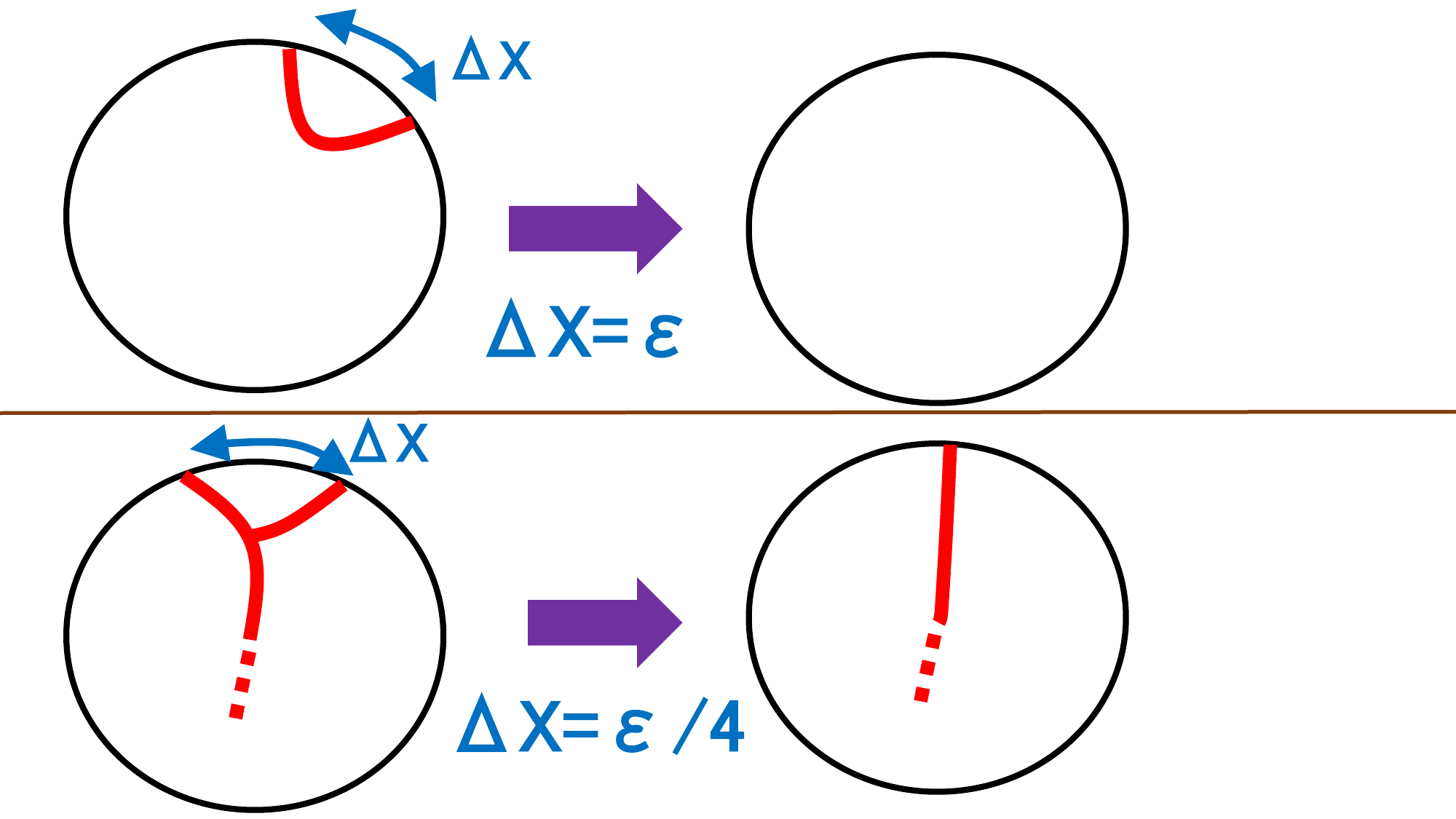}
\caption{Eliminating a simple interval (up) 
and an interval with a vertex (down).}
\label{Reduction}
\end{figure}

When we reduce the Renyi entropy $S^{(2)}_{n}$  for the disconnected intervals $AB$ to a single interval by taking the limit $x_2\to x_3$, we can simply set
\ba
\Delta x=\ep,  \label{yyyw}
\ea
where $\Delta x$ is the length of interval which we would like to eliminate. In the limit $x_2\to x_3$, the 2nd Renyi entropy computed from (\ref{renee}) is evaluated as 
\ba
S^{(2)}_{2}\simeq \frac{c}{4}\log\frac{|x_1-x_4|}{\ep}+\frac{c}{4}\log\frac{|x_2-x_3|}{\ep}.
\ea
If we set $x_3-x_2=\Delta x$, then indeed we can recover the single interval 2nd Renyi entropy result $S^{(2)}_2=\frac{c}{4}\log \frac{|x_4-x_1|}{\ep}$ \cite{Calabrese:2004eu}, as expected. In this way, we find that to eliminate an interval we can set its length to (\ref{yyyw}) as depicted in the upper panel of Fig.\ref{Reduction}.

Let us find an analogous rule which allows us to reduce the multi entropy in the disconnected case to that in the coincident case, and the further to a single interval Renyi entropy. For our $q=3$ multi entropy $S_{ABC}$, we argue that the correct rule is to set the length of interval to the following value:
\ba
\label{eq:limit:eps/4}
\ti{\Delta} x=\frac{\ep}{4},
\ea
as sketched in the lower panel of Fig.\ref{Reduction}. Even though this rule is different from (\ref{yyyw}), this is not surprising because in this case we need to eliminate a vertex which joins three legs, in addition.

If we apply this to (\ref{xxxa}), by substituting $x_{23}=\ti{\Delta}x (=\ep/4)$, then we get 
\ba
S^{(3)}_{2}\simeq \frac{c}{4}\log 2  +\frac{c}{8}\log \left[\frac{x_{12}x_{24}x_{14}}{\ep^3}\right].\label{redmue}
\ea
This suggests $\kappa^{(3)}_2$, which is defined by 
\ba
\kappa^{(3)}_{n}=S^{(3)}_{n}-\frac{1}{2}\left(S^{(2)}_n(A)+S^{(2)}_n(B)+S^{(2)}_n(C)\right).
\ea
and characterizes the tripartite entanglement, is given by 
\ba
\kappa^{(3)}_2=\frac{c}{4}\log 2.
\ea
As a consistency check, let us further reduce (\ref{redmue}) by setting $x_{14}=\ti{\Delta} x(=\ep/4)$. This leads to 
\ba
\frac{c}{4}\log \left[\frac{x_{12}}{\ep}\right]=S^{(2)}_{2},
\ea
which agrees with the single interval Renyi entropy as expected.

\subsection{n=2 q=4}
\begin{figure}[H]
\begin{tabular}{cc}
\centering
\includegraphics[width=.45\textwidth,page=22]{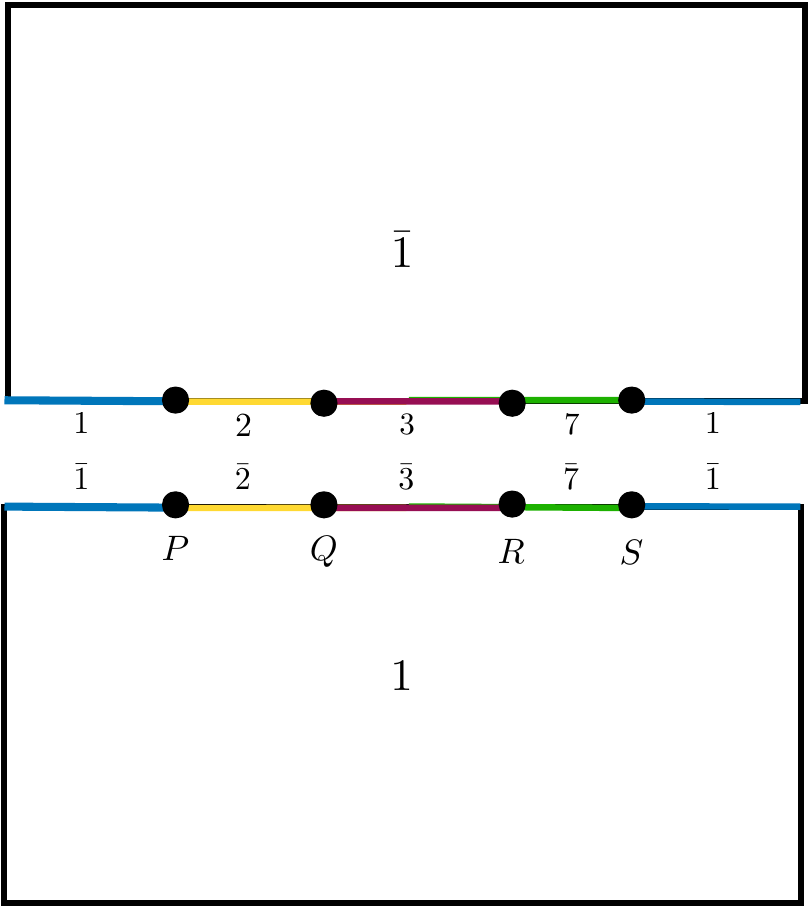}&
\includegraphics[width=.45\textwidth,page=23]{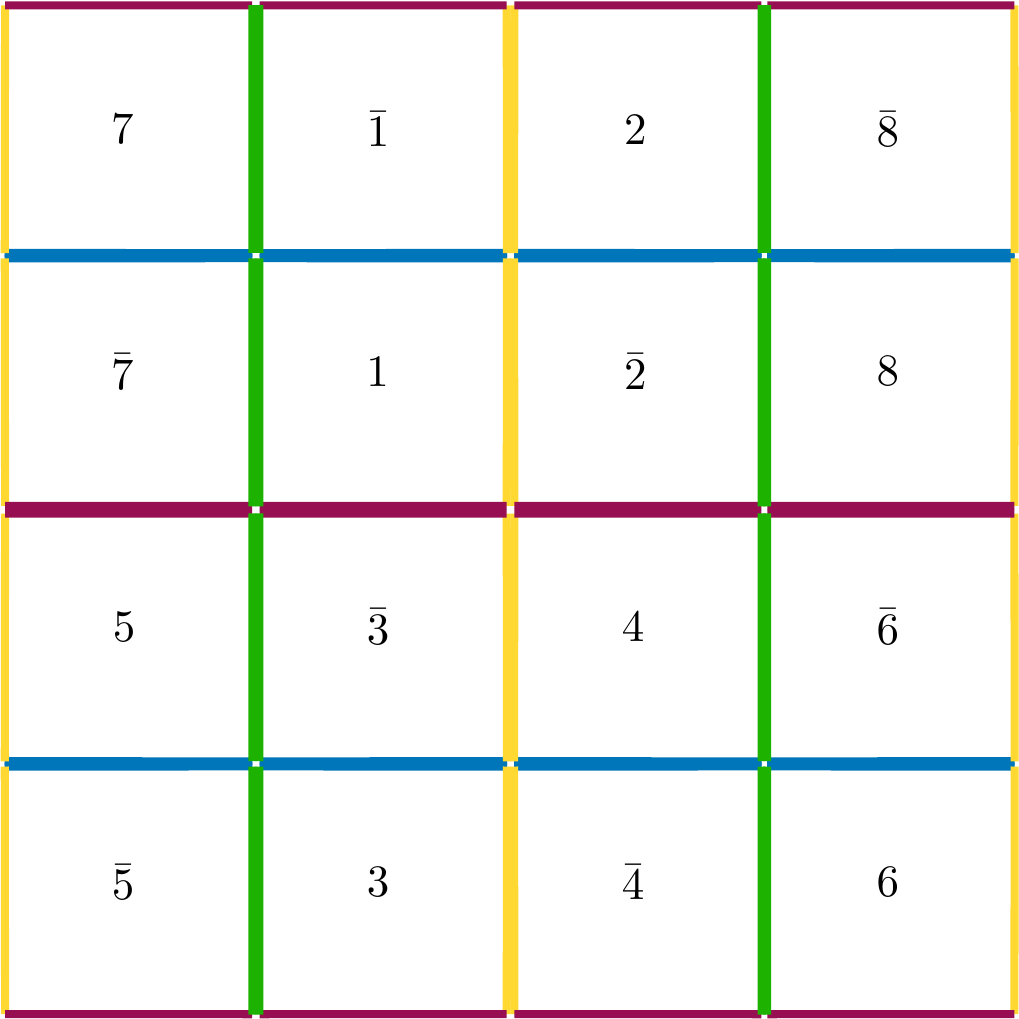}
\end{tabular}
\caption{\label{fig:q4replicated} Replicated geometry for $q=4$ and $n=2$.}
\end{figure}

We now consider the case of four twist operator insertions for $n=2$ with finite group symmetry $\mathbb{Z}_2^3$. with group presentation $\langle a,b,c|a^2=b^2=c^2=e\rangle$. For the three generators $a,b,c$ we pick an explicit representation in terms of permutations to be 
\be
\begin{split}
   a:\quad (12)(34)(56)(78)\\
   b:\quad (13)(24)(57)(68)\\
   c:\quad (17)(28)(35)(46)
\end{split}
\ee
 Now we choose the three twist operators to be charged under the group elements $a,ab,c,bc$ which determines the monodromies:
\be
\begin{split}
\sigma_{a}(x_1):& \quad (12)(34)(56)(78) \\
\sigma_{ab}(x_2):& \quad (14)(23)(58)(67) \\
\sigma_{c}(x_3):& \quad (17)(28)(35)(46) \\
\sigma_{bc}(x_4):& \quad (15)(26)(37)(48)
\end{split}
\ee
It can be verified that replicated geometry is a torus. The explicit construction is shown in figure \ref{fig:q4replicated}.
Because the resulting replica manifold is again a rectangular torus we can use the result
\ba
\la O_1(0)O_2(\eta)O_3(1)O_4(\infty)\lb
=\left[2^8\eta(1-\eta)\right]^{-\frac{c}{3}}\cdot Z_{torus}(\tau),
\ea
and making a mobuis transformation we find
\ba
&& \la \sigma_{a}(x_1)\sigma_{ab}(x_2) \sigma_{c}(x_3)\sigma_{bc}(x_4)\lb \no
&& =\left[x_{12}x_{13}x_{14}x_{23}x_{24}x_{34}\right]^{-\frac{4h}{3}}\cdot\eta^{\frac{4}{3}h}(1-\eta)^{\frac{4}{3}h}\cdot
\la O_1(0) O_2(\eta) O_3(1) O_4(\infty)\lb,\no
&& =2^{-\frac{8c}{3}}\cdot\left[x_{12}x_{13}x_{14}x_{23}x_{24}x_{34}\right]^{-\frac{c}{3}}\cdot Z_{torus}(\tau)
\ea
Because of the cycle structure of the twist operators we now have $h=\frac{c}{4}$. Even though we have more copies the ratio of the sides of the torus remains unchanged thus the moduli $\tau$ is identical to the one in (\ref{modlitet}). The Renyi multi-entropy is given by
\ba
S_2^{(4)}=\frac{2c}{3}\log 2 +\frac{c}{12}\log\left[x_{12}x_{13}x_{14}x_{23}x_{24}x_{34}\right]-\frac{1}{4}\log Z_{torus}(\tau).
\ea

Here since the modular parameter is $\tau$ there will be a phase transition at $\eta'_*=\frac{1}{2}$.

For holographic theories we find that for $\eta <\eta'_*$
\ba
S_2^{(4)}=\frac{2c}{3}\log 2 +\frac{c}{12}\log\left[x_{12}x_{13}x_{14}x_{23}x_{24}x_{34}\right]+\frac{\pi ic\tau}{24}\no
\approx \frac{c}{2}\log 2 +\frac{c}{12}\log\left[\frac{x^{\frac{3}{2}}_{12}x^{\frac{3}{2}}_{34}x_{14}x_{23}x_{13}^{\frac{1}{2}}x_{24}^{\frac{1}{2}}}{\epsilon^6}\right]
\ea
Where we have used
\be
\eta \sim 2^4e^{i\pi  \tau},\quad Z_{torus}\sim 2^{\frac{2c}{3}}\eta^{-\frac{c}{6}}
\ee
\be
-\frac{1}{4}\log(Z_{torus})\sim -\frac{c}{6}\log(2)+\frac{c}{24}\log\left(\frac{x_{12}x_{34}}{x_{13}x_{24}}\right)
\ee
On the other hand, for $\eta>\eta'_*$, we obtain 
\ba
S_2^{(4)}=\frac{2c}{3}\log 2 +\frac{c}{12}\log\left[x_{12}x_{13}x_{14}x_{23}x_{24}x_{34}\right]-\frac{\pi ic}{24\tau}\no
\approx \frac{c}{2}\log 2  +\frac{c}{12}\log\left[\frac{x_{12}x_{34}x_{14}^{\frac{3}{2}}x_{23}^{\frac{3}{2}}x_{13}^{\frac{1}{2}}x_{24}^{\frac{1}{2}}}{\epsilon^6}\right]
\ea
with
\be
\eta \sim 1-2^4e^{-\frac{i\pi}{  \tau}},\quad Z_{torus}\sim 2^{\frac{2c}{3}}(1-\eta)^{-\frac{c}{6}}
\ee
\be
-\frac{1}{4}\log(Z_{torus})\sim -\frac{c}{6}\log(2)+\frac{c}{24}\log\left(\frac{x_{14}x_{23}}{x_{13}x_{24}}\right)
\ee
\subsubsection{Coincident Limit}
In the limit $x_2\rightarrow x_1$ or $\eta\rightarrow0$ the twist operators $\sigma_{a}$ and $\sigma_{ab}$ fuse and we have
\be
\begin{split}
\sigma_b(x_1):& \quad (13)(24)(57)(68)\\
\sigma_{c}(x_3):& \quad (17)(28)(35)(46)\\
\sigma_{bc}(x_4):& \quad (15)(26)(37)(48).
\end{split}
\ee
the resulting replicated manifold has genus $g=-1$ which is consistent with two disjoint sphere. This makes sense as the cycles factorize into those between copies $1,3,5,7$ and $2,4,6,8$. In fact the resulting disconnected manifold is the same as two copies of what was found for $q=3,n=2$.

The limit $x_2\rightarrow x_3$ or $\eta\rightarrow1$ has an identical structure with the factorization now between copies $1,2,5,6$ and $3,4,7,8$.

\begin{figure}[H]
  \centering
     \includegraphics[width=10cm]{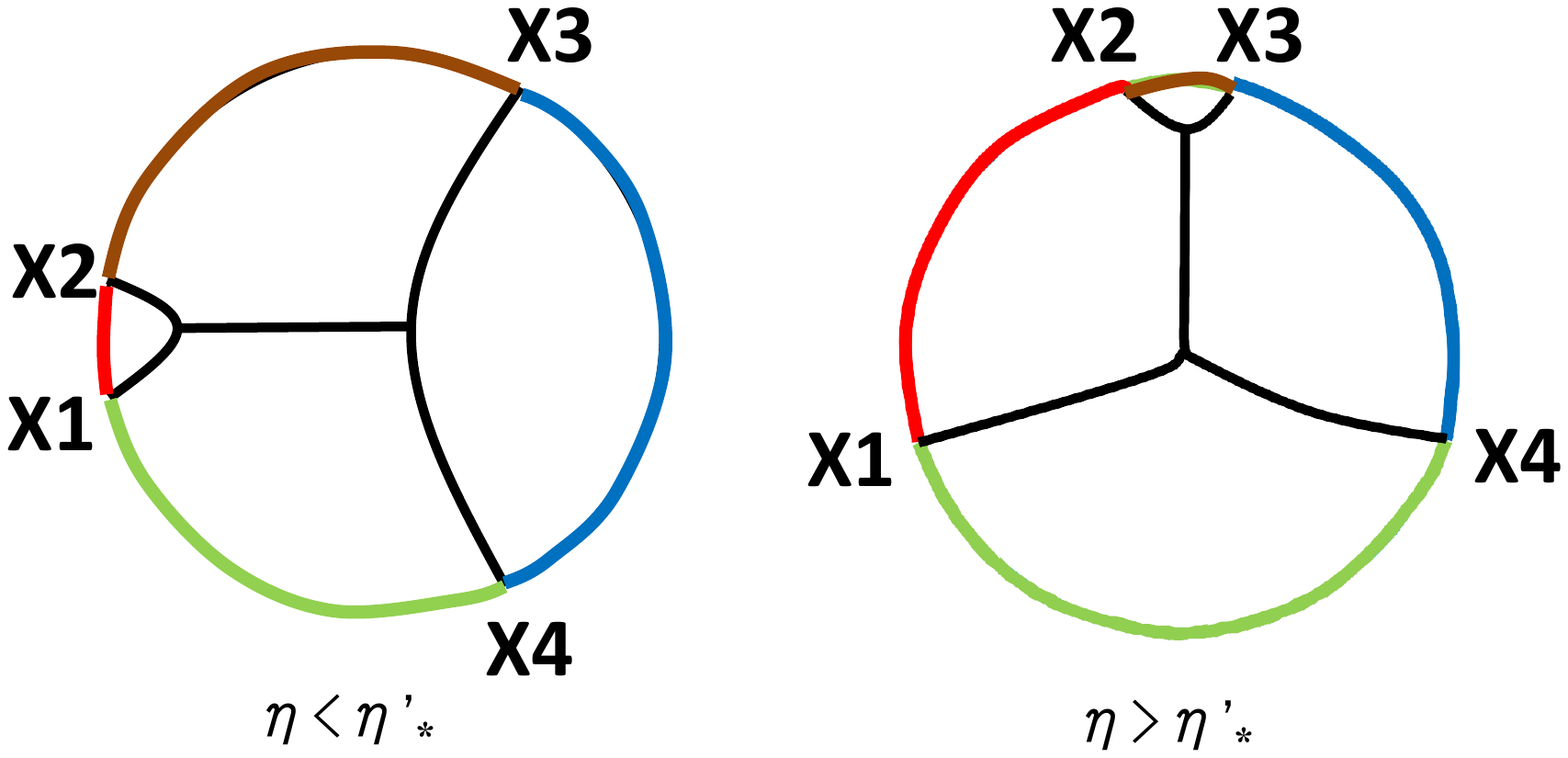}
  \caption{The holographic interpretation of the two phases of the $(n,q)=(2,4)$ multi-entropy $S^{(2)}_{A:B:C:D}$. The left is in the phase $\eta>\eta_*$. The right is in the phase $\eta>\eta_*$.} 
\label{fig:phase2}
\end{figure}

Now we examine the resulting renyi multi-entropy in these limits. To do so we use the expansion of the renyi multi entropy for small(large) $\eta$ and then take $x_2$ to the appropriate coincident limit. We find

\be
\begin{split}
    \eta\rightarrow0,\quad x_2\rightarrow x_1:\quad \frac{c}{2}\log 2+\frac{c}{8}\log\frac{x_{12}}{\epsilon}+\frac{c}{8}\log\frac{x_{13}x_{14}x_{34}}{\epsilon^3}\\
    \eta\rightarrow1,\quad x_2\rightarrow x_3:\quad \frac{c}{2}\log 2+\frac{c}{8}\log\frac{x_{23}}{\epsilon}+\frac{c}{8}\log\frac{x_{13}x_{14}x_{34}}{\epsilon^3}
\end{split}
\ee
which is the expected answer of \eqref{eq:q3_n2_23limit}. A qualitative interpretation in terms of gravity dual is presented in Fig.\ref{fig:phase2}.

It is also straightforward to confirm that by setting $x_{12}=\ti{\Delta} x$ (or 
$x_{23}=\ti{\Delta} x$) in the limit $\eta\to 0$ (or $\eta\to 1$), we can recover 
a half of the $(n,q)=(2,3)$ limit result (\ref{redmue}), as expected.

\subsection{$n=3$ $q=3$}

\begin{figure}[H]
\begin{tabular}{cc}
\centering
\includegraphics[width=.45\textwidth,page=17]{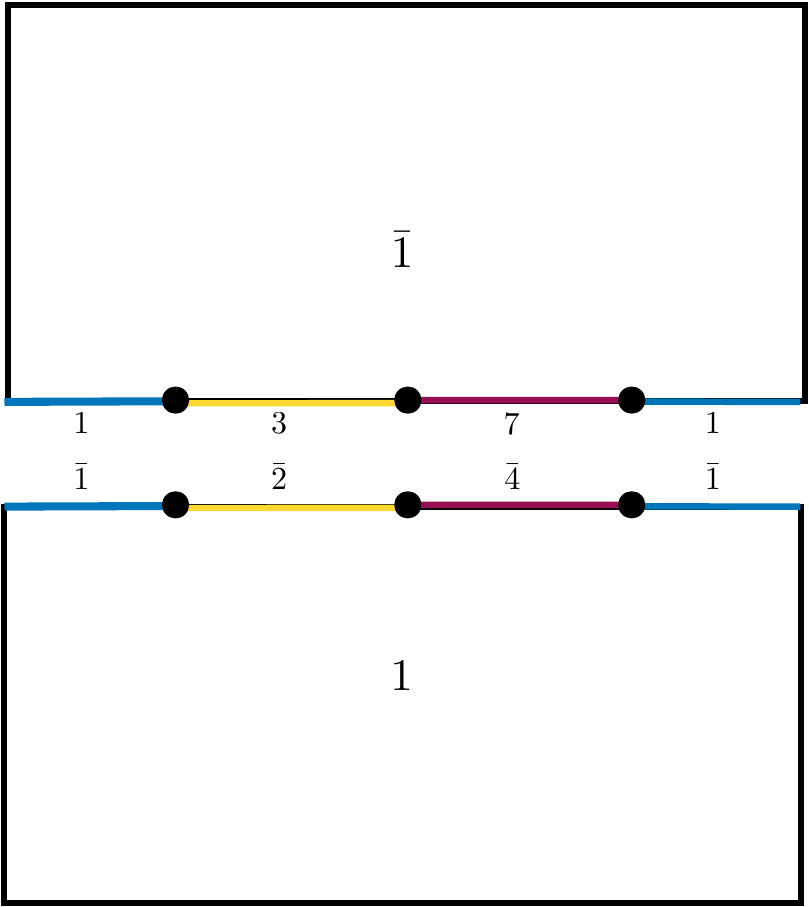}&\qquad\qquad
\includegraphics[width=.25\textwidth,page=18]{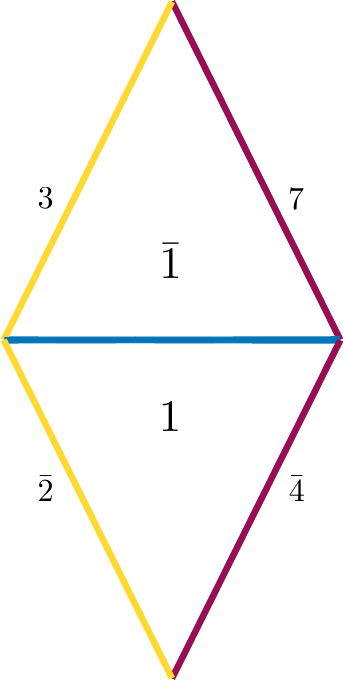}
\end{tabular}
\caption{\label{fig:n3q3replicated} L: A single copy of the CFT split into its bra and ket in the gravitational path integral. The twist operator insertions are shown at the real axis. The twist operators divide the real axis into three regions. We choose to glue the region $O$ (shown in blue) such that the gravitation path integral corresponds to the reduced density matrix $\rho_{AB}$. Along the four remaining intervals are shown the corresponding copy it will be glued to to form the replica manifold. These are done in accordance with the monodromies of the twist operators. R: After applying the map \eqref{eq:schw_tri} each half plane is mapped to an equiangular flat triangle with angle $\frac{\pi}{3}$.  }
\end{figure}

\begin{figure}[H]
  \centering
   \includegraphics[width=.8\textwidth,page=19]{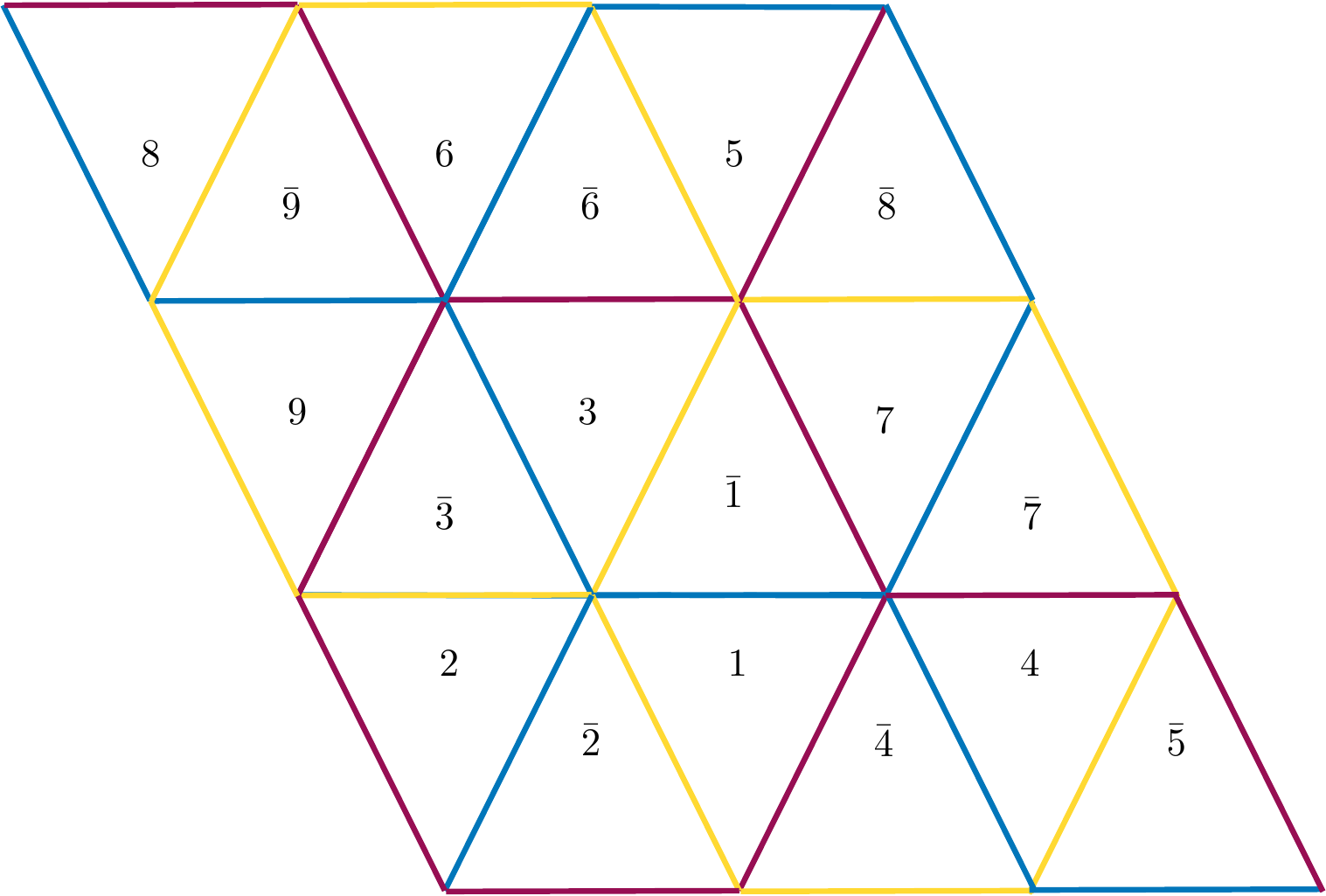}
  \caption{The replica manifold associated with the renyi multi-entropy $S_3^{(3)}$. Each fundamental region consists of two equiangular euclidean triangles each with with internal angles $\frac{\pi}{3}$. The nine fundamental regions are glued according to the explicit permutation representation for the mondromies of the twist operators. They tile a torus with modular parameter $\tau=e^{\frac{i\pi}{3}}$. In the figure opposite sides are identified. } 
\label{fig:q3n3replica}
\end{figure}

We now consider the case of three twist operator insertions for $n=3$ with finite group symmetry $\mathbb{Z}_3^2$ with group presentation $\langle a,b|a^3=b^3=e\rangle$. For the two generators $a,b$ we pick an explicit representation in terms of permutations to be 
\be
\begin{split}
   a:\quad (123)(456)(789)\\
   b:\quad  (159)(267)(348)\\
\end{split}
\ee
 Now we choose the three twist operators to be charged under the group elements $a,ab,ab^2$ which determines the monodromies:
\be
\begin{split}
\sigma_a(x_1):& \quad (123)(456)(789)\\
\sigma_{ab}(x_2):& \quad (168)(249)(357)\\
\sigma_{ab^2}(x_3):& \quad  (174)(285)(396)
\end{split}
\ee
For each copy we can use the map \eqref{eq:schw_tri} which takes each half plane to a flat equilateral triangle with angles $\frac{\pi}{3}$ (see figure \ref{fig:n3q3replicated}). The resulting replica manifold is given by a torus tiled by equilateral triangles (see figure \ref{fig:q3n3replica}). This fixes the modular parameter to $\tau=e^{\frac{i\pi}{3}}$.

In order to carry out the calculation we will use the full machinery of the uniformization method. In particular we found \cite{Lunin:2000yv,2010PhDT.......134A,2023arXiv230714434V} useful for the details presented here.

Up to modular transformations there is unique uniformization map $\Gamma:\Sigma\rightarrow \mathbb{CP}$ from the the torus back to the sphere with the correct properties. We take the images of the original twist operator insertions on the sphere to be located on the torus at
\be
z_1=0,\quad z_2=\frac{1}{3}+\frac{\tau}{3}, \quad  z_3=\frac{2}{3}+\frac{2\tau}{3}.
\ee
On the torus there are three such points of each type separated by periods of the lattice spacing. Each of these should be mapped respectively to the points $x_1,x_2,x_3$ so that $\Gamma(z_i)=x_i$.
We further demand that the series expansion take the form
\be\label{eq:uniexp}
\Gamma(z)=x_i+a_i(z-z_i)^{\omega_i}+\cdots
\ee
where the cycle structure of the twist operators fixes $\omega_i=3$. From these considerations we have the solution \cite{2020arXiv200211729E,Sansone_Gerretsen_1969}
\be\label{eq:map3}
\Gamma(z)=\frac{A\wp'(z|\tau)+B}{C\wp'(z|\tau)+D}, \quad \partial\Gamma(z)=\frac{6(AD-BC)\wp(z|\tau)^2}{(C\wp'(z|\tau)+D)^2}
\ee
which is the inverse of the map \eqref{eq:schw_tri} with all angles $\frac{\pi}{3}$. Here $\wp$ is the Weierstrass elliptic function which has the properties
\be
\wp(z_2|\tau)=\wp(z_3|\tau)=0, \quad \wp'(z_2|\tau)=i\sqrt{g_3},\quad \wp'(z_3|\tau)=-i\sqrt{g_3}
\ee
with the elliptic parameters given by $g_2=0$, $g_3=(2\pi)^{-6}\Gamma(\frac{1}{3})^{18}$.
$\wp$ and $\wp'$ are also related by the relation
\be\label{eq:wpdiffeq}
\wp'(z|\tau)^2=4\wp(z|\tau)^3-g_2\wp(z|\tau)-g_3.
\ee
The parameters $A,B,C,D$ which control the additional mobius transformation are chosen such that $\Gamma(z_i)=x_i$ is true:
\be
\begin{split}
A=(i\sqrt{g_3})^{-1}x_1(x_2-x_3), \quad B=x_1x_2+x_1x_3-2x_2x_3\\
C=(i\sqrt{g_3})^{-1}(x_2-x_3),\quad D=2x_1-x_2-x_3\\
AD-BC=\frac{2}{i\sqrt{g_3}}(x_1-x_2)(x_1-x_3)(x_2-x_3)
\end{split}
\ee
By expanding around each $z_i$ one can confirm that the first and second order terms are zero and that the $a_i$s are given by:
\be
a_1=i\sqrt{g_3}\frac{(x_1-x_2)(x_1-x_3)}{(x_2-x_3)}, \quad a_2=i\sqrt{g_3}\frac{(x_1-x_2)(x_2-x_3)}{(x_1-x_3)}, \quad a_3=i\sqrt{g_3}\frac{(x_1-x_3)(x_2-x_3)}{(x_1-x_2)}.
\ee

\begin{figure}[H]
\begin{tabular}{cc}
\centering
\includegraphics[width=.35\textwidth,page=36]{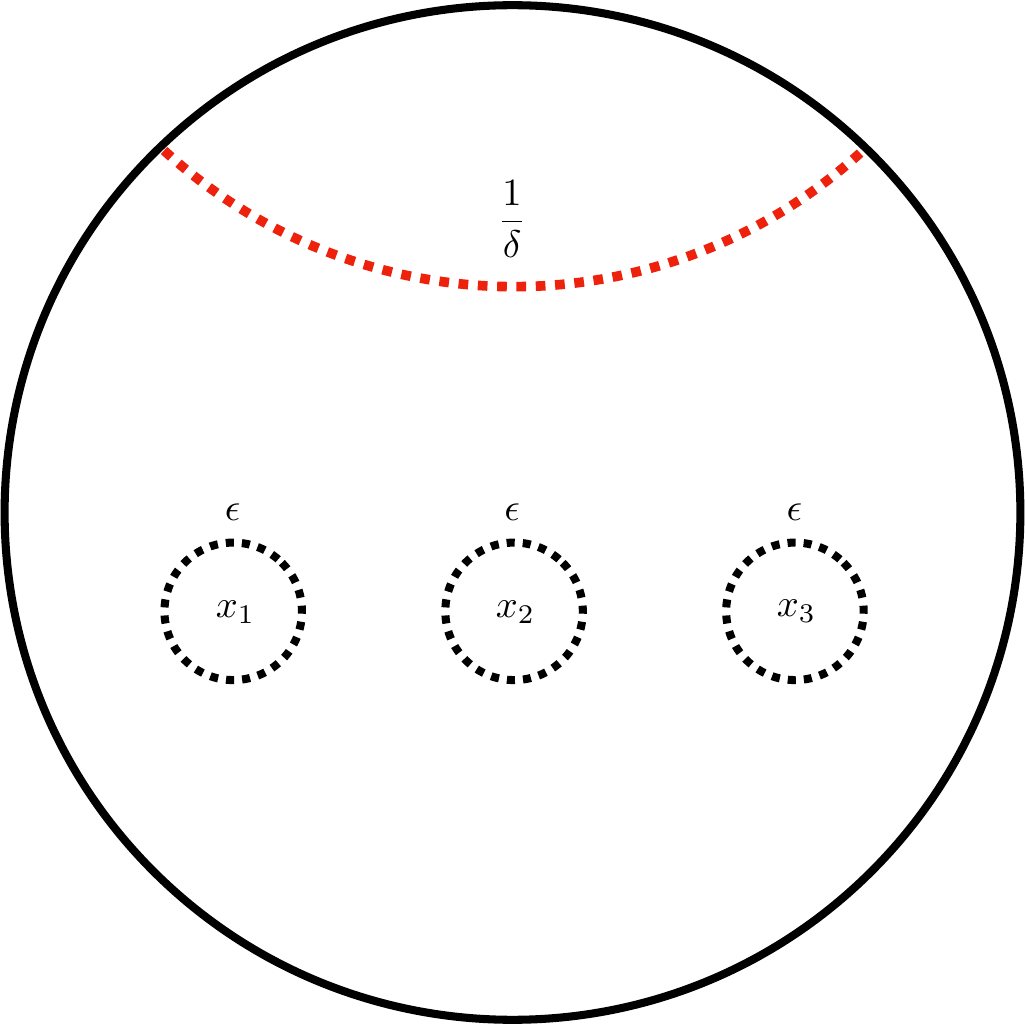}&\qquad\qquad
\includegraphics[width=.5\textwidth,page=35]{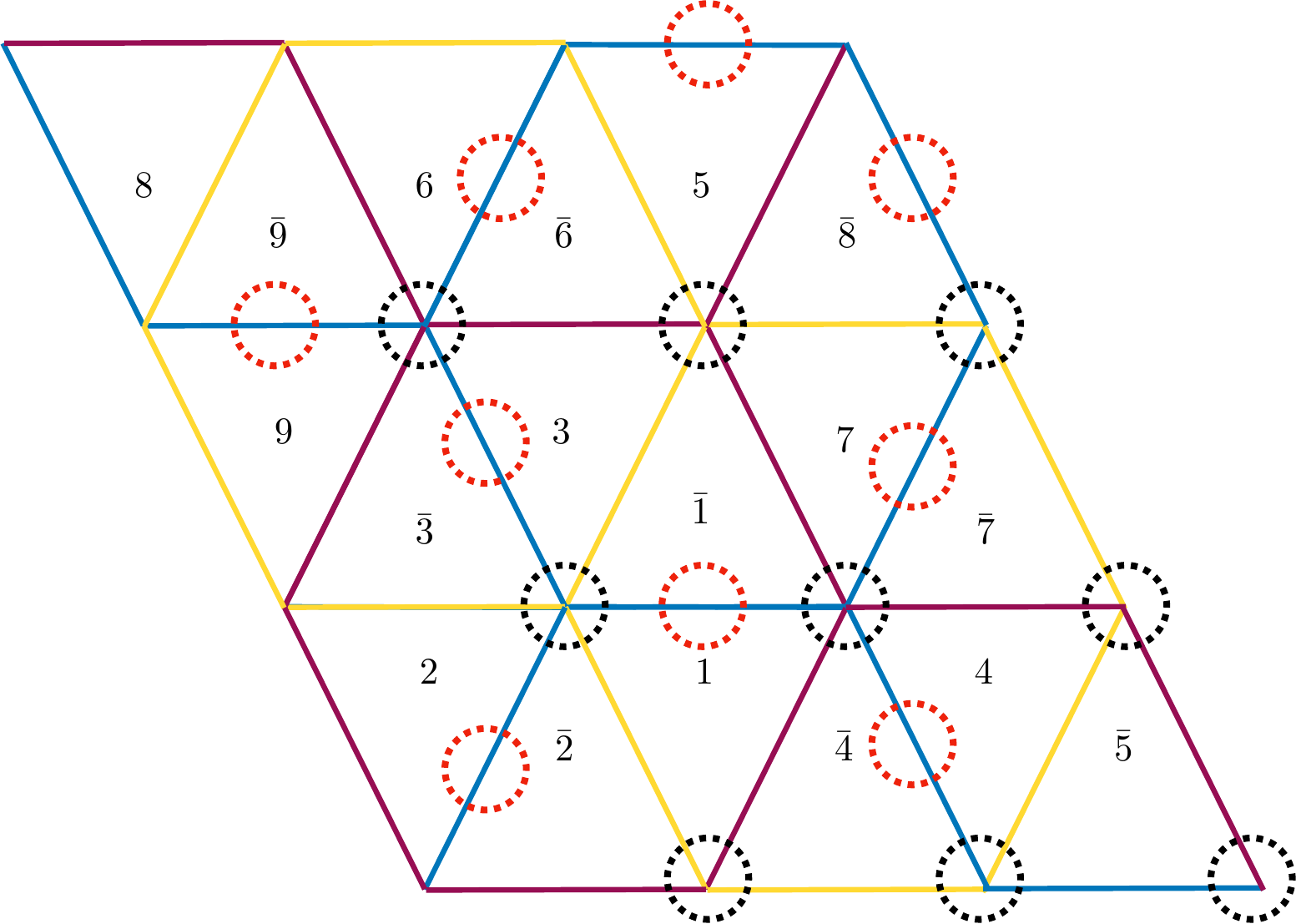}
\end{tabular}
\caption{\label{fig:n3cuttoffs} L: The sphere with twist operators insertions is regularized by introducing cut-offs. At each location we excise a region of radius $\epsilon$. We also impose a cut off at radius $\frac{1}{\delta}$ this allows us to use the flat metric on the sphere. R: The nine regular points and nine images of infinity on the torus which contribute to the Liouville action. At each we show the image of the cut-off. Note that because sides are identified some points are represented multiple times along the boundary. We have only marked each point once. As in the main text we assume here that $D=0$ which fixes the images of infinity to the half periods of the torus. For other values they will move to other locations.}
\end{figure}
From here it is necessary to determine the contributions of the map to the Liouville action \eqref{LVAB}. To do so we make some simplifying choices: On the sphere we choose the flat metric. In order for the manifold to have the correct curvature it is necessary to introduce a singularity which we take at a radius $|x|=\frac{1}{\delta}$. On the torus we also take the flat metric.

There will be two different contributions to the Liouville action: The first are the "regular points" which are the images of the twist operators on the torus. As a result of the cycle structure and the period structure of the torus there will be three images of each twist operator. To regularize these are taken with a cut-off radius of $\epsilon$ around each point on the sphere. The second type ``points of infinity" are the images of $x=\infty$ on the torus. These locations correspond to the solutions $\{z_*\}$ of $C\wp'(z_*|\tau)+D=0$ and result because of the needed curvature on the sphere. They are regularized with radius $|z-z_*|=|\Gamma^{-1}\left(\frac{1}{\delta}\right)|$.

Integrating \eqref{LVAB} by parts and using that the metric is flat the Liouville potential may be written in the form
\be
S_L=\frac{c}{96\pi}i\int_{\partial\Sigma}\phi\partial\phi dz
\ee
where $\partial\Sigma$ are the boundaries created by introducing the cut-offs $\epsilon,\frac{1}{\delta}$ on the sphere. In addition there is a convention that external contours are always taken with opposite orientation. 

The contribution of each regular point depends only on the parameters $a_i,\omega_i$. Using the expansion \eqref{eq:uniexp} we find the potential to be
\be
\partial_z \Gamma \approx \omega_ia_i(z-z_i)^{\omega_i-1} \implies \phi \approx 2\log(\omega_i|a_i| |z-z_i|^{\omega_i-1}).
\ee
Parameterizing the path $|z-z_i|=\left(\frac{\epsilon}{|a_i|}\right)^{\frac{1}{\omega_i}} e^{i\theta}, \theta\in[0,2\pi]$ the total contribution to the action is
\be
\frac{c}{96\pi}i\int_0^{2\pi}4i(\omega_i-1)\log\left(\omega_i|a_i|\left(\frac{\epsilon}{|a_i|}\right)^{\frac{\omega-1}{\omega_i}}\right)d\theta
\ee
so that the Liouville action term for the regular point is given by
\be
S^i_L=-\frac{c}{12}(\omega_i-1)\left(\frac{1}{\omega_i}\log|a_i|+\log(\omega_i)+\frac{\omega_i-1}{\omega_i}\log(\epsilon)\right)
\ee
On the torus we have nine total regular points corresponding to the images of each of the three twist operators. Each image is mapped to the same point $x_i$ by the map $\Gamma$ following directly from the periodic properties of $\wp'$. Accounting for this multiplicity the total contribution is
\be
S^r_{L}=-\frac{c}{6}\log(|a_1||a_2||a_3|)-\frac{3c}{2}\log(3)-c\log(\epsilon)
\ee
where the superscript $r$ indicates these are the contributions from the regular points. We thus find
\be
S^r_L=-\frac{c}{6}\log(x_{12}x_{13}x_{23})-\frac{c}{4}\log(g_3)-\frac{3c}{2}\log(3)-c\log(\epsilon)
\ee

Next we consider the contributions coming from points of infinity. To simplify the calculation we can use properties of the uniformization map without knowledge of the location $\{z_*\}$. To do so we note that $dz\partial_z\phi=dx\partial_x\phi$ and work on the sphere to determine the contribution at $|x|=\frac{1}{\delta}$. We consider the ansatz
\be\label{eq:CoveringAnsatz}
\partial_z \Gamma = \alpha\left[(x-x_1)(x-x_2)(x-x_3)\right]^{\frac{2}{3}}
\ee
making use of \eqref{eq:map3} and \eqref{eq:wpdiffeq} we find $|\alpha|$ to be
\be
|\alpha|=\frac{3g_3^{\frac{1}{6}}}{\left(|x_1-x_2||x_1-x_3||x_2-x_3|\right)^{\frac{1}{3}}}.
\ee
Thus, we find the potential around $|x|=\frac{1}{\delta}$ on the sphere to be
\be
\partial_z \Gamma \approx \alpha x^2 \implies \phi \approx 2\log(|\alpha| |x|^2)
\ee
Parameterizing the path $x=\frac{1}{\delta}e^{i\theta}, \theta\in[0,2\pi]$ the contribution to the potential becomes
\be
-9\frac{c}{96\pi}i\int_0^{2\pi}8i\log\left(\frac{|\alpha|}{\delta^2}\right)d\theta
\ee
so that
\be\label{eq:infycont}
\begin{split}
S^{\infty}_L=&\frac{3c}{2} \log\left(|\alpha|\delta^{-2}\right)\\
=&-\frac{c}{2}\log(x_{12}x_{13}x_{23})+\frac{c}{4}\log(g_3)+\frac{3c}{2}\log(3)-3c\log(\delta).
\end{split}
\ee
Note that there is a factor of 9 accounting for the contributions of each copy and an extra minus to account for the orientation of the contour since it is external on the sphere. The correctness can be validated by verifying that the cut-off $\delta$ vanishes at the end of the calculation.

Alternatively we can work on the torus. The difficulty here is in determining the locations $\{z_*\}$ as well as the expansion around these points. To aid us we will temporarily fix the locations of the twist operators in order to force $\{z_*\}$ to a convenient points. We note if we take $D=0$ then we only need to solve the equation $\wp'=0$. These points are well known to be the half periods of the lattice given explicitly here by $\frac{1}{2}\{1,\tau\}$ as well as other points related by the lattice periods. On figure \ref{fig:n3cuttoffs} these are located at the half-way point on the segment connecting for each copy $I$ and $\bar{I}$. In particular there are nine of them each of which gives an equal contribution to the Liouville action.

At these points to leading order we have expansions \cite{dlmf}
\be
\wp(z_*|\tau)=\frac{ g_3^{\frac{1}{3}}}{2^{\frac{2}{3}}}, \quad \wp'(z_*|\tau)=6\frac{ g_3^{\frac{2}{3}}}{2^{\frac{4}{3}}}(z-z_*)
\ee
so that
\be
\Gamma(z)\approx\beta(z-z_*)^{-1},\quad \beta=\frac{1}{3}\frac{x_{23}}{2^{\frac{2}{3}}g_3^{\frac{1}{6}}}.
\ee
Thus, we find the potential around $|z-z_*|=|\beta|\delta$ to be\footnote{
As another option, $|\beta|$ for arbitrary $x_1,x_2,x_3$ can be determined in the following manner. We introduce the inverse of $x=\Gamma(z)$ as $w$ ($w=1/x=1/\Gamma$), and the corresponding cutoff in the $z$-coodinate as $\delta_z$. 
\begin{align}
    \frac{1}{\Gamma(z_*)}=0,~~~\left|\frac{1}{\Gamma(z_*+\delta_z)}\right|=\delta.
\end{align}
We can rewrite $\delta$ in the integral form as following:
\begin{align}
    \delta = \int_0^\delta dw = \int_{z_*}^{z_*+\delta_z}\frac{dw}{dz}dz.
\end{align}
Here, by using \eqref{eq:CoveringAnsatz}, we obtain
\begin{align}
    \frac{dw}{dz} = \partial_z\frac{1}{\Gamma (z)}= -\frac{\partial_z\Gamma(z)}{\Gamma(z)^2} =&~ -\alpha\left[\left(1-\frac{x_1}{\Gamma(z)}\right)\left(1-\frac{x_2}{\Gamma(z)}\right)\left(1-\frac{x_3}{\Gamma(z)}\right)\right]^{2/3} \nonumber\\
    =&~-\alpha\left[\left(1-x_1w\right)\left(1-x_2w\right)\left(1-x_3w\right)\right]^{2/3}.
\end{align}
When $z$ is in between $z_*$ and $z_*+\delta_z$, $w$ is $O(\delta)$. Thus, the integral above becomes 
\begin{align}
    \int_{z_*}^{z_*+\delta_z}(-\alpha)[1+O(\delta)]dz = -\alpha\delta_z +O(\delta).
\end{align}
As a result, we find $\delta = \frac{1}{|\alpha|}\delta_z$. This leads to $|\beta|=1/|\alpha|$.
}
\be
\partial_z \Gamma \approx -\beta(z-z_*)^{-2} \implies \phi \approx 2\log(|\beta| |z|^{-2})
\ee
Parameterizing the path $|z-z_*|=|\beta|\delta e^{i\theta}, \theta\in[0,2\pi]$ the total contribution to the action to be
\be
\frac{c}{96\pi}i\int_0^{2\pi}8i\log\left(|\beta|\delta^2\right)d\theta
\ee
Accounting for the nine points we find the total contribution
\be
S^{\infty}_L=-\frac{3c}{2}\log(x_{23})+\frac{c}{4}\log(g_3)+\frac{3c}{2}\log(3)+c\log(2)-3c\log(\delta)
\ee
which is the same as \eqref{eq:infycont} under the substitution $D=0\rightarrow 2x_1=x_2+x_3$ as required\footnote{As long as none of the twist operators are placed at a radius larger than the cut off $\frac{1}{\delta}$ (This case can also be a handled with the same final result but requires more care in determining the contributions see \cite{2010PhDT.......134A} for some examples.) the structure of the singularities will remain the same regardless of the choice of locations of the twist operators. The effect of the mobius transformation which moves the twist operators will be to also move the points of infinity to other locations on the torus.}.

Now combining the contributions from all points we find the total Liouville action to be
\be
S_L=-\frac{2c}{3}\log(x_{12}x_{13}x_{23})-c\log(\epsilon)-3c\log(\delta).
\ee
The three point function can now be evaluated. Under the uniformization map we can relate the torus partition function to that of the sphere by taking into account the conformal anomaly. Accounting for the nine copies of the original theory we have
\be
\langle\sigma^{\epsilon}(x_1)\sigma^{\epsilon}(x_2)\sigma^{\epsilon}(x_3)\rangle_{\delta}=e^{S_L}\frac{Z_{torus}}{Z^9_\delta}, \quad Z_{\delta}=Q^c\delta^{-\frac{c}{3}}
\ee
Here $Z_\delta$ is there sphere partition function with curvature located at the radius $|x|=\frac{1}{\delta}$ and $Q$ is a scale related to the size of the sphere which drops out of the final calculation.

To normalize the twist operator we note that the two point function of any twist operator and its inverse factorizes to three independent two points functions. These are spheres with two twist operators insertions and cyclic monodromy with cycle length 3. That is explicitly
\be
\langle\sigma^{\epsilon}(x_1)_g\sigma^{\epsilon}(x_2)_{g^{-1}}\rangle_{\delta}=\left(x_{12}^{-\frac{4}{9}}3^{\frac{1}{3}}\epsilon^{-\frac{2}{9}}Q^{-2}\right)^{3c}
\ee
Using this we can define the normalized twist  operators to be
\be
\sigma_g(x)=\frac{\epsilon^{\frac{c}{3}}Q^{3c}}{3^{\frac{c}{2}}}\sigma_g^{\epsilon}(x).
\ee
Together this gives
\be
\log\left(\langle\sigma(x_1)_{a}\sigma(x_2)_{ab}\sigma(x_3)_{ab^2}\rangle\right)=-\frac{2c}{3}\log(x_{12}x_{13}x_{23})-\frac{3c}{2}\log(3)+\log\left(Z_{torus}\right).
\ee
Note that all length scale and all cut-offs have dropped out of the final result as required. Now finally the Renyi multi-entropy is given by
\be
\begin{split}
S^{(3)}_{3}=&-\frac{1}{6}\log\langle\sigma(x_1)_{a}\sigma(x_2)_{ab}\sigma(x_3)_{ab^2}\rangle\\
=&\frac{c}{9}\log(x_{12}x_{13}x_{23})+\frac{c}{4}\log(3)-\frac{1}{6}\log(Z_{torus}).
\end{split}
\ee
For holographic theories we can evaluate the torus partition function by taking the leading semiclassical approximation \cite{1998JHEP...12..005M,2020JHEP...02..170D}
\be
S_{\text{min}}(\tau)=\min_{a,b,c,d\in\mathbb{Z},
\;ad-bc=1}\left[\frac{i\pi c}{12}\left(\frac{a\tau+b}{c\tau+d}-\frac{a\overline{\tau}+b}{c\overline{\tau}+d}\right)\right], \quad Z_{\text{torus}}(\tau)=e^{-S_{\text{min}}(\tau)}.
\ee
In particular evaluating for  $\tau=e^{\frac{i\pi}{3}}$ we have
\be
S_{\text{min}}\left(e^{\frac{i\pi}{3}}\right)=-\max_{a,b,c,d\in\mathbb{Z},
\;ad-bc=1}\frac{\pi c\sqrt{3}}{12}\left[\frac{ad-bc}{c^2+d^2+cd}\right]
\ee
so that
\be
Z_{torus}\simeq e^{\frac{\pi c \sqrt{3}}{12}}.
\ee
After restoring the uv-cutoff we find the Renyi multi-entropy to be
\be
\begin{split}
S^{(3)}_{3}&=\frac{c}{9}\log\left(\frac{x_{12}x_{13}x_{23}}{\epsilon^3}\right)+\frac{c}{4}\log(3)-\frac{c}{6}\frac{\pi \sqrt{3}}{12}\\
&\approx \frac{c}{9}\log\left(\frac{x_{12}x_{13}x_{23}}{\epsilon^3}\right) +.199c
\end{split}
\ee

\section{Renyi multi-entropy in free fermion CFT and relation to BTZ background}\label{sec:FF}
In this section we would like to analyze the multi-entropy of the massless free Dirac fermion CFT in two dimensions as an explicit and solvable example.

First, we review the ordinary bosonization trick in free fermion to calculate Renyi and entanglement entropy i.e. $q=2$ (Renyi) multi-entropy. 
After, we construct the twist operators to calculate the $(n,q)=(2,3)$ Renyi multi-entropy on the plane,
we apply the result to thermal setup. 
Finally, we briefly discuss the relation to $q=3$ multi-entropy in high-temperature holographic CFT i.e. RT surface in BTZ background.

\subsection{Review: calculation of entanglement entropy via bosonization}
First we would like to briefly review an analytical calculation of entanglement entropy in the massless Dirac Fermion CFT \cite{Casini:2005rm,Azeyanagi:2007bj,Takayanagi:2010wp}, using the bosonization method. 

We start with $n$-replica of $c=1$ Dirac fermion CFT. Here we have $n$ massless Dirac fermions
$\Psi^{(k)}$, labeled by $k = 0,\cdots, n-1$. The space is divided into two intervals, $A=[x_1,x_2]$ and $\Bar{A} = [x_2, \infty) \cup (-\infty, x_1]$.
The fermion fields transform by turning around the point $z=x_1,z=x_2$ as
\begin{align}
    \begin{cases}
        z=x_1: & \Psi^{(k)}(x_1+e^{2\pi i }y) = \Psi^{(k-1)}(x_1+y),\\
        z=x_2: & \Psi^{(k)}(x_2+e^{2\pi i }y) = \Psi^{(k+1)}(x_2+y).
    \end{cases}
\end{align}
This twist-boundary condition can be diagonalized by performing the discrete fourier transformation
\begin{align}
\label{eq:FermionDiscreteFT}
    \psi^{(l)} = \frac{1}{\sqrt{n}}\sum_{k=0}^{n-1} e^{2\pi i \frac{kl}{n}} \Psi^{(k)},
\end{align}
here we should note that the range of $l$ is $-\frac{n-1}{2},\cdots,\frac{n-1}{2}$, because we need to respect the (anti-)periodic boundary condition\footnote{This boundary condition comes from the local conformal transformation $(w-w_0)^n=(z-z_0)$: 
\begin{align}
    \Psi(w) = \left(\frac{dz}{dw}\right)^{\frac{1}{2}}\Psi^{(k)}(z)=\sqrt{n}(z-z_0)^{\frac{n-1}{2n}}\Psi^{(k)}(z).
\end{align}
In the $w$-plane $\Psi(w)$ do not have monodromy i.e. $\Psi(w_0 + e^{2\pi i }\epsilon)=\Psi(w_0 + \epsilon)$. However, in the $z$-plane language (the right-hand side of the above equation), we have $\Psi^{(k)}(z_0+e^{2n\pi i }y)=\Psi^{(k + n)}(z_0+y)$ and $(e^{2n\pi i }y)^{\frac{n-1}{2n}}=e^{(n-1)\pi i}y^{\frac{n-1}{2n}}$. These lead to (\ref{eq:FermionPeriodicity}).
}
\begin{align}
\label{eq:FermionPeriodicity}
    e^{(n-1)\pi i}\Psi^{(k+n)}=\Psi^{(k)}.
\end{align}
This transformation (\ref{eq:FermionDiscreteFT}) leads to the following twist-boundary condition:
\begin{align}
    \begin{cases}
        z=x_1: & \psi^{(l)}(x_1+e^{2\pi i }y) = e^{2\pi i \frac{l}{n}}\psi^{(l)}(x_1+y),\\
        z=x_2: & \psi^{(l)}(x_2+e^{2\pi i }y) = e^{-2\pi i \frac{l}{n}}\psi^{(l)}(x_2+y). \label{qqqqw}
    \end{cases}
\end{align}
By using the bosonization technique we can construct twist-operators that mimic the twist-boundary condition. We set 
\begin{gather}
    \psi^{(l)}(z)=e^{i\phi^{(l)}_L}(z), \\
    \phi^{(l_1)}_L(z_1)\phi^{(l_2)}_L(z_2) \sim -\delta_{l_1,l_2}\log (z_1-z_2), \\
    \psi^{(l)}(z_1) \Bar{\psi}^{(l)}(z_2) \sim \frac{\delta_{l_1,l_2}}{z_1-z_2}.
\end{gather}
The anti-chiral part is defined in the same manner and we set $\phi^{(l)}(z,\Bar{z})=\phi^{(l)}_L(z)+\phi^{(l)}_R(\Bar{z})$. The twist-operators can be constructed as
\begin{align}
\begin{cases}
    z=x_1: & \displaystyle \sigma(x_1) = \prod_{l=-\frac{n-1}{2}}^{\frac{n-1}{2}}e^{ i \frac{l}{n}\phi^{(l)}}(x_1),\\
    z=x_2: & \displaystyle \Bar{\sigma}(x_2) = \prod_{l=-\frac{n-1}{2}}^{\frac{n-1}{2}}e^{ -i \frac{l}{n}\phi^{(l)}}(x_2).
\end{cases}
\end{align}
These operators have the same conformal dimension $h=\Bar{h} = \frac{1}{24}\left(n-\frac{1}{n}\right)$. As we can verify in the explicit calculation, we get $\operatorname{Tr}_A \left[(\rho_A)^n\right]$ as
\begin{align}
    \operatorname{Tr}_A \left[(\rho_A)^n\right] 
    =&~ \left\langle \sigma(x_1) \Bar{\sigma}(x_2)\right\rangle \nonumber\\
    =&~ \prod_{l=-\frac{n-1}{2}}^{\frac{n-1}{2}}\left\langle e^{ i \frac{l}{n}\phi^{(l)}}(x_1) e^{ -i \frac{l}{n}\phi^{(l)}}(x_2)\right\rangle \nonumber\\
    =&~ \prod_{l=-\frac{n-1}{2}}^{\frac{n-1}{2}} \left|x_1-x_2\right|^{-2\left(\frac{l}{n}\right)^2} \nonumber\\
    =&~ \left|x_1-x_2\right|^{-\frac{1}{6}\left(n-\frac{1}{n}\right)}.
\end{align}
This result reproduces the well-known result of the usual $n$-th Renyi entropy and entanglement entropy (here we introduced the UV cut-off $\epsilon$)
\begin{align}
    S_n =&~ \frac{1}{1-n} \log\operatorname{Tr}_A \left[(\rho_A)^n\right] \nonumber\\
    =&~ \frac{n+1}{6n} \log\left|\frac{x_1-x_2}{\epsilon}\right|, \\
    S =&~ \lim_{n\to 1} S_n \nonumber\\
    =&~ \frac{1}{3} \log\left|\frac{x_1-x_2}{\epsilon}\right|.
\end{align}


\subsection{Multi-entropy for $n=2$, $q=3$}
\label{sec:Dirac Fermion CFT Case}
As explained in Appendix \ref{sec:failure}, naive application of the bosonization trick reviewed in the previous subsection does not work, and this is the case with $(n,q)=(2,3)$ as well. 
Instead, we consider the disconnected subsystem setup as done in section \ref{sec:q=3n=2general} to make use of the general formula 
(\ref{twisft}). 
We estimate the twist operators in the bosonized form.

\begin{figure}[h]
    \begin{center}
    \includegraphics[width=8cm]{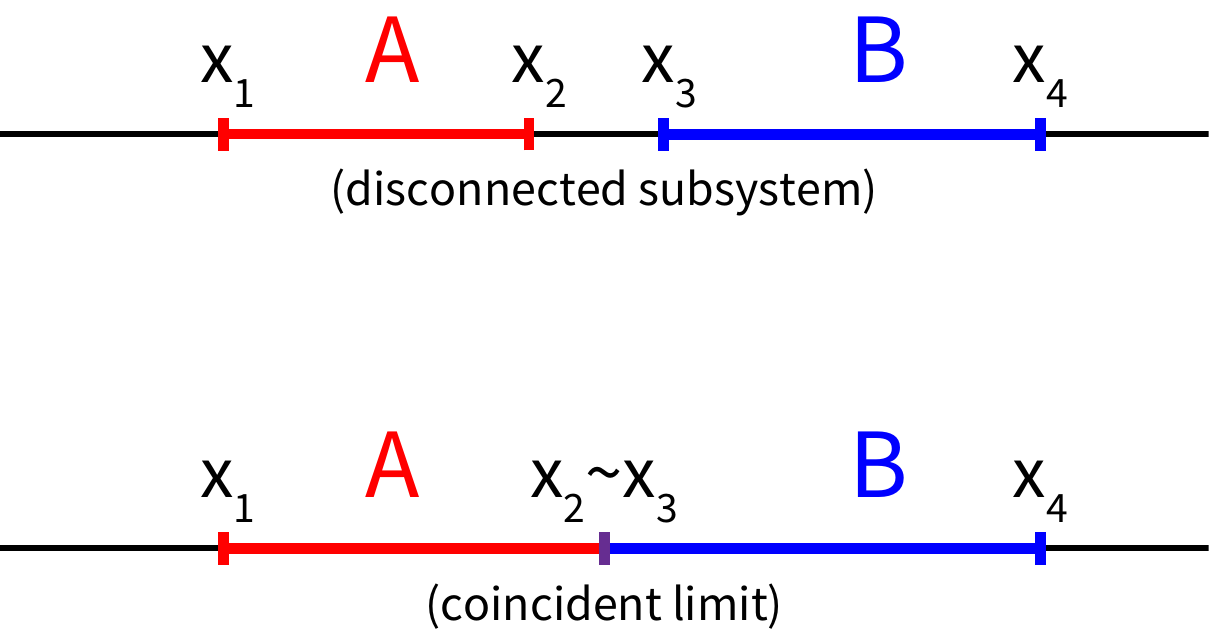}
    \caption{The setup of subsystem in this section. Firstly we take it disjoint, and after that we consider coincident limit.}
    \label{pic:intervalsetupFF}
    \end{center}
\end{figure}

After we obtain the putative form of twists in the disconnected $(n,q)=(2,3)$ case, we take the coincidence limit $x_2\to x_3$ and construct the correct form for connected $(n,q)=(2,3)$. See figure \ref{pic:intervalsetupFF}.

\subsubsection{Twist operators in disconnected subsystem setup}
Here we consider the disconnected subsystem case in free fermion, in parallel with section \ref{sec:q=3n=2general}.
To make use of the general formula 
(\ref{twisft}), we first recall that the torus partition function of a single-copy of the free fermion depends on the spin structure:
\ba
&& Z^{\text{NS}}_{\text{torus}}=\mbox{Tr}_{\text{NS}}e^{-\beta H}=\left|\frac{\theta_3(\tau)}{\eta(\tau)}\right|^2
=2^{\frac{4}{3}}\cdot \frac{1}{2}\left(\eta^{-\frac{1}{3}}(1-\eta)^{-\frac{1}{12}}+\eta^{-\frac{1}{3}}(1-\eta)^{\frac{5}{12}}\right),\no
&& Z^{\text{NS}(-1)^F}_{\text{torus}}=\mbox{Tr}_{\text{NS}}(-1)^F e^{-\beta H}=\left|\frac{\theta_4(\tau)}{\eta(\tau)}\right|^2
=2^{\frac{4}{3}}\cdot \eta^{-\frac{1}{3}}(1-\eta)^{\frac{1}{6}},\no
&& Z^{\text{R}}_{\text{torus}}=\mbox{Tr}_{\text{R}}e^{-\beta H}=\left|\frac{\theta_2(\tau)}{\eta(\tau)}\right|^2
=2^{\frac{4}{3}}\cdot \frac{1}{2}\left(\eta^{-\frac{1}{3}}(1-\eta)^{-\frac{1}{12}}-\eta^{-\frac{1}{3}}(1-\eta)^{\frac{5}{12}}\right).
\ea
These lead to the following four point function in each spin structure:
\ba
&&\la \sigma_1(x_1)\sigma_2(x_2) \sigma_3(x_3) \sigma_4(x_4)\lb_{\text{NS}}\no
&=&\frac{1}{2}
\left(x_{12}^{-\frac{1}{2}}x_{34}^{-\frac{1}{2}}x_{14}^{-\frac{1}{4}}x_{23}^{-\frac{1}{4}}x_{13}^{\frac{1}{4}}x_{24}^{\frac{1}{4}}
+x_{12}^{-\frac{1}{2}}x_{34}^{-\frac{1}{2}}x_{13}^{-\frac{1}{4}}x_{24}^{-\frac{1}{4}}x_{14}^{\frac{1}{4}}x_{23}^{\frac{1}{4}}
\right),\label{fpfa}\\
&&\la \sigma_1(x_1)\sigma_2(x_2) \sigma_3(x_3) \sigma_4(x_4)\lb_{\text{NS}(-1)^F} \no
&=& x_{12}^{-\frac{1}{2}}x_{34}^{-\frac{1}{2}},
\label{fpfb}\\
&&\la \sigma_1(x_1)\sigma_2(x_2) \sigma_3(x_3) \sigma_4(x_4)\lb_{\text{R}}\no
&=&\frac{1}{2}
\left(x_{12}^{-\frac{1}{2}}x_{34}^{-\frac{1}{2}}x_{14}^{-\frac{1}{4}}x_{23}^{-\frac{1}{4}}x_{13}^{\frac{1}{4}}x_{24}^{\frac{1}{4}}
-x_{12}^{-\frac{1}{2}}x_{34}^{-\frac{1}{2}}x_{13}^{-\frac{1}{4}}x_{24}^{-\frac{1}{4}}x_{14}^{\frac{1}{4}}x_{23}^{\frac{1}{4}}
\right).\label{fpfc}
\ea

Let us interpret these results in terms of explicit constructions of twist operators.
First We expect that the natural choice of twisting boundary conditions at $x=x_1,x_2,x_3,x_4$ are given by
\begin{align}\label{monogr}
     \sigma_1\ \mbox{at}\ x=x_1:&\ \ (\psi_1,\psi_2,\psi_3,\psi_4)\to (i\psi_2,i\psi_1,i\psi_4,i\psi_3),\nonumber\\
     \sigma_2\ \mbox{at}\ x=x_2:&\ \ (\psi_1,\psi_2,\psi_3,\psi_4)\to (-i\psi_2,-i\psi_1,-i\psi_4,-i\psi_3),\nonumber\\
    \sigma_3\ \mbox{at}\ x=x_3:&\ \ (\psi_1,\psi_2,\psi_3,\psi_4)\to (i\psi_3,i\psi_4,i\psi_1,i\psi_2),\nonumber\\
     \sigma_4\ \mbox{at}\ x=x_4:&\ \ (\psi_1,\psi_2,\psi_3,\psi_4)\to (-i\psi_3,-i\psi_4,-i\psi_1,-i\psi_2).
\end{align}
This Z$_2\times$ Z$_2$ twist action above can be simultaneously diagonalized by taking the following linear combination of fermions which are bosonized in terms of the four scalar fields 
$\phi^{(1,2,3,4)}$:
\begin{align}\label{twistopaq}
     e^{i\phi^{(1)}}=&~\frac{1}{2}(\psi_1+\psi_2+\psi_3+\psi_4),\nonumber\\
     e^{i\phi^{(2)}}=&~\frac{1}{2}(\psi_1-\psi_2+\psi_3-\psi_4),\nonumber\\
     e^{i\phi^{(3)}}=&~\frac{1}{2}(\psi_1-\psi_2-\psi_3+\psi_4),\nonumber\\
     e^{i\phi^{(4)}}=&~\frac{1}{2}(\psi_1+\psi_2-\psi_3-\psi_4).
\end{align}
This leads to the following identification of twist operators: 
\ba
&& \sigma_1=e^{\frac{i}{4}(\phi^{(1)}-\phi^{(2)}-\phi^{(3)}+\phi^{(4)})},\ \ \  \no 
&& \sigma_2=e^{-\frac{i}{4}(\phi^{(1)}-\phi^{(2)}-\phi^{(3)}+\phi^{(4)})},\no
&& \sigma_3=e^{\frac{i}{4}(\phi^{(1)}+\phi^{(2)}-\phi^{(3)}-\phi^{(4)})},\ \ \no 
&&\sigma_4=e^{-\frac{i}{4}(\phi^{(1)}+\phi^{(2)}-\phi^{(3)}-\phi^{(4)})},
\ea
then we find the four point function is given by $\la \sigma_1\sigma_2\sigma_3\sigma_4\lb=x_{12}^{-1/2}x_{34}^{-1/2}$ and thus coincides with $\la \sigma_1(x_1)\sigma_2(x_2) \sigma_3(x_3) \sigma_4(x_4)\lb_{\text{NS}(-1)^F}$. 

On the other hand, if we choose $\sigma_3, \sigma_4$ as the following unusual form 
\ba
\ti{\sigma}_3=e^{-\frac{i}{4}(\phi^{(1)}+\phi^{(2)}+\phi^{(3)}-\phi^{(4)})},\ \ \  \ \ti{\sigma}_4=e^{+\frac{i}{4}(\phi^{(1)}+\phi^{(2)}+\phi^{(3)}-\phi^{(4)})},
\ea
then the correlation function of twists yields
$\la\sigma_1\sigma_2\ti{\sigma}_3\ti{\sigma}_4\lb=x_{12}^{-\frac{1}{2}}x_{34}^{-\frac{1}{2}}x_{14}^{-\frac{1}{4}}x_{23}^{-\frac{1}{4}}x_{13}^{\frac{1}{4}}x_{24}^{\frac{1}{4}}$, which coincides with $\la \sigma_1\sigma_2\sigma_3\sigma_4\lb_{\text{NS}}+\la \sigma_1\sigma_2\sigma_3\sigma_4\lb_{\text{R}}$. 
In a similar way, if we take another choice of twist operators as 
\ba
\sigma'_3=e^{+\frac{i}{4}(\phi^{(1)}+\phi^{(2)}+\phi^{(3)}-\phi^{(4)})},\ \ \  \ \sigma'_4=e^{-\frac{i}{4}(\phi^{(1)}+\phi^{(2)}+\phi^{(3)}-\phi^{(4)})},
\ea
we find $\la\sigma_1\sigma_2\sigma'_3\sigma'_4\lb=x_{12}^{-\frac{1}{2}}x_{34}^{-\frac{1}{2}}x_{13}^{-\frac{1}{4}}x_{24}^{-\frac{1}{4}}x_{14}^{\frac{1}{4}}x_{23}^{\frac{1}{4}}=\la \sigma_1\sigma_2\sigma_3\sigma_4\lb_{\text{NS}}-\la \sigma_1\sigma_2\sigma_3\sigma_4\lb_{\text{R}}$.

In summary, we find the following relations:
\ba
&& \la\sigma_1\sigma_2\sigma_3\sigma_4\lb=x_{12}^{-1/2}x_{34}^{-1/2}=\la\sigma\sigma\sigma\sigma\lb_{\text{NS}(-1)^F},
\no
&& \la\sigma_1\sigma_2\ti{\sigma}_3\ti{\sigma}_4\lb
=x_{12}^{-\frac{1}{2}}x_{34}^{-\frac{1}{2}}x_{14}^{-\frac{1}{4}}x_{23}^{-\frac{1}{4}}x_{13}^{\frac{1}{4}}x_{24}^{\frac{1}{4}}
=\la\sigma\sigma\sigma\sigma\lb_{\text{NS}}+\la\sigma\sigma\sigma\sigma\lb_{\text{R}},\no
&& \la\sigma_1\sigma_2\sigma'_3\sigma'_4\lb=
x_{12}^{-\frac{1}{2}}x_{34}^{-\frac{1}{2}}x_{13}^{-\frac{1}{4}}x_{24}^{-\frac{1}{4}}x_{14}^{\frac{1}{4}}x_{23}^{\frac{1}{4}}
=\la\sigma\sigma\sigma\sigma\lb_{\text{NS}}-\la\sigma\sigma\sigma\sigma\lb_{\text{R}}.
\ea
Recall that $Z^{\text{total}}_{\text{torus}}=\frac{1}{2}\left(Z^{\text{NS}}_{\text{torus}}+Z^{\text{NS}(-1)^F}_{\text{torus}}+Z^{\text{R}}_{\text{torus}}\right)$, $(n,q)=(2,3)$ Renyi multi-entropy for disconnected subsystem is
\begin{align}
    S^{(3)}_2 =&~ -\frac{1}{2} \log \langle\sigma\sigma\sigma\sigma\rangle_{\text{total}} \nonumber\\
    =&~ -\frac{1}{2} \log \frac{1}{2}\left(
    \langle\sigma\sigma\sigma\sigma\rangle_{\text{NS}}
    +\langle\sigma\sigma\sigma\sigma\rangle_{\text{NS}(-1)^F}
    +\langle\sigma\sigma\sigma\sigma\rangle_{\text{R}}
    \right) \nonumber\\
    =&~ -\frac{1}{2} \log \frac{1}{2}\left(
    \la\sigma_1\sigma_2\sigma_3\sigma_4\lb
    +\la\sigma_1\sigma_2\ti{\sigma}_3\ti{\sigma}_4\lb
    \right),
\end{align}
where each twist operators are 
\ba
&& \sigma_1=e^{+\frac{i}{4}(\phi^{(1)}-\phi^{(2)}-\phi^{(3)}+\phi^{(4)})},\ \ \sigma_2=e^{-\frac{i}{4}(\phi^{(1)}-\phi^{(2)}-\phi^{(3)}+\phi^{(4)})},\no
&& \sigma_3=e^{+\frac{i}{4}(\phi^{(1)}+\phi^{(2)}-\phi^{(3)}-\phi^{(4)})},\ \ \sigma_4=e^{-\frac{i}{4}(\phi^{(1)}+\phi^{(2)}-\phi^{(3)}-\phi^{(4)})},\ \ \no 
&&\ti{\sigma}_3=e^{-\frac{i}{4}(\phi^{(1)}+\phi^{(2)}+\phi^{(3)}-\phi^{(4)})},\ \ \ti{\sigma}_4=e^{+\frac{i}{4}(\phi^{(1)}+\phi^{(2)}+\phi^{(3)}-\phi^{(4)})}.
\ea
Also, it is explicitly written as
\begin{align}
\label{eq:disjointintervalRE}
    S^{(3)}_2 =&~ \frac{1}{2} \log 2 - \frac{1}{2} \log \left( \frac{\epsilon}{x_{12}^{\frac{1}{2}}x_{34}^{\frac{1}{2}}} + \frac{\epsilon\cdot x_{13}^{\frac{1}{4}}x_{24}^{\frac{1}{4}}}
    {x_{12}^{\frac{1}{2}}x_{34}^{\frac{1}{2}}x_{14}^{\frac{1}{4}}x_{23}^{\frac{1}{4}}} \right).
\end{align}

\subsubsection{Twist operators in connected subsystem setup}
Here we take the coincident limit $x_2\to x_3$, $x_{23}\to \epsilon/4$ as done in (\ref{eq:limit:eps/4}). One can find that the relevant contribution in (\ref{eq:disjointintervalRE}) is 
\begin{align}
    S^{(3)}_2 =&~ - \frac{1}{2} \log \frac{1}{2}\left( \frac{\epsilon\cdot x_{13}^{\frac{1}{4}}x_{24}^{\frac{1}{4}}}
    {x_{12}^{\frac{1}{2}}x_{34}^{\frac{1}{2}}x_{14}^{\frac{1}{4}}x_{23}^{\frac{1}{4}}} \right),
\end{align}
and this comes from $\la\sigma_1\sigma_2\ti{\sigma}_3\ti{\sigma}_4\lb$.
In this limit, the set of twist operators $\sigma_2\ti{\sigma}_3$ becomes a single operator $\sqrt{2}\sigma_{AB}$ as\footnote{
Merging $\sigma_2$ and $\ti{\sigma}_3$ is carried in the following way. Firstly we consider the OPE of them:
\begin{align}
    \sigma_2(x_2)\ti{\sigma}_3(x_3) =&~
    :e^{-\frac{i}{4}(\phi^{(1)}-\phi^{(2)}-\phi^{(3)}+\phi^{(4)})}(x_2)::e^{-\frac{i}{4}(\phi^{(1)}+\phi^{(2)}+\phi^{(3)}-\phi^{(4)})}(x_3): \nonumber\\
    =&~|x_{23}|^{-1/4}
    :e^{-\frac{i}{4}(\phi^{(1)}-\phi^{(2)}-\phi^{(3)}+\phi^{(4)})}(x_2)e^{-\frac{i}{4}(\phi^{(1)}+\phi^{(2)}+\phi^{(3)}-\phi^{(4)})}(x_3):.
\end{align}
Substituting $x_{23}=\epsilon/4$ we find
\begin{align}
    \sigma_2(x_3+\epsilon/4)\ti{\sigma}_3(x_3) = \frac{\sqrt{2}}{\epsilon^{1/4}}\left(:e^{-\frac{i}{2}\phi^{(1)}}(x_3):+O(\epsilon)\right).
\end{align}
In the text we ignore the subleading terms and drop $\epsilon^{-1/4}$ factor.
}
\begin{align}
\label{eq:connectedintervaltwist}
    \sigma_A =&~ e^{+\frac{i}{4}\left(\phi^{(1)}-\phi^{(2)}-\phi^{(3)}+\phi^{(4)}\right)} ~~~~~(=\sigma_1),
    \nonumber\\
    \sigma_{AB} =&~ e^{-\frac{i}{2}\phi^{(1)}}, \nonumber\\
    \sigma_B =&~ e^{+\frac{i}{4}\left(\phi^{(1)}+\phi^{(2)}+\phi^{(3)}-\phi^{(4)}\right)} ~~~~~(=\tilde{\sigma}_4),
\end{align}
and the overall factor becomes $1/\sqrt{2}$. 
This construction precisely reproduces (\ref{redmue}):
\begin{align}
    S^{(3)}_2 =&~ -\frac{1}{2} \log \frac{1}{\sqrt{2}} \langle \sigma_A \sigma_{AB} \sigma_B\rangle \nonumber\\
    =&~ \frac{1}{4}\log 2  +\frac{1}{8}\log \left[\frac{x_{12}x_{24}x_{14}}{\ep^3}\right].
\end{align}

\subsection{Thermal states and BTZ black hole}\label{sec:BTZ}

Here we calculate the second Renyi multi-entropy for a thermal state in the free fermion theory by evaluating the correlation function of twists on torus with modulus $\tau$. We use the set of twist operators (\ref{eq:connectedintervaltwist}) constructed in previous section.
The partition function is
\begin{align}
    Z(\tau)=\frac{1}{2}\left(\left|\frac{\theta_2(\tau)}{\eta(\tau)}\right|^2+\left|\frac{\theta_3(\tau)}{\eta(\tau)}\right|^2+\left|\frac{\theta_4(\tau)}{\eta(\tau)}\right|^2\right).
\end{align}
Additionally, we introduce the following new function $Z(z|\tau)$:
\begin{align}
    Z(z|\tau)=\frac{1}{2}\left(-\left|\frac{\theta_1(z|\tau)}{\eta(\tau)}\right|^2+\left|\frac{\theta_2(z|\tau)}{\eta(\tau)}\right|^2+\left|\frac{\theta_3(z|\tau)}{\eta(\tau)}\right|^2+\left|\frac{\theta_4(z|\tau)}{\eta(\tau)}\right|^2\right).
\end{align}
Then, the correlation function of twist operators, $\frac{1}{\sqrt{2}}\langle \sigma_A (z_1) \sigma_{AB} (z_2) \sigma_B (z_3) \rangle$, is 
\begin{align}
\label{eq:thermaltripartcorr}
    \frac{1}{\sqrt{2}}
    \left|\frac{\partial\theta_1(\tau)}{\theta_1(z_{12}|\tau)}\right|^{1/4}
    \left|\frac{\partial\theta_1(\tau)}{\theta_1(z_{23}|\tau)}\right|^{1/4}
    \left|\frac{\partial\theta_1(\tau)}{\theta_1(z_{13}|\tau)}\right|^{1/4}
    \frac{1}{\left(Z(\tau)\right)^4}
    Z\left(\frac{1}{4}z_1-\frac{1}{2}z_2+\frac{1}{4}z_3|\tau\right)
    \left(
    Z\left(\frac{1}{4}z_1-\frac{1}{4}z_3|\tau\right)
    \right)^3.
\end{align}
We set the $z$s to be real, $ z_1 < z_2 < z_3 < z_1 +1$. In the limit of $\tau\to i\infty$, we find
\begin{align}
    \frac{1}{\left(Z(\tau)\right)^4}
    Z\left(\frac{1}{4}z_1-\frac{1}{2}z_2+\frac{1}{4}z_3|\tau\right)
    \left(
    Z\left(\frac{1}{4}z_1-\frac{1}{4}z_3|\tau\right)
    \right)^3\to 1.
\end{align}
On the other hand, in the limit of $\tau\to i\delta$ where $0<\delta \ll 1$, we find
\begin{align}
    &~ \frac{1}{\left(Z(\tau)\right)^4}
    Z\left(\frac{1}{4}z_1-\frac{1}{2}z_2+\frac{1}{4}z_3|\tau\right)
    \left(
    Z\left(\frac{1}{4}z_1-\frac{1}{4}z_3|\tau\right)
    \right)^3 \nonumber\\
    \to &~  e^{-\frac{2\pi}{\delta}\left(\frac{1}{4}z_1-\frac{1}{2}z_2+\frac{1}{4}z_3\right)^2}\left(e^{-\frac{2\pi}{\delta}\left(\frac{1}{4}z_1-\frac{1}{4}z_3\right)^2}\right)^3 \nonumber\\
    =&~ \exp\left[-\frac{\pi}{4\delta}\left((z_1-z_2)^2+(z_2-z_3)^2+(z_3-z_1)^2\right)\right],
\end{align}
and if we impose $\delta\ll|z|<1$, we obtain
\begin{align}
    \left|\frac{\partial\theta_1(\tau)}{\theta_1(z|\tau)}\right|\to\frac{2\pi}{\delta} \exp\left[\frac{\pi}{\delta}\left(z^2-|z|\right)\right].
\end{align}
We find the $q=3,n=2$ Renyi multi-entropy for thermal state as\footnote{Because of the periodicity of torus $z\sim z+1$, we can relabel $z$s, as for example $z^{\text{new}}_1=z^{\text{old}}_2$, $z^{\text{new}}_2=z^{\text{old}}_3$, $z^{\text{new}}_3=z^{\text{old}}_1+1$. Although it changes the result of high-temperature limit, we shall not go into further detail here.
}
\begin{align}
\label{eq:fermiontorusmulti}
    S^{(3)}_2 = 
    \begin{cases}
        \frac{1}{4}\log 2  +\frac{1}{8}\log \left[\left|\frac{\sin\pi z_{21}}{\pi}\right|
        \left|\frac{\sin\pi z_{32}}{\pi}\right|
        \left|\frac{\sin\pi z_{31}}{\pi}\right|
        \right] & (\text{low-temperature limit}), \\
        \frac{1}{4}\log 2  -\frac{3}{8}\log \frac{2\pi}{\delta} + \frac{\pi}{8\delta} \left(z_{21}+z_{32}+z_{31}\right) & (\text{high-temperature limit}).
    \end{cases}
\end{align}

\subsubsection{Comparison to holographic setup: BTZ background}
\begin{figure}[h]
    \begin{center}
    \includegraphics[width=3.8cm]{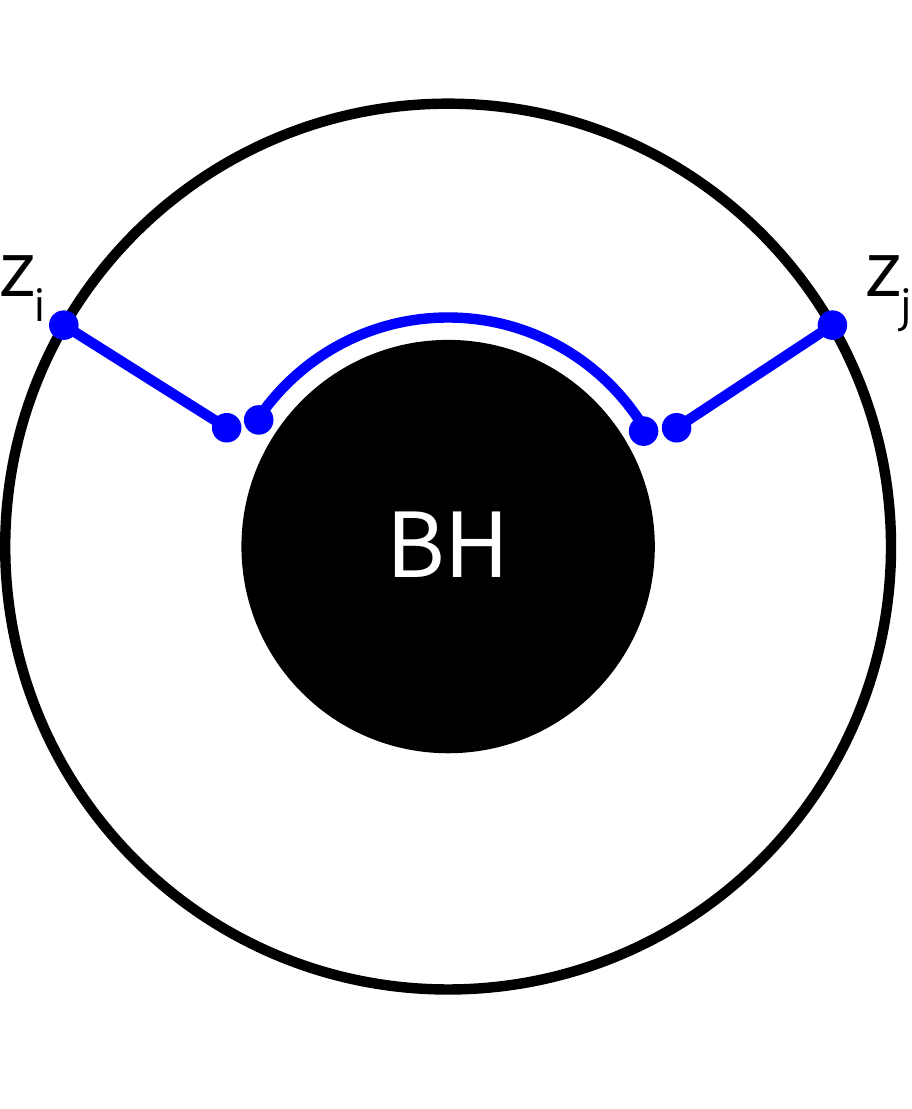}
    \caption{Geodesic approximation in high-temperature limit. The straight lines correspond to the log-divergent contribution and the curved line is $2\pi l z_{ij}/\delta$.}
    \label{pic:GeodesicApprox}
    \end{center}
\end{figure}

Here we consider the Euclidean BTZ geometry:
\begin{align}
    ds^2=\frac{r^2-{r_+}^2}{l^2}dt^2+\frac{l^2}{r^2-{r_+}^2}dr^2+r^2d\phi^2.
\end{align}
This coordinates correspond to the coordinates $z$ and $\bar{z}$ in the boundary holographic CFT as $2\pi z = \phi + it/l$. In the high-temperature limit $\tau = i\delta ~(0<\delta \ll 1)$, 
all the geodesics in the BTZ geometry approaches the horizon and its length is approximated as follows (see figure \ref{pic:GeodesicApprox}):
\begin{align}
    (\text{length}) \simeq (\text{log-divergent part}) + \frac{2\pi l}{\delta}z_{ij}.
\end{align}

\begin{figure}[h]
    \begin{center}
    \includegraphics[width=3.8cm]{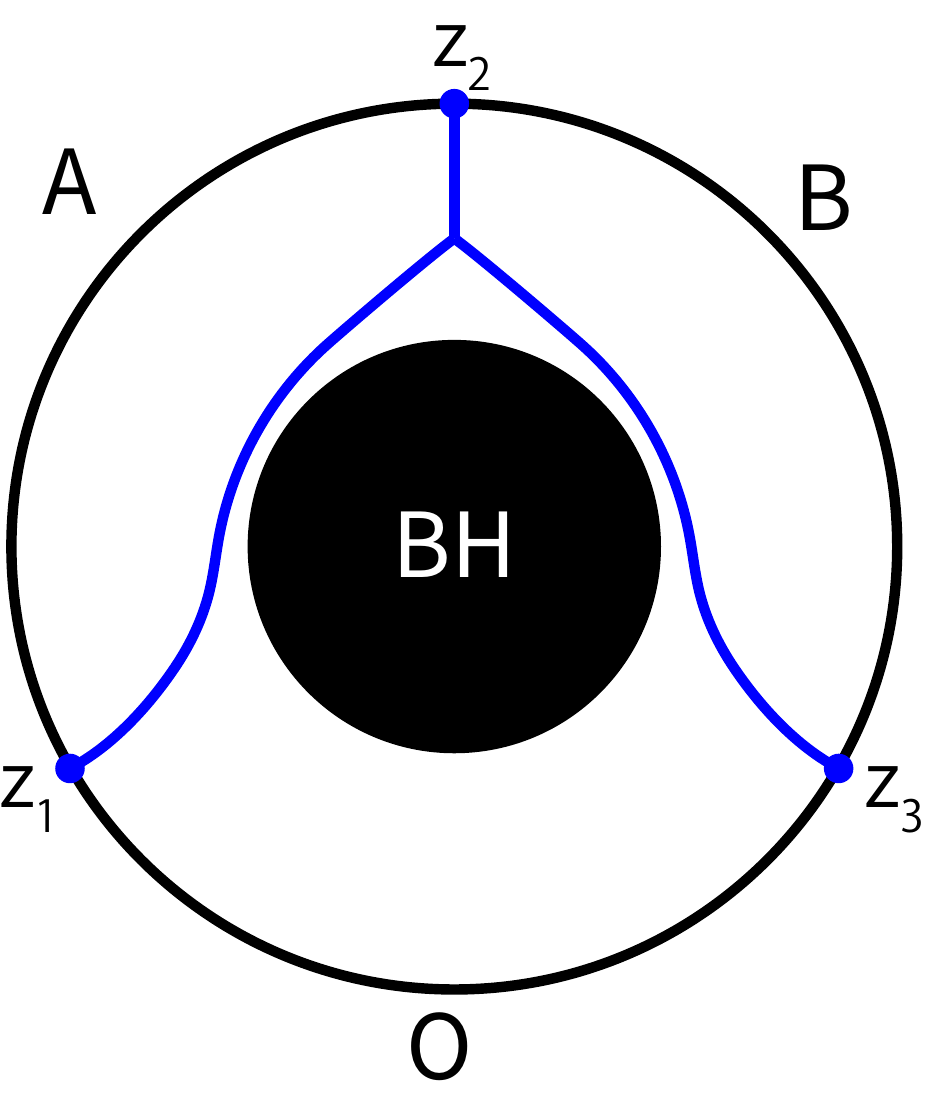}
    \caption{RT surface for $q=3$ multi-entropy in the BTZ background.}
    \label{pic:Geodesics}
    \end{center}
\end{figure}

Once we assume that the corresponding RT surface for $q=3$ multi-entropy in the BTZ background is pictorially given by figure \ref{pic:Geodesics},
the multi-entropy can be estimated as
\begin{align}
\label{eq:BTZmuitl}
    S^{(3)} \simeq&~ 
    (\text{log-divergent part})
    +\frac{\pi l}{\delta \cdot 4G_N}\left(z_{21}+z_{32}+z_{31}\right) \nonumber\\
    =&~(\text{log-divergent part})
    +\frac{\pi c}{6\delta}\left(z_{21}+z_{32}+z_{31}\right).
\end{align}
Although both results (\ref{eq:fermiontorusmulti}) and (\ref{eq:BTZmuitl}) are for the different Renyi-index $n$ and
the non-universal contribution is highly dependent on the details of the CFT,
in the high-temperature limit they have a very similar contribution of $\frac{1}{\delta}$ times the sum of $z_{ij}$.
This may be one evidence for that the choice of RT surface (figure \ref{pic:Geodesics}) is correct. 

\section{Locally excited states}\label{sec:LQ}

In this section, as an example of multi-entropy for excited states, we would like to compute the $(q,n)=(3,2)$ multi-entropy  (\ref{sthtw}) in a two dimensional massless free scalar CFT for a locally excite state. We consider the following time evolution of locally excited state:
\ba
|\Psi(t)\lb\propto e^{-iHt}e^{-\ep H}O(x_0)|0\lb,\label{pppt}
\ea
where the local operator $\mathcal{O}(x_0)$ inserted at $x=x_0$ and $t=0$. The infinitesimally small parameter $\ep$ is a UV regulator. 

We introduce the complex coordinate $(w,\bar{w})$ to describe the Euclidean plane by setting $w=x+i\tau$, where we analytically continue the Euclidean time 
$\tau$ to the Lorentzian time $t$ via $\tau=it$.

In terms of the free real scalar field $\phi(w,\bar{w})=\phi_L(w)+\phi_R(\Bar{w})$, which has the OPEs $\phi(w,\Bar{w})\phi(0,0)\sim - \log |w|^2$, we choose the local operator $O(x_1)$ to be
\ba
\mathcal{O}(x_1)=\f{1}{\s{2}}\left( e^{\frac{i}{2}\phi(w_0,\bar{w}_0)}+e^{-\frac{i}{2}\phi(w_0,\bar{w}_0)}\right),
\label{loco}
\ea
where we take  $w_0=x_0-i\ep+t$ to describe the time evolution of (\ref{pppt}).
We choose the subsystem $A$ and $B$ to be the interval $x_1< x\leq x_2$ and 
$x_2< x\leq x_3$ along the $x$-axis i.e. $\tau=0$, respectively and the subsystem $C$ is the complement of $AB$ on $\tau=0$.

\begin{figure}[H]
  \centering
   \includegraphics[width=.65\textwidth,page=1]{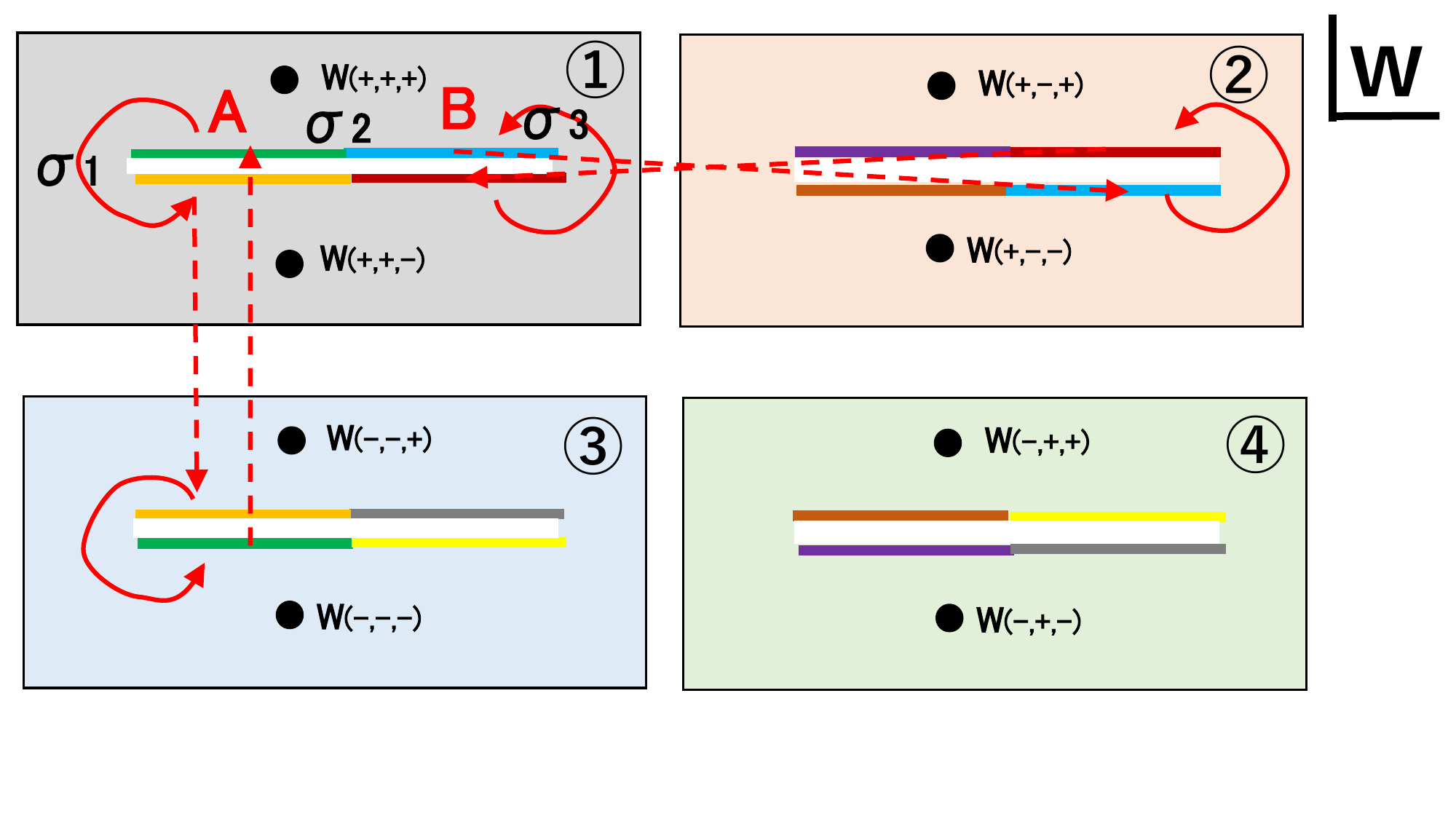}
  \caption{The replica manifold in the $w$ coordinate which computes the multi-entropy $(q,n)=(3,2)$ in the presence of local operator excitation. There are eight insertions of the local operator $\mathcal{O}(w,\bar{w})$ at $w=w_{\pm,\pm,\pm}$.} 
\label{fig:replicaLQ}
\end{figure}

As in the case of the entanglement entropy (i.e. $q=2$) \cite{Nozaki:2014hna,Nozaki:2014uaa,He:2014mwa}, we can compute the multi-entropy for the above locally excited state via the replica method by inserting the local operators in the Euclidean path-integral over the replicated space. This is obtained by gluing four complex planes along $A$ and $B$ an by inserting eight local operators at $w=w_{\pm,\pm,\pm}$ as depicted in Fig.\ref{fig:replicaLQ}.

In terms of the twist operators, inserted at $x=x_1,x_2$ and $x_3$ on the $x$ axis, we can write the partition function on the replicated space as follows:
\begin{align}
 &~\mbox{MTr}
\left[|\Psi\lb\la \Psi|_{ABC}^{\otimes 4}\right]\no
=&~\left\langle \prod_{\eta_1=\pm}\prod_{\eta_2=\pm}\prod_{\eta_3=\pm} \mathcal{O}(w_{\eta_1,\eta_2,\eta_3},\Bar{w}_{\eta_1,\eta_2,\eta_3})    \sigma_1(x_1)\sigma_2(x_2)\sigma_3(x_3)\right\rangle_w \nonumber\\
    =&~ 
    \prod_{\eta_1=\pm}\prod_{\eta_2=\pm}\prod_{\eta_3=\pm}  
\left(\left.\frac{dz}{dw}\right|_{w=w_{\eta_1,\eta_2,\eta_3}}\right)^{h_O}
    \left(\left.\frac{d\Bar{z}}{d\Bar{w}}\right|_{\Bar{w}=\Bar{w}_{\eta_1,\eta_2,\eta_3}}\right)^{\bar{h}_O}
   \left\la \prod_{\eta_1=\pm}\prod_{\eta_2=\pm}\prod_{\eta_3=\pm} \mathcal{O}(z_{\eta_1,\eta_2,\eta_3},\Bar{z}_{\eta_1,\eta_2,\eta_3}) \right\lb_z \label{eightp}
\end{align}
In the final expression we map the eight-point function on the replicated manifold into that on the $z$ plane via the map (\ref{cmapr}).

For our explicit calculation, we set $x_0=-1$, $x_1=0$, $x_2=1$ and $x_3=2$ for simplicity, though the extension to generic points is straightforward. Then the eight points for the operator insertions on the $w$ sheets are given as follows:
\begin{align}
    w_{(\pm,\pm,\pm)}=&~-1+t\pm i\epsilon, \nonumber\\
    \Bar{w}_{(\pm,\pm,\pm)}=&~-1-t\mp i\epsilon, 
    \label{wppp}
\end{align}
where the last sign corresponds to the sign of $i\epsilon$, and the other signs to the choice of replica sheets, as in Fig.(\ref{fig:replicaLQ}). We take $\epsilon>0$ to be an infinitesimally small.

\subsection{Results of Multi-entropy}\label{finalrf}
The evaluation of the eight-point function can be done via the standard Wick contraction in the field scalar field theory. The non-trivial point is to carefully take the limit $\ep\to 0$ of
$z_{\pm,\pm,\pm}$ and $\Bar{z}_{\pm,\pm,\pm}$ as explained in the appendix \ref{detailLQ}.

The 8-point function turns out to be
\begin{align}
    &~\prod _{(\pm,\pm,\pm)} 
    \left[
    \left(\left.\frac{dz}{dw}\right|_{z=z_{(\pm,\pm,\pm)}}\right)^{1/8}
    \left(\left.\frac{d\Bar{z}}{d\Bar{w}}\right|_{\Bar{z}=\Bar{z}_{(\pm,\pm,\pm)}}\right)^{1/8}
    \right]
    \left\langle \prod _{(\pm,\pm,\pm)} \mathcal{O}\left(z_{(\cdot,\cdot,\cdot)},\Bar{z}_{(\cdot,\cdot,\cdot)}\right) 
    \right\rangle_z \nonumber\\
    =&~\begin{cases}
        \frac{1}{(2\epsilon)^2} & (0<t<1), \\
        \frac{1}{4(2\epsilon)^2} & (1<t<2), \\
        \frac{1}{4(2\epsilon)^2} & (2<t<3), \\
        \frac{1}{(2\epsilon)^2} & (3<t). \\
    \end{cases}
\end{align}
The normalization factor $1/\langle\mathcal{O}\mathcal{O}\rangle_w^4$ removes $\frac{1}{(2\epsilon)^2}$ above, where 
\begin{align}
    \langle\mathcal{O}\mathcal{O}\rangle_w =&~ \left\langle 
    \mathcal{O}\left(w_{(\cdot,\cdot,+)},\Bar{w}_{(\cdot,\cdot,+)}\right) 
    \mathcal{O}\left(w_{(\cdot,\cdot,-)},\Bar{w}_{(\cdot,\cdot,-)}\right)
    \right\rangle_w \nonumber\\
    =&~ \frac{1}{\sqrt{2\epsilon}}.
\end{align}
Finally we can calculate the difference between the multi-entropy for the locally excited state and that for the CFT vacuum as follows: 
\begin{align}
    \Delta S^{(3)}_{2}
    =&~ S^{(3)}_{2}(\text{excited state})-S^{(3)}_{2}(\text{ground state}) \nonumber\\
    =&~ \begin{cases}
        0 & (0<t<1), \\
        \log{2} & (1<t<2), \\
        \log{2} & (2<t<3), \\
        0 & (3<t). \\
    \end{cases}\label{resLQ}
\end{align}

\subsection{Quasi-particle Interpretation}
Now, let us interpret this result (\ref{resLQ}) in terms of a simple quantum system, which consists of four qubits, each of them denoted by $A$,$B$,$C_L$ and $C_R$. These model the subsystems in the CFT on an infinite line, where $C_L$ and $C_R$ belong to the subsystem $C$.
We can regard the local operator excitation (\ref{loco}) as an Bell pair which entangles $C_L$ qubit and $C_R$ qubit. This is because by decomposing the scalar field in terms of the left and right moving part, the excited state can be expressed as
\ba
\mathcal{O}|0\lb=\frac{1}{\s{2}}e^{\frac{i}{2}\phi_L}e^{\frac{i}{2}\phi_R}|0\lb+\frac{1}{\s{2}}e^{-\frac{i}{2}\phi_L}e^{-\frac{i}{2}\phi_R}|0\lb= \frac{1}{\s{2}}|+\lb_L |+\lb_R+\frac{1}{\s{2}}|-\lb_L |-\lb_R.
\ea
In this way we expect that the locally excited state created by inserting $\mathcal{O}$ at $x=-1$ is described by 
\ba
|\Psi(t_0)\lb=|0\lb_A|0\lb_B
\left(\frac{1}{\s{2}} |+\lb_{C_L}  |+\lb_{C_R} +\frac{1}{\s{2}}
|-\lb_{C_L} |-\lb_{C_R}\right).  \label{trivst}
\ea
Under the time evolution of one part of the entangled pair propagates 
in the right direction and the other does in the left direction. For the time period $0<t_0<1$, the state is still given by (\ref{trivst}) because each of the pair is inside the subsystem $C$. It is 
straightforward to confirm that the multi-entropy (\ref{sthtw}) using (\ref{Mtra}) is trivial $S^{(3)}_2=0$. This is because $\rho_{AB}$ is given by 
\ba
\rho_{AB}=|0\lb\la 0|_A\otimes  |0\lb\la 0|_B,
\ea
which leads to 
$\mbox{MTr}\left[|\Psi\lb\la \Psi|_{ABC}^{\otimes 4}\right]=1$.

However, at $t=1$, one of the pair reaches at one end point of subsystem $A$.
Thus, for the time period $1<t_1<2$, the state looks like
\ba
|\Psi(t_1)\lb=\frac{1}{\s{2}}|+\lb_A|0\lb_B |+\lb_{C_L} |0\lb_{C_R}+
\frac{1}{\s{2}}|-\lb_A|0\lb_B |-\lb_{C_L} |0\lb_{C_R}.
\ea
In this case the reduced density matrix now becomes the mixed state
\ba
\rho_{AB}=\frac{1}{2}\Bigl[|+\lb\la+|+|-\lb\la-|\Bigr]_A\otimes |0\lb\la 0|_B. \label{rabred}
\ea
This leads to 
\ba
\mbox{MTr}\left[|\Psi\lb\la \Psi|_{ABC}^{\otimes 4}\right]=\frac{1}{4},
\ea
and thus we find
\ba
S^{(3)}_2=\log 2. \label{emyksfa}
\ea

For $2<t_2<3$, it is in the subsystem B and we have 
\ba
|\Psi(t_2)\lb=\frac{1}{\s{2}}|0\lb_A|+\lb_B |+\lb_{C_L} |0\lb_{C_R}+
\frac{1}{\s{2}}|0\lb_A|-\lb_B |-\lb_{C_L} |0\lb_{C_R}.
\ea
The reduced density matrix takes the same result as (\ref{rabred}), which leads to the same value of multi-entropy (\ref{emyksfa}).

On the other hand, for the later time $t_3>3$, the entangled pair are both in the subsystem $C$ again. Thus the state looks like (\ref{trivst}) and the multi-entropy becomes vanishing. 

Indeed, the above evolution of multi-entropy perfectly reproduces the result  (\ref{resLQ}) in the free scalar CFT. This confirms the validity of our replica method calculations of multi-entropy.

\section{Conclusion}\label{sec:con}
In this paper we have furthered the program initiated by \cite{2022PhRvD.106l6001G,2023JHEP...08..202G,2022arXiv221116045P}.
Our contributions are two-fold:
\begin{itemize}
    \item For several tractable examples in two dimensional CFTs, we have explicitly calculated the Renyi multi-entropy. This was accomplished by construction of the relevant replica manifold and application of the uniformization method. Since our results do not cover the von-Neumann like ($n=1$) limit, we cannot compare our results with the holographic proposal of multi-entropy. However we have observed from CFT calculations that our results for holographic CFTs at $n=2$ show phase transition phenomena which qualitatively agree with gravity dual expectations.
    
    \item For the free Dirac fermion CFT in two dimensions, we identified an outstanding problem with the explicit construction of the twist operators used for the calculation of Renyi multi-entropy. We also computed the multi-entropy at finite temperature in this CFT using the twist operators we found and confirmed a qualitative agreement with its gravity dual. As another solvable example, we also analyzed a free scalar CFT in two dimensions and calculated the multi-entropy in the presence of a local operator excitation. The result perfectly reproduces what we expect from a quasi-particle interpretation of qubits.
    
\end{itemize}
There are a number of interesting future directions one could consider:
\begin{itemize}
    \item As a check of the proposed duality one would like to be able to calculate $\kappa$ in the boundary theory. As mentioned in the main text this is a significant challenge as the replica surfaces associated with the Renyi multi-entropy grow in genus. The current techniques available for exact calculation of the three-point coefficient are limited to the uniformization method. As such direct calculation is untenable because both the partition function (especially for large c holographic CFTs) and unformization map are unknown. One would prefer alternate methods which would allow for more direct computations of the three-point coefficients which would allow for the exact computation of $\kappa$.
    
    \item The multi-entropy generalizes the entanglement entropy by changing the finite group symmetry of the monodromies of the twist operators. In \cite{2023JHEP...08..202G} this was further extended to all Abelian groups which can always be expressed a direct product of cyclic groups (now including the possibility of different orders). The proposed bulk dual consists of weighted Steiner trees where geodesics meet in trivalent intersections, but with angles dependent on the relative weights. It would be worthwhile to repeat the analysis performed in the paper to these quantities.
    
    In addition it would also be interesting to consider information measures defined from twist operator with non-Abelian mondromies. This gives another possible rich set of examples to further investigate the connection between the geometry of holographic spacetimes and information measures of boundary states.
    \item As we saw for the free fermion CFT it is currently unknown how to explicitly construct the twist operators used for the calculation of multi-entropy. In particular the bosonization method used for entanglement entropy implicitly relies on the correlation function of twist operators being a two-point function. This is because it forces the resulting two-point functions of vertex operators after bonsonization to be charge conserving. As soon as one considers higher-point functions this is no longer true and this naive calculation will fail to reproduce the correct conformal dimensions of the twist operators. Though we were able to supply a definition for the specific example of $n=2$ which we considered, one would desire a more general and robust method for their definition in a systematical way.

    Furthermore, our method of deriving the bosonized twist operator, which we did in section \ref{sec:FF}, is in some sense incomplete. The resulting twist operators are not symmetric under the permutations, and their relation to the representation of bosonized fermion is not so clear. Elucidating this detail is one possible next step.
\end{itemize}

\section*{Acknowledgements}
We are grateful to Abhijit Gadde, Onkar Parrikar, Bartek Czech,  Dongsheng Ge, Matt Headrick, Veronika Hubeny  Vineeth Krishna, Luca Lionni, Ali Mollabashi, Shinsei Ryu, and  Wayne Wang for discussions. We are grateful to the long term workshop YITP-T-23-01 ``Quantum Information, Quantum Matter and Quantum Gravity'', held at YITP, Kyoto University, where a part of this work was done. J.H. would like to especially thank Abhijit Gadde and Onkar Parrikar for hospitailty during a long term visit to TIFR Mumbai where this work was finalized.

\paragraph{Funding information}
This work is supported by the Simons Foundation through the ``It from Qubit'' collaboration and by MEXT KAKENHI Grant-in-Aid for Transformative Research Areas (A) through the ``Extreme Universe'' collaboration: Grant Number 21H05187. This work is also supported by Inamori Research Institute for Science and by JSPS Grant-in-Aid for Scientific Research (A) No.~21H04469.
Takashi Tsuda is supported by JST SPRING, Grant Number JPMJSP2110.

\paragraph{Author contributions}
All authors contributed equally to the research and writing of this article.

\begin{appendix}

\section{Failure of naive bosonization for free fermion multi-entropy}
\label{sec:failure}
Here we demonstrate that the bosonization method used for the definition of twist operators in the 2d free fermion CFT fails when applied to the Renyi multi-entropy. To do so we focus on the case $n=3,q=3$ in which case we have nine fermion fields $\psi_i$. We consider the three point function
\be
\langle \sigma_a(x_1)\sigma_{ab}(x_2)\sigma_{ab^2}(x_3)\rangle
\ee
where the monodromies of the twist operators are given by
\be
\begin{split}
\sigma_a(x_1):& \quad (123)(456)(789)\\
\sigma_{ab}(x_2):& \quad (168)(249)(357)\\
\sigma_{ab^2}(x_3):& \quad  (174)(285)(396)
\end{split}
\ee
Each of these permutations can be represented as a matrix $T_{g}$ with entry $ij$ 1 only if $i\rightarrow j$ is in the cycle structure of $g$ and all other entries zero.
In particular this implies analogous transformation properties of the fermion fields around each of the twist operators
\be
\begin{split}
\sigma_{a}(x_1): (\psi_1,\psi_2,\psi_3,\psi_4,\psi_4,\psi_6,\psi_7,\psi_8,\psi_9) \longrightarrow (\psi_2,\psi_3,\psi_1,\psi_5,\psi_6,\psi_4,\psi_8,\psi_9,\psi_7)\\
\sigma_{ab}(x_3): (\psi_1,\psi_2,\psi_3,\psi_4,\psi_4,\psi_6,\psi_7,\psi_8,\psi_9) \longrightarrow (\psi_6,\psi_4,\psi_5,\psi_9,\psi_7,\psi_8,\psi_3,\psi_1,\psi_2)\\
\sigma_{ab^2}(x_3): (\psi_1,\psi_2,\psi_3,\psi_4,\psi_4,\psi_6,\psi_7,\psi_8,\psi_9) \longrightarrow (\psi_7,\psi_8,\psi_9,\psi_1,\psi_2,\psi_3,\psi_4,\psi_5,\psi_6)
\end{split}
\ee
To proceed we diagonlize the mondromies. This can be done by defining new fermion fields $\tilde{\psi}_i$:
\be
\begin{pmatrix}
\tilde{\psi}_1\\
\tilde{\psi}_2\\
\tilde{\psi}_3\\
\tilde{\psi}_4\\
\tilde{\psi}_5\\
\tilde{\psi}_6\\
\tilde{\psi}_7\\
\tilde{\psi}_8\\
\tilde{\psi}_9
\end{pmatrix}=\frac{1}{3}
\begin{pmatrix}
1 & 1 & 1 & 1 & 1 & 1 & 1 & 1 & 1 \\
\xi_3 & \xi_3^2 & 1 & 1 & \xi_3 & \xi_3^2 & \xi_3^2 & 1 & \xi_3 \\
\xi_3^2 & \xi_3 & 1 & 1 & \xi_3^2 & \xi_3 & \xi_3 & 1 & \xi_3^2 \\
\xi_3^2 & \xi_3 & 1 & \xi_3^2 & \xi_3 & 1 & \xi_3^2 & \xi_3 & 1 \\
\xi_3 & \xi_3^2 & 1 & \xi_3 & \xi_3^2 & 1 & \xi_3 & \xi_3^2 & 1 \\
\xi_3 & 1 & \xi_3^2 & 1 & \xi_3^2 & \xi_3 & \xi_3^2 & \xi_3 & 1 \\
\xi_3^2 & 1 & \xi_3 & 1 & \xi_3 & \xi_3^2 & \xi_3 & \xi_3^2 & 1 \\
\xi_3^2 & \xi_3^2 & \xi_3^2 & \xi_3 & \xi_3 & \xi_3 & 1 & 1 & 1 \\
\xi_3 & \xi_3 & \xi_3 & \xi_3^2 & \xi_3^2 & \xi_3^2 & 1 & 1 & 1 
\end{pmatrix} 
\begin{pmatrix}
\psi_1\\
\psi_2\\
\psi_3\\
\psi_4\\
\psi_5\\
\psi_6\\
\psi_7\\
\psi_8\\
\psi_9
\end{pmatrix}
\ee
where $\xi_n=e^{\frac{2\pi i}{n}}$ are the $n$th roots of unity\footnote{The exponents are always taken mod $2\pi$ such that $|\arg(\xi_n^j)|\leq\pi$. This ensures that the resulting twist operators will be of lowest conformal weight.}. This is the basis for which all three matrices $T_g$ are simultaneously diagonalized.

Next we bosonize the fermions by introducing boson fields $\phi_i$ and taking $\tilde{\psi}_i=e^{i\phi_i}$.
The twist operators are then determined by demanding the correct monodromies with the fermion fields. This accomplished by taking a product of vertex operators where the charges are directly determined by the eigenvalues of the corresponding permutation matrix $T_{g}$\footnote{For Abelian groups of odd order this procedure generalizes. In particular because the group is Abelian the matrices $T_g$ can be simultaneously diagonalized. The correct diagonal basis is found by taking a discrete group-valued Fourier transformation \cite{Wikipedia_2023} of the fermion field which generalizes the transformation \eqref{eq:FermionDiscreteFT}. The eigenvalues of $T_g$ are the characters of the group element $g$. As such the group elements of the twist operators completely fix the charges of the resulting vertex operators in terms of the characters. More care need to be taken when considering even $n$ as there are additional subtleties in the definitions of the mondromies of the fermion fields (see e.g. the main text \eqref{monogr}). Even still similar issues of charge conservation can generically arise especially in the case of odd $q$.}. Let
\be
T_g\tilde{\psi}_i=\lambda_{gi}\tilde{\psi}_i
\ee
then
\be
\sigma_g = \prod_{l=1}^{n}e^{ i q_{gl}\phi_l}, \quad q_{gl}= \frac{1}{2\pi i}\log{\lambda_{gl}}.
\ee
Since the boson fields are commuting the calculation of the three point function of twist operators factorizes into a number of separate three-point functions of vertex operators \cite{Yellow}
\be
V_{q}(x)\coloneqq e^{iq\phi(x)}, \quad \langle V_i(x_i)\cdots V_k(x_k)\rangle=\delta_{i+j+\cdots k,0}\prod_{i< j}^k|x_i-x_j|^{ij}.
\ee
Defining 
\be
V_{abc}=\langle V_a(x_1)V_b(x_2)V_c(x_3)\rangle
\ee
we find
\be
\langle \sigma_a(x_1)\sigma_{ab}(x_2)\sigma_{ab^2}(x_3)\rangle =V_{000}V_{\frac{1}{3}\frac{1}{3}\frac{1}{3}}V_{-\frac{1}{3}-\frac{1}{3}-\frac{1}{3}}\prod_{S^3}V_{0\frac{1}{3}-\frac{1}{3}}
\ee
where the final product is over the six possible permutations of the indices.
In particular $n$-point functions of vertex operators are non-zero only if they are charge conserving. Thus the presence of the terms $V_{\frac{1}{3}\frac{1}{3}\frac{1}{3}}$ and $V_{-\frac{1}{3}-\frac{1}{3}-\frac{1}{3}}$ which are not charge conserving implies the entire 3-point function is zero which is incorrect.

Supposing that these non-charge conserving terms can be regulated or removed we can proceed to calculate
\be
V_{000}\prod_{S^3}V_{0\frac{1}{3}-\frac{1}{3}}=\left(|x_1-x_2||x_2-x_3||x_3-x_1|\right)^{-\frac{2}{9}}
\ee
which predicts the twist operators to be of conformal dimension $\frac{2}{9}$. This should be compared with the actual value of
\be
h=\frac{c}{24}(n^2-1)=\frac{1}{3}
\ee
As such the standard procedure of defining the twist operators via bosonization fails.

\section{Details of Multi-entropy Calculations with Local Operator Excitation}\label{detailLQ}

Below we show details of replica computations of multi-entropy with local excitation whose results were presented in section (\ref{sec:LQ}). We again set $x_0=-1$, $x_1=0$, $x_2=1$ and $x_3=2$ for simplicity. 

 In the $z$-coordinate, the eight points (\ref{wppp}) on the $w$-sheet are given by 
\begin{align}
    z_{(\pm,\pm,+)} =&~ \pm \sqrt{ \frac{-1\pm \sqrt{1-(-2+t+i\epsilon)^2}}
    {-2+t+i\epsilon} } = \pm \sqrt{ \frac{-1\pm \sqrt{(3-t)(t-1)-2i\epsilon(-2+t)}}
    {-2+t+i\epsilon} }, \nonumber\\
    \Bar{z}_{(\pm,\pm,+)}=&~ \pm \sqrt{ \frac{-1\pm \sqrt{1-(-2-t-i\epsilon)^2}}
    {-2-t-i\epsilon} } = \pm \sqrt{ \frac{-1\pm \sqrt{(3+t)(-t-1)+2i\epsilon(-2-t)}}
    {-2-t-i\epsilon} }, \nonumber\\
    z_{(\pm,\pm,-)} =&~ \pm \sqrt{ \frac{-1\pm \sqrt{1-(-2+t-i\epsilon)^2}}
    {-2+t-i\epsilon} } = \pm \sqrt{ \frac{-1\pm \sqrt{(3-t)(t-1)+2i\epsilon(-2+t)}}
    {-2+t-i\epsilon} }, \nonumber\\
    \Bar{z}_{(\pm,\pm,-)}=&~ \pm \sqrt{ \frac{-1\pm \sqrt{1-(-2-t+i\epsilon)^2}}
    {-2-t+i\epsilon} } = \pm \sqrt{ \frac{-1\pm \sqrt{(3+t)(-t-1)-2i\epsilon(-2-t)}}
    {-2-t+i\epsilon} }.
\end{align}
To be precise, for example,
\begin{align}
    z_{(+,-,+)} =&~ + \sqrt{ \frac{-1- \sqrt{1-(-2+t+i\epsilon)^2}}
    {-2+t+i\epsilon} } = + \sqrt{ \frac{-1- \sqrt{(3-t)(t-1)-2i\epsilon(-2+t)}}
    {-2+t+i\epsilon} }. \nonumber
\end{align}
In the $t \to 0$ limit $\Bar{z}_{(\pm,\pm,\pm)}$ is exactly the complex conjugate of $z_{(\pm,\pm,\pm)}$. When taking $t$ to be large, we need to choose the appropriate branch such that $z_{(\pm,\pm,\pm)}$s change continuously.
The differentiation of $z$ is 
\begin{align}
    \frac{dz}{dw}=&~\frac{(z^4+1)^2}{4z(z^4-1)}, \nonumber\\
    \frac{d\Bar{z}}{d\Bar{w}}=&~\frac{(\Bar{z}^4+1)^2}{4\Bar{z}(\Bar{z}^4-1)}.
\end{align}
It is straightforward to calculate the eight-point function
\begin{align}
    \prod _{(\pm,\pm,\pm)} 
    \left[
    \left(\left.\frac{dz}{dw}\right|_{z=z_{(\cdot,\cdot,\cdot)}}\right)^{1/8}
    \left(\left.\frac{d\Bar{z}}{d\Bar{w}}\right|_{\Bar{z}=\Bar{z}_{(\cdot,\cdot,\cdot)}}\right)^{1/8}
    \right]
    \left\langle \prod _{(\pm,\pm,\pm)} \mathcal{O}\left(z_{(\cdot,\cdot,\cdot)},\Bar{z}_{(\cdot,\cdot,\cdot)}\right) 
    \right\rangle_z .
\end{align}

\subsection*{Careful analysis of $\ep\to 0$ limit}
\subsubsection*{Anti-holomorphic part}
By taking $\epsilon \to 0$ limit, we get
\begin{align}
    \Bar{z}_{(+,+,+)}=&~
    \frac{+i}{\sqrt{2+t}}\sqrt{-1-i\sqrt{(3+t)(t+1)}}\left(1-\frac{\epsilon}{2(2+t)\sqrt{(3+t)(t+1)}}\right), \nonumber\\
    \Bar{z}_{(+,+,-)}=&~
    \frac{-i}{\sqrt{2+t}}\sqrt{-1+i\sqrt{(3+t)(t+1)}}\left(1-\frac{\epsilon}{2(2+t)\sqrt{(3+t)(t+1)}}\right), \nonumber\\
    \Bar{z}_{(+,-,+)}=&~
    \frac{+i}{\sqrt{2+t}}\sqrt{-1+i\sqrt{(3+t)(t+1)}}\left(1+\frac{\epsilon}{2(2+t)\sqrt{(3+t)(t+1)}}\right), \nonumber\\
    \Bar{z}_{(+,-,-)}=&~
    \frac{-i}{\sqrt{2+t}}\sqrt{-1-i\sqrt{(3+t)(t+1)}}\left(1+\frac{\epsilon}{2(2+t)\sqrt{(3+t)(t+1)}}\right).
\end{align}
Here, $\Bar{z}_{(-,\cdot,\cdot)}$s are just $-1$ times $\Bar{z}_{(+,\cdot,\cdot)}$. These results are obtained by picking up the correct branch of square root by smoothly following the time evolution from $t=0$. The pair of two points whose distance is of the order of $\epsilon$ is as follows:
\begin{align}
    (+,+,+)\leftrightarrow &~(-,-,-),& 
    (+,+,-)\leftrightarrow &~(-,-,+),\nonumber\\
    (+,-,+)\leftrightarrow &~(-,+,-),& 
    (+,-,-)\leftrightarrow &~(-,+,+).
\end{align}

\subsubsection*{Holomorphic part: $0<t<1$ case}
The holomorphic part has several case divisions depending on the value of $t$. In the case of $0<t<1$, we have
\begin{align}
    z_{(+,+,+)}=&~
    \frac{-i}{\sqrt{2-t}}\sqrt{-1+i\sqrt{(3-t)(1-t)}}\left(1-\frac{\epsilon}{2(2-t)\sqrt{(3-t)(1-t)}}\right), \nonumber\\
    z_{(+,+,-)}=&~
    \frac{+i}{\sqrt{2-t}}\sqrt{-1-i\sqrt{(3-t)(1-t)}}\left(1-\frac{\epsilon}{2(2-t)\sqrt{(3-t)(1-t)}}\right), \nonumber\\
    z_{(+,-,+)}=&~
    \frac{-i}{\sqrt{2-t}}\sqrt{-1-i\sqrt{(3-t)(1-t)}}\left(1+\frac{\epsilon}{2(2-t)\sqrt{(3-t)(1-t)}}\right), \nonumber\\
    z_{(+,-,-)}=&~
    \frac{+i}{\sqrt{2-t}}\sqrt{-1+i\sqrt{(3-t)(1-t)}}\left(1+\frac{\epsilon}{2(2-t)\sqrt{(3-t)(1-t)}}\right).
\end{align}
Here, $z_{(-,\cdot,\cdot)}$s are just $-1$ times $z_{(+,\cdot,\cdot)}$. These results are obtained by picking up the correct branch of square root by smoothly following the time evolution from $t=0$. The pair of two points whose distance is of the order of $\epsilon$ is the same as anti-holomorphic case, as
\begin{align}
    (+,+,+)\leftrightarrow &~(-,-,-),& 
    (+,+,-)\leftrightarrow &~(-,-,+),\nonumber\\
    (+,-,+)\leftrightarrow &~(-,+,-),& 
    (+,-,-)\leftrightarrow &~(-,+,+).
\end{align}

\subsubsection*{Holomorphic part: $1<t<2$ case}
In the case of $1<t<2$, we have
\begin{align}
    z_{(+,+,+)}=&~
    \frac{1}{\sqrt{2-t}}\sqrt{1-\sqrt{(3-t)(t-1)}}\left(1-\frac{i\epsilon}{2(2-t)\sqrt{(3-t)(1-t)}}\right), \nonumber\\
    z_{(+,+,-)}=&~
    \frac{1}{\sqrt{2-t}}\sqrt{1-\sqrt{(3-t)(t-1)}}\left(1+\frac{i\epsilon}{2(2-t)\sqrt{(3-t)(1-t)}}\right) \nonumber\\
    z_{(+,-,+)}=&~
    \frac{-1}{\sqrt{2-t}}\sqrt{1+\sqrt{(3-t)(t-1)}}\left(1+\frac{i\epsilon}{2(2-t)\sqrt{(3-t)(t-1)}}\right), \nonumber\\
    z_{(+,-,-)}=&~
    \frac{-1}{\sqrt{2-t}}\sqrt{1+\sqrt{(3-t)(t-1)}}\left(1-\frac{i\epsilon}{2(2-t)\sqrt{(3-t)(t-1)}}\right).
\end{align}
Here, $z_{(-,\cdot,\cdot)}$s are just $-1$ times $z_{(+,\cdot,\cdot)}$ as before. These results are obtained by picking up the correct branch of square root by smoothly following the time evolution from $t=0$. The pair of two points whose distance is of the order of $\epsilon$ as follows:
\begin{align}
    (+,+,+)\leftrightarrow &~(+,+,-),& 
    (+,-,+)\leftrightarrow &~(+,-,-),\nonumber\\
    (-,+,+)\leftrightarrow &~(-,+,-),& 
    (-,-,+)\leftrightarrow &~(-,-,-).
\end{align}

\subsubsection*{Holomorphic part: $2<t<3$ case}
In the case of $2<t<3$, we have
\begin{align}
    z_{(+,+,+)}=&~
    +\frac{-i}{\sqrt{t-2}}\sqrt{1-\sqrt{(3-t)(t-1)}}\left(1+\frac{i\epsilon}{2(t-2)\sqrt{(3-t)(t-1)}}\right), \nonumber\\
    z_{(+,+,-)}=&~
    +\frac{+i}{\sqrt{t-2}}\sqrt{1-\sqrt{(3-t)(t-1)}}\left(1-\frac{i\epsilon}{2(t-2)\sqrt{(3-t)(t-1)}}\right) \nonumber\\
    z_{(+,-,+)}=&~
    -\frac{+i}{\sqrt{t-2}}\sqrt{1+\sqrt{(3-t)(t-1)}}\left(1-\frac{i\epsilon}{2(t-2)\sqrt{(3-t)(t-1)}}\right), \nonumber\\
    z_{(+,-,-)}=&~
    -\frac{-i}{\sqrt{t-2}}\sqrt{1+\sqrt{(3-t)(t-1)}}\left(1+\frac{i\epsilon}{2(t-2)\sqrt{(3-t)(t-1)}}\right).
\end{align}
Here, $z_{(-,\cdot,\cdot)}$s are just $-1$ times $z_{(+,\cdot,\cdot)}$ as before. These results are obtained by picking up the branch of square root as continuous to the $1<t<2$ case. Here, $z_{(\cdot,-,\cdot)}$s have the overall sign opposite to the $(-\pi,\pi)$ branch. The pair of two points whose distance is of the order of $\epsilon$ as follows:
\begin{align}
    (+,+,+)\leftrightarrow &~(-,+,-),& 
    (+,+,-)\leftrightarrow &~(-,+,+),\nonumber\\
    (+,-,+)\leftrightarrow &~(-,-,-),& 
    (+,-,-)\leftrightarrow &~(-,-,+).
\end{align}

\subsubsection*{Holomorphic part: $3<t$ case}
In the case of $3<t$, we have
\begin{align}
    z_{(+,+,+)}=&~
    +\frac{1}{\sqrt{t-2}}\sqrt{-1-i\sqrt{(t-3)(t-1)}}\left(1-\frac{\epsilon}{2(t-2)\sqrt{(t-3)(t-1)}}\right), \nonumber\\
    z_{(+,+,-)}=&~
    +\frac{1}{\sqrt{t-2}}\sqrt{-1+i\sqrt{(t-3)(t-1)}}\left(1-\frac{\epsilon}{2(t-2)\sqrt{(t-3)(t-1)}}\right) \nonumber\\
    z_{(+,-,+)}=&~
    -\frac{1}{\sqrt{t-2}}\sqrt{-1+i\sqrt{(t-3)(t-1)}}\left(1+\frac{\epsilon}{2(t-2)\sqrt{(t-3)(t-1)}}\right), \nonumber\\
    z_{(+,-,-)}=&~
    -\frac{1}{\sqrt{t-2}}\sqrt{-1-i\sqrt{(t-3)(t-1)}}\left(1+\frac{\epsilon}{2(t-2)\sqrt{(t-3)(t-1)}}\right).
\end{align}
Here, $z_{(-,\cdot,\cdot)}$s are just $-1$ times $z_{(+,\cdot,\cdot)}$ as before. These results are obtained by picking up the branch of square root as continuous to the $2<t<3$ case. Here, $z_{(\cdot,-,\cdot)}$s have the overall sign opposite to the $(-\pi,\pi)$ branch. The pair of two points whose distance is of the order of $\epsilon$ the same as anti-holomorphic case, as
\begin{align}
    (+,+,+)\leftrightarrow &~(-,-,-),& 
    (+,+,-)\leftrightarrow &~(-,-,+),\nonumber\\
    (+,-,+)\leftrightarrow &~(-,+,-),& 
    (+,-,-)\leftrightarrow &~(-,+,+).
\end{align}
Finally the evaluation of eight-point function (\ref{eightp}) is straightforward with the above result of the location of operators in the $\ep\to 0$ limit, employing the standard Wick contractions in free scalar field theory. This leads to the final result in subsection (\ref{finalrf}).

\end{appendix}

\bibliography{main.bib}

\nolinenumbers

\end{document}